\documentclass[preprint%
]{aastex63}
\usepackage{color}      
\usepackage{soul,xcolor}
\newcommand{\add}[1]{{#1}}

\begin{document}
\setstcolor{red}

\title{
Shear-Driven Transition to Isotropically Turbulent Solar Wind Outside the 
Alfv\'en Critical Zone}

\author[0000-0003-3414-9666]{D. Ruffolo}
\email{david.ruf@mahidol.ac.th}
\affiliation{Department of Physics, Faculty of Science, Mahidol University, Bangkok 10400, Thailand}

\author[0000-0001-7224-6024]{W.~H. Matthaeus}
\affiliation{Department of Physics and Astronomy and Bartol Research Institute, University of Delaware, Newark, DE 19716, USA}

\author[0000-0002-7174-6948]{R. Chhiber}
\affiliation{Department of Physics and Astronomy and Bartol Research Institute, University of Delaware, Newark, DE 19716, USA}
\affiliation{Heliophysics Science Division, NASA Goddard Space Flight Center, Greenbelt MD 20771, USA}

\author[0000-0002-0209-152X]{A.~V.~Usmanov}
\affiliation{Department of Physics and Astronomy and Bartol Research Institute, University of Delaware, Newark, DE 19716, USA}
\affiliation{Heliophysics Science Division, NASA Goddard Space Flight Center, Greenbelt MD 20771, USA} 

\author[0000-0003-2965-7906]{Y. Yang
}
\affiliation{\add{Department of Mechanics and Aerospace Engineering,} Southern University of Science and Technology, Shenzhen, Guangdong 518055, People's Republic of China}

\author[0000-0002-6962-0959]{R. Bandyopadhyay}
\affiliation{Department of Astrophysical Sciences, Princeton University, Princeton, NJ 08544, USA}

\author[0000-0003-0602-8381]{T.~N. Parashar}
\affiliation{Department of Physics and Astronomy and Bartol Research Institute, University of Delaware, Newark, DE 19716, USA}
\affiliation{School of Chemical and Physical Sciences, Victoria University of Wellington, Wellington 6012, New Zealand}

\author[0000-0002-5317-988X]{M. L. Goldstein}
\affiliation{Goddard Planetary Heliophysics Institute, University of Maryland Baltimore County, Baltimore, MD 21250, USA}

\author[0000-0002-7164-2786]{C. E. DeForest}
\affiliation{Southwest Research Institute, 1050 Walnut Street, Suite 300, Boulder, CO 80302, USA}

\author[0000-0003-2965-7906]{M. Wan
}
\affiliation{\add{Department of Mechanics and Aerospace Engineering,} Southern University of Science and Technology, Shenzhen, Guangdong 518055, People's Republic of China}


\author[0000-0001-8478-5797]{A. Chasapis}
\affiliation{Laboratory for Atmospheric and Space Physics, University of Colorado Boulder, Boulder, CO 80303, USA}

\author[0000-0002-2229-5618]{B.~A. Maruca}
\affiliation{Department of Physics and Astronomy and Bartol Research Institute, University of Delaware, Newark, DE 19716, USA}










\author[0000-0002-2381-3106]{M. Velli}
\affiliation{Department of Earth, Planetary, and Space Sciences, University of California, Los Angeles, CA 90095, USA}


\author[0000-0002-7077-930X]{J.~C. Kasper}
\affiliation{Climate and Space Sciences and Engineering, University of Michigan, Ann Arbor, MI 48109, USA}
\affiliation{Smithsonian Astrophysical Observatory, Cambridge, MA 02138 USA}








\begin{abstract}
Motivated by  prior remote  observations  of  a  transition  from  striated solar coronal structures  to  more  isotropic  ``flocculated''  fluctuations,  we propose that the dynamics of the inner solar wind just outside the Alfv\'en critical zone, and in the vicinity of the first $\beta=1$ surface, is powered by the relative velocities of adjacent coronal magnetic flux tubes.  
We suggest that large amplitude flow contrasts are magnetically constrained at lower altitude but shear-driven dynamics are triggered as such constraints are released above the Alfv\'en critical zone, as suggested by global magnetohydrodynamic (MHD) simulations that include self-consistent turbulence transport.  
We argue that this dynamical evolution accounts for features observed by {\it Parker Solar Probe} ({\it PSP}) near initial perihelia, including magnetic ``switchbacks'', and large transverse velocities that are partially corotational and saturate near the local Alfv\'en speed.  
Large-scale magnetic increments are more longitudinal than latitudinal, a state unlikely to originate in or below the lower corona. We attribute this to preferentially longitudinal velocity shear from varying degrees of corotation.  
Supporting evidence includes comparison with a high Mach number three-dimensional compressible MHD simulation of nonlinear shear-driven turbulence, reproducing several observed diagnostics, including characteristic  distributions  of  fluctuations that are qualitatively similar to {\it PSP} observations near the first perihelion.  
The concurrence of evidence from remote sensing observations, {\it in situ} measurements, and both global and local simulations supports the idea that the dynamics just above the Alfv\'en critical zone boost low-frequency plasma turbulence to the level routinely observed throughout the explored solar system.
\end{abstract}

\section{Introduction}
The solar atmosphere originates 
in the highly dynamic photosphere 
and expands outward, generating the magnetically dominated corona.
The outward acceleration eventually causes the velocity to exceed the local Alfv\'en speed, and in that super-Alfv\'enic regime, embedded magnetic fluctuations will only propagate outward.
Consequently, a feature distinguishing the inner
corona from the super-Alfv\'enic solar wind is that 
magnetohydrodynamic (MHD) fluctuations 
can propagate both upwards and downwards in the inner corona, but in the 
solar wind such signals cannot propagate back into the corona. 
The hypothetical boundaries between these regions are the Alfv\'en critical surface, where the Alfv\'en speed of magnetic fluctuations equals the solar wind speed,
and the sonic critical surface, where the speed of sound equals the solar wind speed.
In view of the highly 
dynamic, or turbulent, nature of both the solar wind and corona, these boundaries are  
almost certainly 
better described as critical {\it zones} \citep{DeForestEA18}. 
For many years there has been discussion and speculation regarding what happens near and at these zones. 
In the simplest picture (the surface version) Alfv\'en or sound waves can propagate only outward at the surface.
\add{Downward propagating fluctuations below the Alfv\'en surface cannot reach the solar wind.
Downward propagating fluctuations above the Alfv\'en critical zone in the solar wind cannot propagate back into the sub-Alfv\'enic corona.}
It would not be unreasonable to imagine 
that with such stagnation of downward-moving fluctuations and their interaction with upward-moving fluctuations, turbulence levels build up in the critical zone, a possibility that has also been suggested based on remote sensing observations \citep{LotovaEA85,LotovaEA11}. 
This reasoning has also long been offered  as 
explaining why the inner solar wind is dominated by a broad band spectrum of {\it outward traveling} waves \citep{BelcherDavis71}.
This ``Alfv\'enic'' property of the fluctuations is characteristic of
the inner heliosphere, where it
forms a power-law inertial range observed from the correlation scale to the ion inertial scale \citep{BavassanoEA82a,
BrunoCarbone13}. 
The corona is presumed to be turbulent  and the solar wind is observed to be turbulent from the distance of 
the latest {\it Parker Solar Probe (PSP)}~\citep{FoxSSR16} perihelia measurements out to the boundary of the heliosphere beyond 100 au.
But how this turbulence changes
in character across the critical zones is not well understood. 
Furthermore,  the 
nature of the transitions 
in all plasma properties 
from coronal to solar wind conditions
remains to be discovered.
Fortunately, {\it PSP} 
data are revealing these properties 
at progressively lower altitudes, 
and more information will soon be forthcoming
from the recently launched
{\it Solar Orbiter}~\citep{Muller2013SP} mission. Together, these two missions are 
expected to unravel
many mysteries of the inner solar wind and outer corona, including the 
issues we investigate here.

Initial results from {\it PSP} have revealed magnetic reversals and velocity 
spikes \citep{BaleEA19Nature,KasperEA19Nature,DudokDeWitEA20} similar to previous observations
at 0.3 AU and beyond \citep{Michel_1967,KahlerEA96,BaloghEA99,Crooker_2004,Borovsky16,HorburyEA18,LockwoodEA19}.
One explanation 
is that the reversals arise from outward propagation of large amplitude 
remnants of magnetic reconnection that occurred at lower altitudes
\citep{AxfordMcKenzie92,AxfordEA99,SamantaEA19,FiskKasper20}. 
Such a mechanism is plausible and difficult to rule out. 
However, another possibility is that the reversals reflect an onset of strong shear-driven 
turbulence that began just outside the Alfv\'en critical zone where the solar wind 
speed first exceeded the Alfv\'en speed. Such shears could produce magnetic reversals through 
large-scale perturbations of the flow.
For example, such perturbations could result from excitation of the Kelvin-Helmholtz instability \citep{MalagoliEA96}.

This scenario is consistent with a suite of observable effects already apparent in 
imaging \citep{DeForestEA16} and {\it in situ} datasets \citep{Borovsky16,HorburyEA18}. 
In particular, \citet{DeForestEA16} interpreted the transition from elongated striae to 
relatively isotropic flocculae as a signature of the onset of shear-driven turbulent activity 
some 20-80 solar radii from the photosphere, where the magnetic field ceases to be a 
dominant constraint on transverse motions. 
In the present work, following \citet{DeForestEA16}, we refer to this process as flocculation.
This interpretation is supported by results from turbulence-driven global simulations of the solar wind
\citep{ChhiberEA18-global-floc}. The presence of velocity shears is also strongly suggested by coronal imaging at lower altitudes \citep{DeForestEA18}. 

Here we use 
{\it PSP} observations in its first two orbits, 
along with supporting simulations, 
to examine the character of the plasma 
dynamics in the solar wind at heliocentric 
distance $r$ as low as 36 $R_\odot$ (0.17 au), along with 
evidence that {\it PSP} approached 
the Alfv\'en critical zone.
Our presentation incorporates 
global heliospheric MHD simulations,
local three-dimensional (3D) compressible MHD simulation, 
and {\it PSP} observations.
The goals are to understand the 
region where flocculation is believed to start, 
to identify signatures of the process of flocculation in {\it PSP} data, and to 
evaluate the hypothesis 
that the transition from striation to flocculation is
a consequence of 
velocity shears. We are therefore led to consider physics 
related to nonlinear shear instabilities, essentially nonlinear
Kelvin-Helmholtz dynamics or mixing layer dynamics, 
appropriately generalized to an MHD or plasma environment.
While we cannot definitively 
resolve these questions based on the 
current observational evidence,
global simulation, and supporting 
simulation of local physics, we nonetheless are able to conclude that the available evidence is consistent with
\add{our hypothesis
of} the role of velocity shear in the inner solar wind.

\begin{figure}
\begin{centering}
\includegraphics[width=.7\columnwidth]{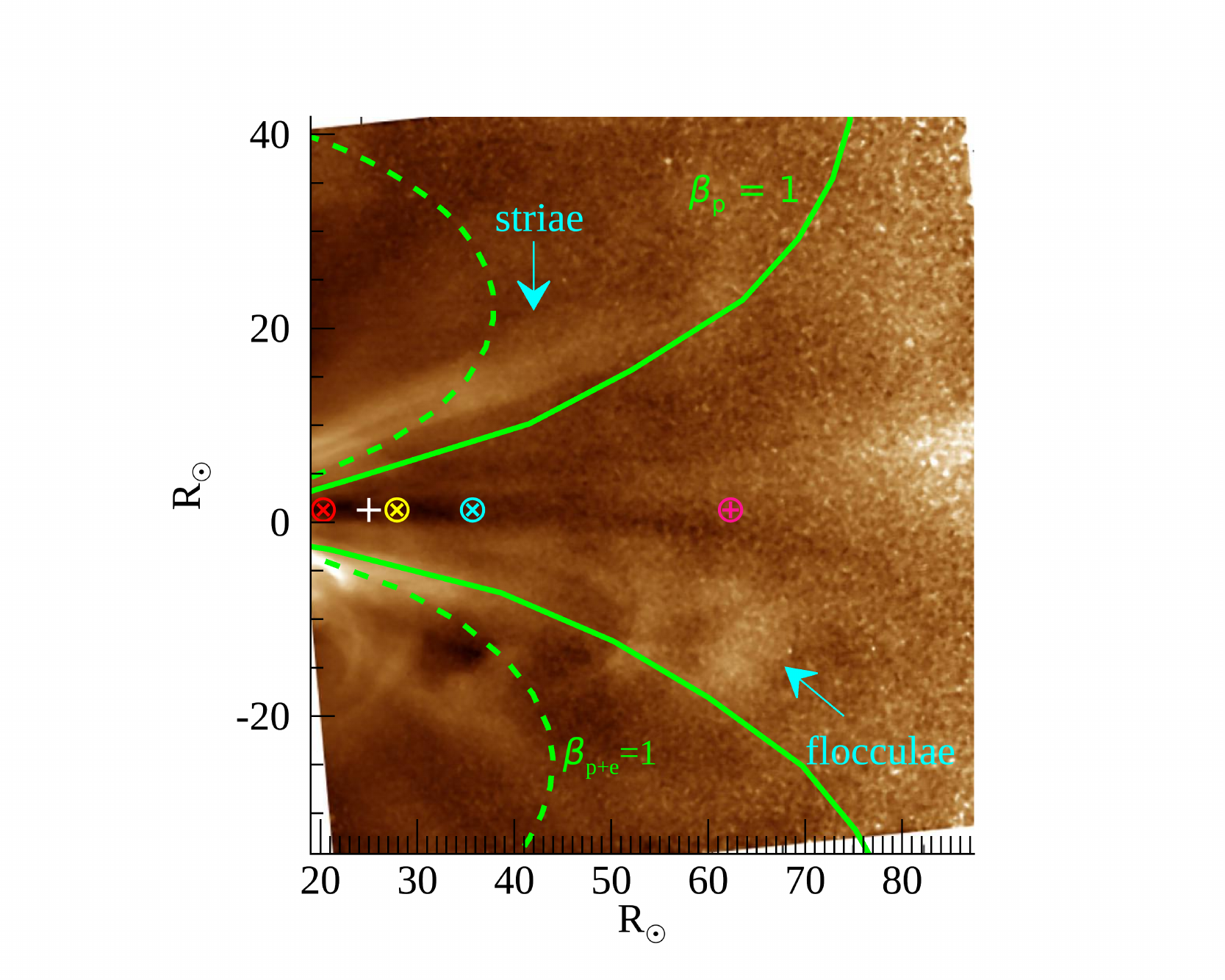}
        \caption{{\it STEREO} Heliospheric Imager snapshot from 16 December 2008, 
        analyzed as in 
        \citet{DeForestEA16} and \citet{ChhiberEA18-global-floc},
showing the transition with radial evolution from a highly anisotropic striated conformation closer to the Sun, to a more isotropic flocculated conformation at greater distances. 
        Also indicated are first
        $\beta=1$ surfaces (green curves) from a global heliospheric MHD simulation, based on a magnetogram corresponding to December 2008.
        For this simulation, the Alfv\'en surface lies below 15 \(R_\odot\) \citep[see][]{ChhiberEA18-global-floc}.
From right to left, circled symbols indicate the {\it Helios}
perihelion ($\oplus$) and the first, fourth, and sixth {\it PSP} perihelia ($\otimes$).
The white plus sign shows
the location of enhanced turbulence inferred by \citet{LotovaEA85}.
        In the present paper we 
        develop the hypothesis that this transition is fueled
        by nonlinear shear instability outside the Alfv\'en critical zone. 
    \label{fig:floc}
    }
    \end{centering}
\end{figure}

\section{Motivation and Hypothesis}

\subsection{Heliospheric Imaging from STEREO\label{stereo}}

\citet{DeForestEA16}
analyzed image sequences recorded by the 
inner Heliospheric Imager instrument
on board the {\it Solar-Terrestrial Relations Observatory} ({\it STEREO}/HI1) in December, 2008. 
The analysis 
covered 
angular distances of approximately 4$^\circ$ to 24$^\circ$ from the center of the Sun.
An observed systematic transition in the images was noted that 
consisted of 
anomalous fading of the radial striae that characterize the corona, along with an 
anomalous relative
brightening of locally dense 
puffs of solar wind, which were described as 
``flocculae.''
This transition was interpreted as the onset of dynamical activity associated with velocity shear present in the nascent solar wind plasma  coming from near-radial corotating flux tubes in the corona. The flux tubes 
confine the plasma, 
magnetic structures, and fluctuations
that were injected at lower altitudes. 
Moving radially outward, the magnetic field progressively loses control of the plasma, which allows for additional physical processes to dominate, including those that give rise to the striation-flocculation transition.  Significant stages of this transition are indicated by 
passage through regions where the flow speed exceeds the Alfv\'en speed (the Alfv\'en critical zone) and where the mechanical pressure approaches or exceeds the magnetic pressure (first $\beta=1$ zones).

Figure \ref{fig:floc}
illustrates a frame of the \citet{DeForestEA16} analysis, 
in which the transition
from striation to flocculation is clear. The tendency for the structures in the image to become more isotropic with increasing heliocentric distance was quantified by 
\citet{DeForestEA16} by computations of radial and transverse-to-radial second-order structure functions of the 
signal. 
Closer to the Sun, the
striation is due to 
more intense gradients in the transverse direction and weaker radial  
gradients, 
indicated by values of 
transverse structure functions greater than 
radial structure functions at a
given lag. 
Moving outward, the corresponding values of the two structure functions 
become more equal, indicating an 
evolution towards isotropy. 

Figure \ref{fig:floc}
is also annotated with approximate 
equivalent positions of the first, fourth, and sixth {\it PSP} perihelia, and the perihelion of the {\it Helios} mission at 0.29 au. 
Based on global MHD simulations that include turbulence transport \citep{ChhiberEA19-1}, the figure also shows the first surfaces where the plasma beta is unity when considering protons and electrons $(\beta_{p+e}=1)$ and when considering only protons $(\beta_{p}=1)$.
For the simulation considered here, the Alfv\'en surface lies below 15 $R_\odot$, to the left of the view in this image.

\begin{figure}
\plotone{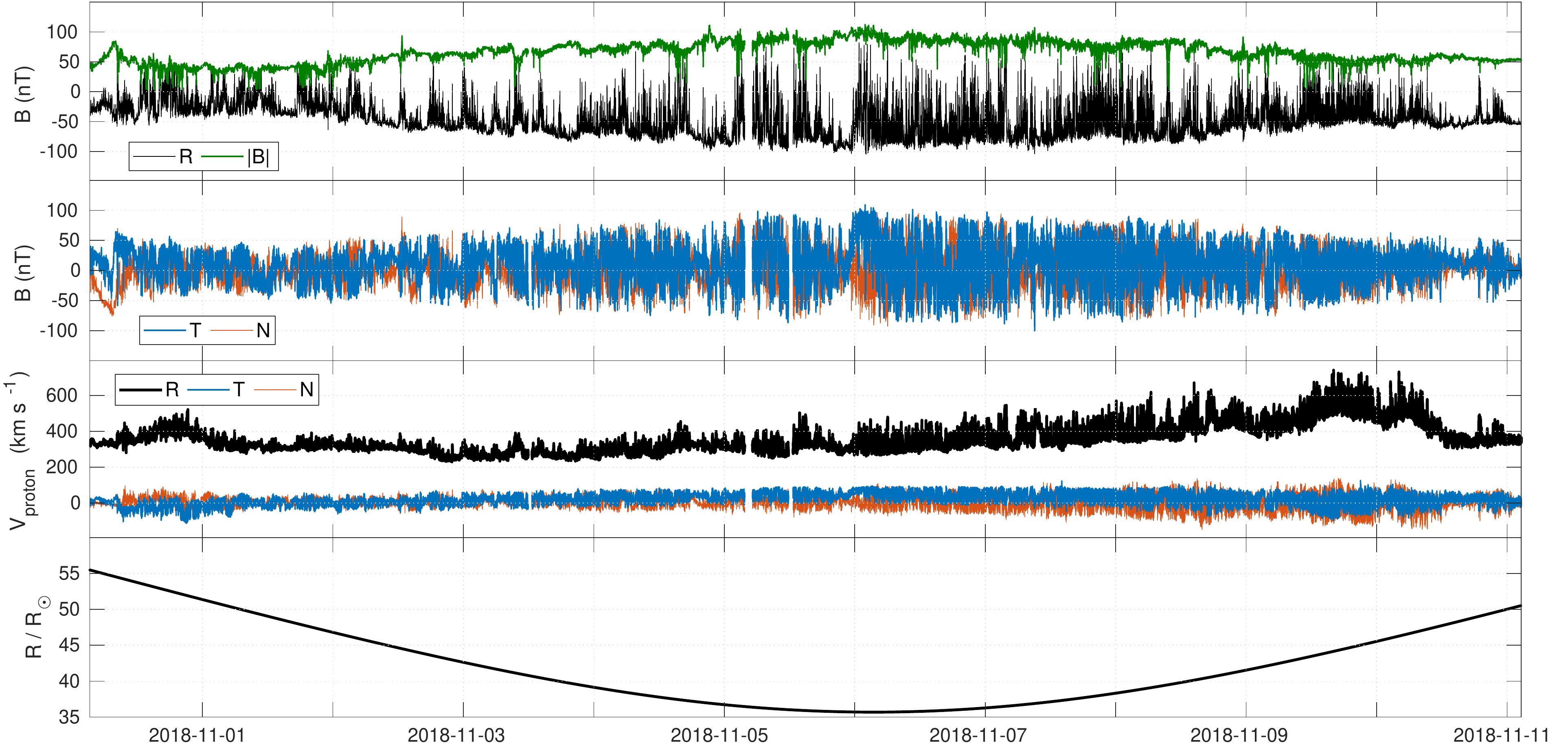}
\caption{Overview of part of the first {\it PSP} encounter. Top panel: Radial component (black) and magnitude (red) of the magnetic field; occurrence of numerous switchbacks (radial component reversals) is evident, although the magnitude exhibits much less variation. Second panel: Tangential (blue) and normal (red) components of the magnetic field. Third panel: Proton velocity components, with the same colors.  The data plotted in (a)-(c) all have a cadence of 1 NYs $\approx$ 0.87 s.  Bottom panel: Radial distance of {\it PSP} from the center of the Sun in units of solar radii.}
\label{fig:stacked}
\end{figure}

\subsection{Parker Solar Probe: Review of Prior Results}
During its first few orbits, the {\it Parker Solar Probe} ({\it PSP}) mission has made pioneering observations in the inner solar wind that bear directly on the questions we explore here. 
{\it PSP} is currently in its 
\add{sixth} orbit and
may be approaching the Alfv\'en 
critical zone\add{, and will thus be} directly examining the region of interest. 
We will delve into the observations in more detail below, but to begin 
the discussion, Figure 
\ref{fig:stacked} illustrates some of the 
important and relevant measurements 
made by {\it PSP} 
from 2018 Oct 31 to Nov 11, an 11-day period surrounding 
its first perihelion at 35.7 solar radii (0.17 au)
on 2018 Nov 6 at 0327 UT.
Cartesian components of magnetic and 
velocity fields and the density are shown
for this period at a cadence of 1 ``New York second'' (NYs)
$\approx 0.87$ s, the fundamental cadence of solar wind velocity measurements. 
These data will be discussed in greater detail below,
and our hypothesis will be evaluated in 
terms of a number of details of these
observations. 
For the moment we wish to call attention to a particular feature 
that has been written about in a number of the early {\it PSP} publications
\citep{BaleEA19Nature,KasperEA19Nature,MozerEA20,DudokDeWitEA20}, 
\add{{\it viz.},} 
the phenomenon of {\it switchbacks},
which has particular relevance to our
proposed model. 
{\it PSP} observations during most time periods
near the first and second perihelia indicated a mean magnetic field that was nearly radially inward.
However, the data 
frequently indicate 
weakening and sometimes reversals (i.e., switchbacks) in 
the radial magnetic field $B_R$. 
The weakening is accompanied by the appearance of substantial transverse components,
i.e., T and N components in the standard spacecraft-centered orthogonal RTN coordinate system,
where +R is radial (antisunward), +T is tangential (toward increasing heliolongitude), and +N is normal (toward increasing heliolatitude). 
Note that vector velocities in this frame are measured relative to the fixed stars, that is, the spacecraft velocity has been subtracted out of the solar wind velocity 
measurement.\footnote{See the \href{http://sweap.cfa.harvard.edu/Data.html}{SWEAP Data User's Guide.}}
These features are apparent in the corresponding panels 
of Figure \ref{fig:stacked} and are discussed further below.  

In what follows we will argue that 
switchbacks and related 
features of the complex dynamics observed by 
{\it PSP} in this region can be explained by {\it in situ} shear-driven dynamics
and are also consistent with the 
striation-flocculation transition 
described in \cite{DeForestEA16}.
Previous arguments in favor of an {\it in situ}  origin of switchbacks and large-amplitude magnetic fluctuations were made by \citet{SquireEA20}, based on expanding-box compressible MHD simulations, and \citet{MacneilEA20}, based on {\it Helios} observations of switchbacks, which were more frequent at greater distance from the Sun.

\subsection{Hypothesis of Shear 
Driving: Cartoon\label{hypothesis}}

Our initial hypothesis is an extrapolation of the ideas in \cite{DeForestEA16}, in which 
the morphological transition 
between striation and 
\add{flocculation} 
that is apparent 
in the STEREO images reflects a transition from largely collimated elongated structures, 
relatively slowly varying in radius, to 
more disordered shapes suggestive of a more isotropic distribution of fluctuations.
This was attributed to an isotropization of turbulence 
as the magnetic field above the Alfv\'en critical zone (and later above a first $\beta=1$ zone) gives 
up much of the control over the plasma that it maintained in the highly magnetized sub-Alfv\'enic corona.

\begin{figure}
\begin{centering}
\includegraphics[width=.75\columnwidth]{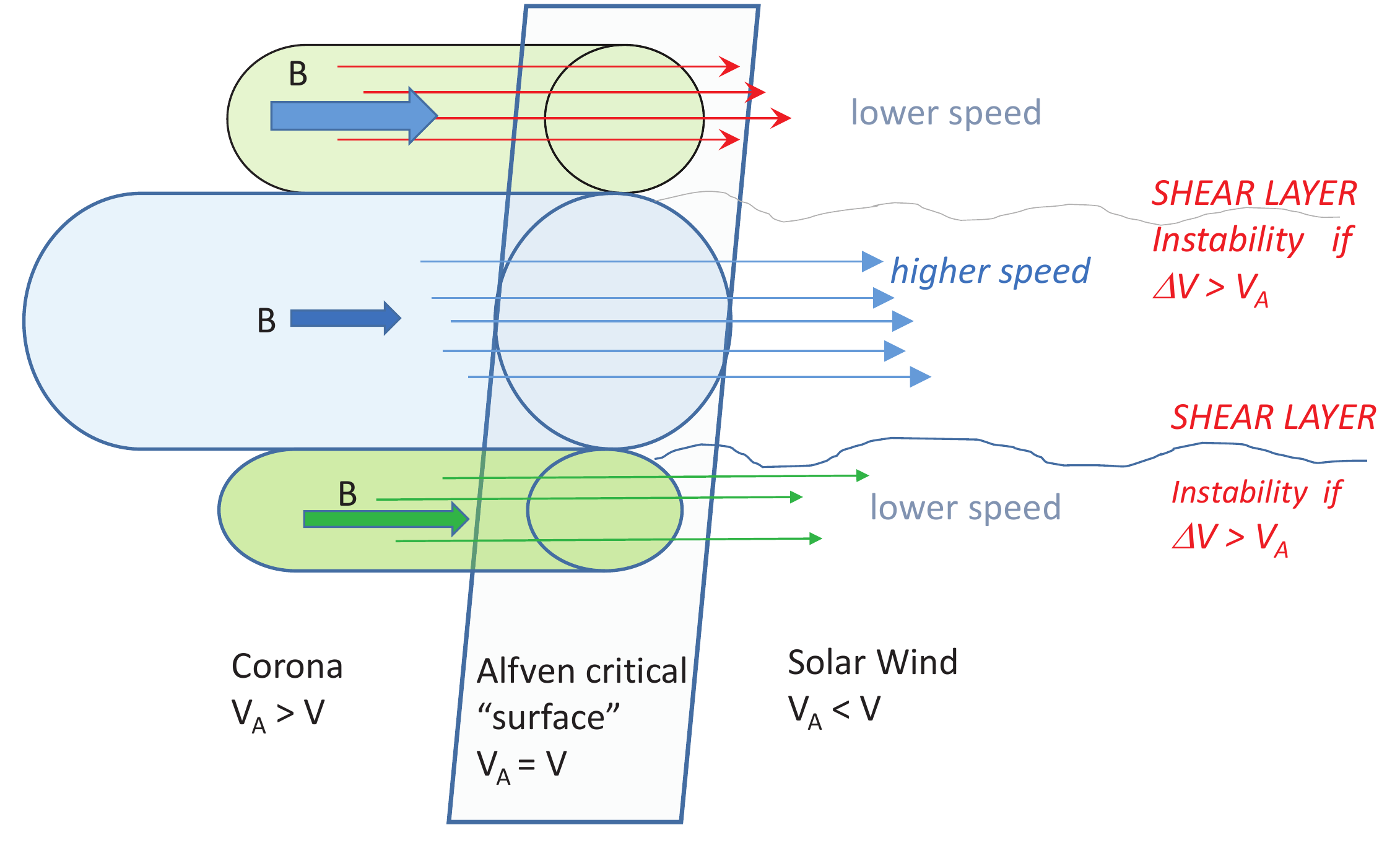}
        \caption{Sketch describing 
        our proposed hypothesis.
        In the corona, the strong magnetic field regulates the dynamics of the nascent solar wind. 
        Each flux tube may contain differing radial speeds and different radial field strengths 
        due to processes at lower altitudes. 
        Beyond the Alfv\'en critical zone the magnetic field is no longer capable of constraining the dynamics and the energy in the velocity contrasts becomes available to drive
        nonlinear magnetized Kelvin-Helmholtz-like dynamics, including  magnetic field amplification and directional change, with associated deflection of velocities into the transverse directions.    
        This may explain the transition
        from striation to flocculation in {\it STEREO} images such as that in Figure \ref{fig:floc},
        and in the present work we point out characteristics of {\it Parker Solar Probe} data that are consistent with this picture of how shear-driven dynamics at and above the Alfv\'en critical zone boosts low frequency turbulence to the levels observed throughout the heliosphere. }
    \label{fig:cartoon}
    \end{centering}
\end{figure}

Here we pursue a particular form of that hypothesis, in which above the Alfv\'en critical surface or zone,
 the velocity differences between adjacent flux tubes may be tapped to supply energy for a more isotropic form of turbulence.
 The physical picture we have in mind involves 
 essentially nonlinear magnetized Kelvin-Helmholtz  dynamics, perhaps better described as a magnetized mixing layer. 
 A highly simplified sketch of the 
scenario we propose is provided in Figure \ref{fig:cartoon}. 

\subsection{Hypothesis of Shear Driving: Expectations and Background}

In examining whether shear-driven 
dynamics are responsible for 
prominent features observed by {\it PSP} near perihelion, we are led to consider
basic physics suggested in 
Figure \ref{fig:cartoon}
that is similar to the classic 
hydrodynamic problem of a mixing layer
\citep{RogersMoser92}.
When two co-linear streams having different velocities come into contact, 
the early part of the dynamics can resemble a linear Kelvin-Helmholtz instability, quickly evolving into a nonlinear mixing layer characterized by vortex rollup.

The addition of a uniform parallel 
magnetic field into the problem presents
the complication that the
transverse displacements needed to produce rollup are inhibited by magnetic field line tension.
The linear theory of stability of 
planar MHD shear layers -- 
a magnetized Kelvin-Helmholtz instability -- was considered by 
\citet{Chandrasekhar},
who stated the
important condition that 
the instability is suppressed when the velocity contrast does not exceed the 
Alfv\'en speed, that is when $\Delta V < V_A$. This instability 
was subsequently considered in greater detail  by \citet{LauLiu80} and \citet{MiuraPritchett82} who refined instability criteria for particular assumptions.
More generally one does not expect that the mixing layer dynamics will be described by linear theory, particularly in cases in which the initial state is not an equilibrium and turbulence, possibly broadband, is present within the velocity streams. 

To examine the nonlinear evolution 
of MHD mixing layer/Kelvin-Helmholtz dynamics, \citet{Miura82} appealed to numerical simulation of compressible MHD. In the nonlinear regime rollup of both 
vortices and magnetic field occurs with a substantial component of transient amplification of magnetic energy. Similar configurations were investigated using an 
incompressible MHD model  
\citep{GoldsteinEA87-eslab,
GoldsteinEA89}
with the goal of understanding  magnetospheric boundary effects and 
reduction of cross helicity (Alfv\'enicity) in the solar wind
\citep{RobertsJGR92}.
Further study using a compressible model
\citep{MalagoliEA96}
revealed additional details of
the rollup process, which 
involves strong coupling between flows and magnetic field structure.  In particular, for cases in which the magnetic field is not strong enough to stabilize the dynamics, the
fully developed state is substantially 
influenced by the presence of the (amplified) magnetic field,
which exhibits distinctive  structure within the vortex
rollups. 

We should note that related studies
\citep{DahlburgEA98,EinaudiEA99}
employed an incompressible MHD model to examine the linear and nonlinear evolution of a 
radial jet confined within a neutral sheet. 
This system, in effect a simplified model of a coronal streamer,
also exhibits Kelvin-Helmholtz-like dynamics at distances large enough that the jet speed exceeds a multiple of the Alfv\'en speed. 
Instability and topological changes in the magnetic field 
at the tip of such a streamer were considered in \citet{RappazzoEA05}.
The effect of shear and expansion on a spectrum of Alfv\'enic fluctuations,
previously examined by incompressible 
simulation \citep{RobertsJGR92}
and in turbulence transport theory \citep{BreechEA08},
was recently considered 
using expanding box 
simulation in the context of {\it Parker Solar Probe} data in \citet{ShiEA20}.

Below we will examine our hypothesis by 
comparison of 
features of 
{\it PSP} observations near 
perihelion to 
computed features of 3D compressible
MHD mixing layer simulations.
We note that similar comparisons
were employed to explain polarity reversals
seen in data from the {\it Ulysses} spacecraft
\citep{LandiEA05,LandiEA06}.

\section{Methods}

\subsection{Observations by the {\it Parker Solar Probe}
\label{pspdata}}

In the present paper we 
make use of publicly available data from 
the first two orbits of {\it Parker Solar Probe}, from two instrument suites, FIELDS \citep{BaleEA16} and SWEAP \citep{KasperEA16}.\footnote{All data were downloaded from https://cdaweb.gsfc.nasa.gov/pub/data/psp/}  
We use Level 2 magnetic field data from FIELDS, which typically have a data rate of 299 Hz, and Level 3 plasma data from SWEAP, with solar wind speed typically available at a cadence of 1 NYs.  
We then usually resample both types of data to either 1-NYs or 1-s cadence, and plasma data from SWEAP are processed to remove spurious spikes. The latter procedure makes use of a time-domain Hampel filter \citep{Davies1993Hampel}, with a filtering interval of 120 seconds and outliers identified as values more than three times larger than the local standard deviation.
We designate the 
inner segments of these orbits (about two months surrounding perihelion) as the first
(solar) encounter (E1),
roughly during
2018 October-November, 
and the second encounter (E2),
roughly during 2019 March-April. 
Observations during E1 were well described in the first results papers 
\citep{BaleEA19Nature,KasperEA19Nature}
and in relevant papers in the special Astrophysical Journal Supplement issue 
\citep[e.g.,][]{DudokDeWitEA20,MozerEA20}.

\subsection{Global Solar Wind Simulations \label{globalsims}}

We employ global 3D MHD modeling to compare 
with several 
features of {\it PSP} observations that 
will be discussed in following sections. 
The fully 3D model
that we employ is 
based on 
mean-field (Reynolds-averaged) solar wind equations, which are solved simultaneously with turbulence transport equations \citep{UsmanovEA18}.
The equations also include electron heat conduction, Coulomb collisions, Reynolds stresses, and electron/proton
heating by a turbulent cascade.
The resolved equations and 
unresolved (subgrid-scale) turbulence equations are solved self-consistently.\footnote{We use Reynolds averaging
based on an ensemble average $\langle \dots\rangle$ \citep{McComb}.
One decomposes 
a variable, such as the 
velocity $\bf u$, as
${\bf u} = \langle {\bf u} \rangle + {\bf u'}$ where
$\langle {\bf u} \rangle$ is the mean and ${\bf u}'$ is the unresolved or fluctuating component. Reynolds averaging 
of nonlinear terms involves contributions from fluctuations 
such as $\langle u'_iu'_j\rangle$, known as the Reynolds stress, which is particularly prominent when fluctuations are incompressible.} 
The turbulence model includes 
three equations: for turbulence energy, normalized cross helicity, and correlation length.

The current computation has evolved from previous work \citep{UsmanovEA14,UsmanovEA16,UsmanovEA18} and is carried out in 
four regions: (1) corona, 1-20 $R_\odot$, (2) inner heliosphere, 20 $R_\odot$-5 au, (3) middle heliosphere, 5-40 au, and (4) outer heliosphere, 40-1200 au.
Boundary conditions are specified at the coronal base (just above the transition region) using ADAPT (Air Force Data Assimilative Photospheric Flux Transport) solar synoptic magnetic field maps
\citep{Arge_et_al_2010, Hickmann_et_al_2015}, in which flux evolution models for the photospheric magnetic field are assimilated with photospheric magnetic field observations. We use the ADAPT map, which is based on the GONG (Global Oscillation Network Group) magnetogram, with the central meridian time 2018 November 6 at 12:00 UTC. The ADAPT map values are scaled by a factor of 2 and are smoothed using a spherical harmonic decomposition up to 15th order.

This global solar wind simulation model has been under 
continual development \citep{UsmanovEA18}
and was recently employed to provide
context for the STEREO observations described in 
\S \ref{stereo} and also 
to generate contextual predictions for the {\it PSP} mission \citep{ChhiberEA19-1, ChhiberEA19-2}.


\subsection{Compressible MHD Simulations of Mixing Layer Dynamics
\label{mhdsims}}

To demonstrate the basic physics associated with 
our hypothesis, we have carried
out a series of nonlinear simulations of fluid-scale turbulence triggered by strong velocity shear in the presence of a moderately strong uniform DC magnetic field.
We solve the compressible MHD equations via a hybrid compact-weighted essentially
non-oscillatory (WENO) scheme, which couples a sixth-order compact scheme for smooth
regions and a 
fifth-order WENO scheme for shock regions. The time marching is performed
by the third-order Runge-Kutta scheme.

The numerical simulations are conducted either 
in two dimensions (2D) 
in a $(2\pi)^2$ domain with $256^2$ resolution, or 
in three dimensions (3D) 
in a $(2\pi)^3$ domain with $256^3$ resolution, all 
with periodic boundary conditions. 
For simplicity, equal viscosity and resistivity are
used, i.e., the magnetic Prandtl number is set equal to unity
and an ideal gas equation of state is adopted. 
While we have varied some initial parameters
to investigate the robustness of our conclusions, 
here we report results from one representative 
3D simulation. 
Initially flow reversal occurred across two thin layers at $y=L_y/4$ and $y=3L_y/4$, where $L_y=2\pi$.  
Initially the velocity and magnetic field are only in the $x$-direction. 
The $x$-direction velocity is given by
\begin{equation}
u_x=U_0\left[1-\tanh{\left(\frac{y-L_y/4}{d}\right)}+\tanh{\left(\frac{y-3L_y/4}{d}\right)}\right],    
\end{equation}
where $U_0=0.27$ and $d=0.025L_y$ is half the thickness of the shear layer.
\add{Similarly, the $x$-direction magnetic field is given by
\begin{equation}
B_x=C_1\left[1-\tanh{\left(\frac{y-L_y/4}{d}\right)}+\tanh{\left(\frac{y-3L_y/4}{d}\right)}\right]+C_2,
\end{equation}
where $C_1=0.05$ and $C_2=0.13$ in our simulation.
Therefore,} the 
velocity streams 
have $u_x= +U_0$ or $u_x = -U_0$,
while the magnetic field in the same stream regions
has $B_x \add{= C_1+C_2}=0.18$ or $B_x \add{= C_2-C_1}=0.08$, respectively, in Alfv\'en speed units.
The magnetic field is initially entirely toward $+x$, stronger in the top and bottom regions and weaker in the middle region.
\add{In this arrangement, qualitatively consistent with the diagram in Figure \ref{fig:cartoon}, the current layers between flux tubes are collocated with the vorticity layers separating streams.}

\begin{table}
\begin{center}
\caption{Initial parameters for shear-driven 3D compressible MHD simulation by methods of \citet{YangEAJCP16}}
\label{tab:parameters}
\begin{tabular}{lc}
$\Delta U/V_A$ & 3.0\\
Plasma $\beta$ & 1\\
Mach number $M_t$ &  3.6 \\
Flow speeds $u_x$ &  $\pm$0.27\\
$B_x$ (strong region) & 0.18\\
$B_x$ (weak region) & 0.08\\
Code resolution& 256$^3$\\
\end{tabular}
\end{center}
\end{table}

Some of the parameters 
\add{of} 
the representative 3D simulation are shown in Table \ref{tab:parameters}.
The simulation initially has uniform density $\rho= 1$
and a velocity difference between streams of
$\Delta U/V_A = 2U_0/V_A= 3$, 
\add{where $V_A$ is derived from the strong field region.}
The plasma beta, i.e., the ratio of plasma pressure to magnetic field pressure, 
is $\beta=1$.
The initial 
turbulent Mach number is $M_t = \Delta U/c_s = 3.6$
for sound speed $c_s$. 
The initial temperature pattern is set to achieve uniform total pressure. 
\add{The value of the polytropic index is $\gamma=1.4$, a value for which extensive testing of the code was carried out 
for high Mach number MHD turbulence
as described by 
\citet{YangEAJCP16}.
Details of the implementation, transport coefficients, and other numerical details are given 
in the same reference.
}

\section{Results: {\it PSP} Observations}

\subsection{
Expectation of High Turbulence Levels Near the
Alfv\'en critical zone}

Global simulations 
of the same type as those 
described in \S \ref{globalsims}
predict \citep{ChhiberEA19-1} 
a high level of turbulence 
in the neighborhood
around and near the Alfv\'en 
critical surface (or zone). 
Physically, this relates to outward traveling Alfv\'enic fluctuations and the stagnation of inward fluctuations as described earlier.
To some degree this may also be anticipated by 
simpler, non-self consistent 
treatments such as
WKB theory \citep{Hollweg74},
turbulence 
transport theory \citep{ZankEA96} 
and, more recently, expanding box simulations \citep{SquireEA20}.
From the self-consistent 
simulations, 
the predicted value of the ratio of magnetic fluctuation intensity to 
the strength of the resolved magnetic field, 
$\delta B/B_0$, is expected 
to be near unity or even above in this zone~\citep{ChhiberEA19-2}.
This is the same region in which 
remote radio observations 
have found enhanced scattering, from which enhanced turbulence levels are inferred 
\citep{LotovaEA11}.

\begin{figure}
\begin{centering}
\includegraphics[width=.85\columnwidth]{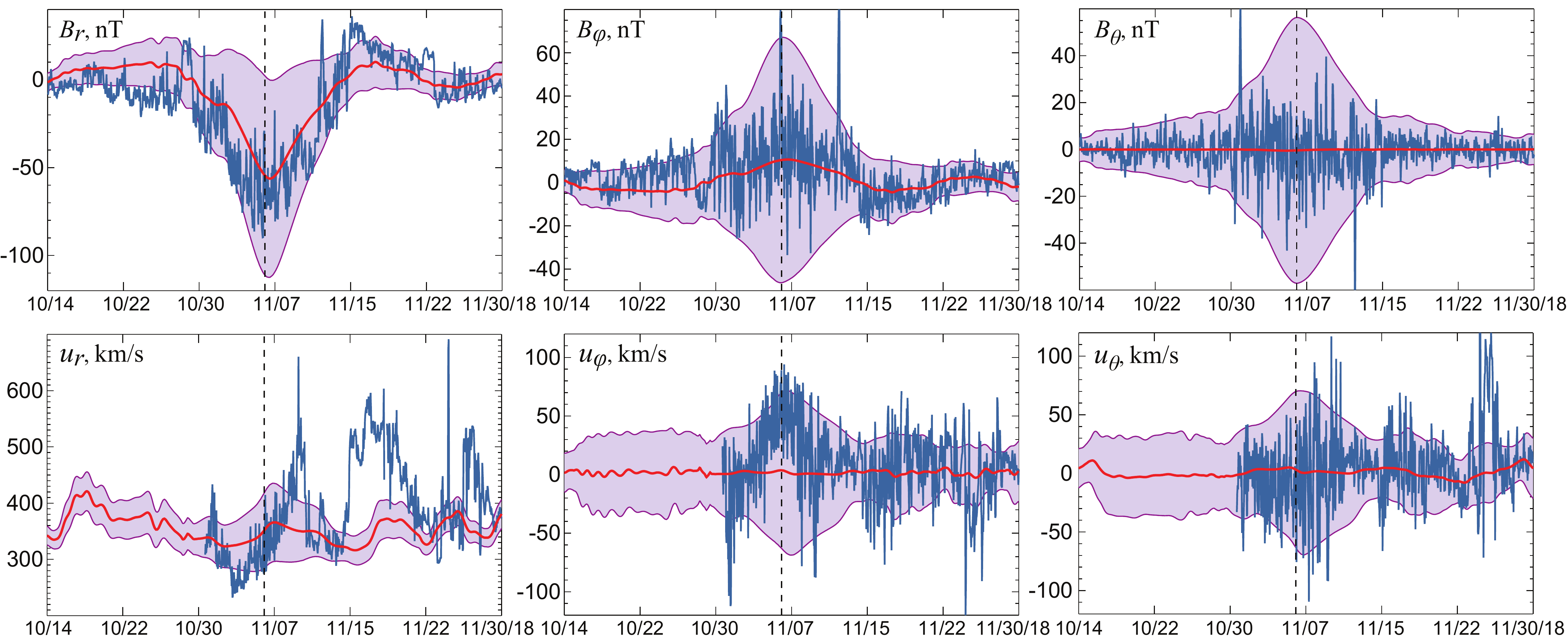}
        \caption{Magnetic field (top) and velocity (bottom) components (left: radial, center: azimuthal, right: meridional) measured by {\it PSP} during the 1st encounter, shown as hour averages (blue line). Also shown are resolved (mean field) solutions for the corresponding magnetic field components from the \citet{UsmanovEA18} global simulation employing an ADAPT magnetogram corresponding to the period of the encounter. Finally the shaded background is an envelope corresponding to an estimate of the turbulence amplitude relative to the mean field
        computed in the simulation using self-consistent turbulence transport equations 
        as explained in the text.}
    \label{fig:E1B}
    \end{centering}
\end{figure}

Further evidence is presented by 
new global simulation results shown in Figure \ref{fig:E1B} 
that address more specifically 
the likelihood of large polarity reversing 
magnetic fluctuations. 
Such simulations provide 
valuable insights into the 
behavior of the solar wind velocity and magnetic field 
components along the first three {\it PSP} orbits.

At the cadences shown in Figure \ref{fig:E1B}, neither the observations nor the simulations 
capture the full fluctuation amplitudes (which can be seen in Figure \ref{fig:stacked}).
The global code 
includes a self-consistent transport model for the turbulence amplitude, which can be used to 
estimate a likely range of turbulent fluctuations. This
range of turbulence values, when superposed on the resolved simulation variables, provides an estimate of the full range of
likely magnetic and velocity components along the {\it PSP} trajectory
during the first encounter. 
The resulting range of predicted values agrees reasonably well with the 
range of fluctuations suggested by the averaged {\it PSP} data. 
For example, focusing on the 
radial component of ${\bf B}$
in  Figure \ref{fig:E1B}, the range of 
expected values accounts for the possibility of numerous switchbacks. 

Based on these considerations and prior evidence, 
one may anticipate that 
the fluctuations become large in and near the 
Alfv\'en critical zone.
Sufficiently large fluctuations, particularly in the Alfv\'en
mode \citep{MatteiniEA18}
can produce large
deflections including reversals of magnetic polarity,
i.e., switchbacks.
\add{Previous} studies focused mainly on the presence of large amplitude fluctuations\add{, while here} we point out a number of other characteristics of {\it PSP} data that require further \add{study}.

\begin{figure}
   \centering
    \includegraphics[width=\linewidth]{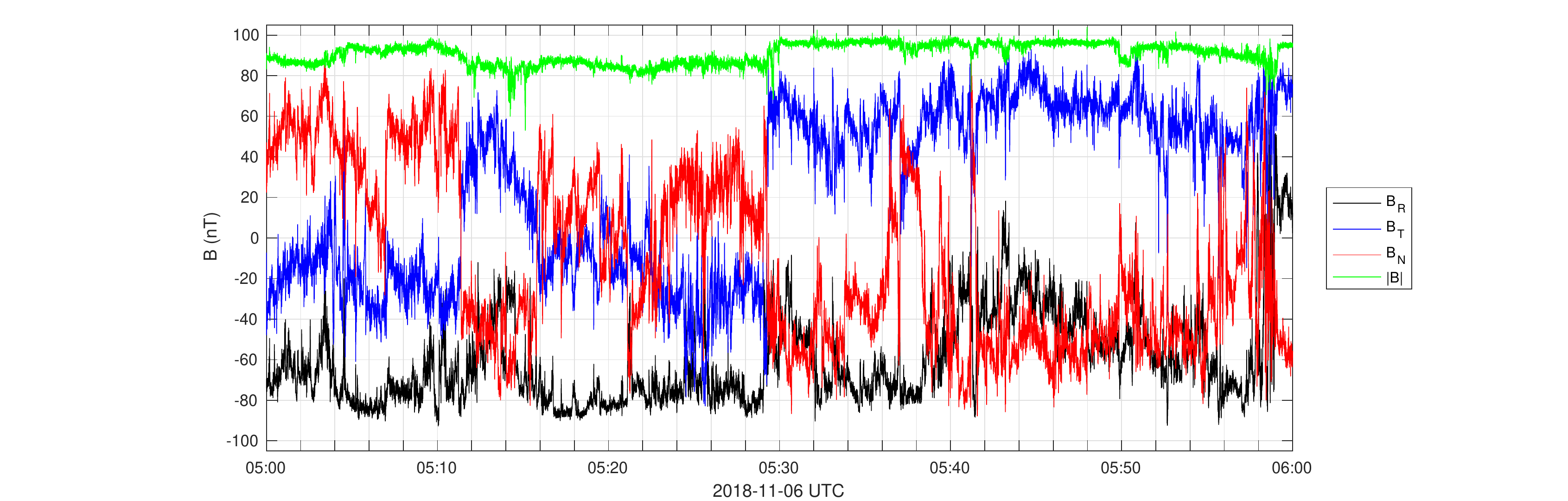}
\caption{Magnetic field components and magnitude as measured by {\it PSP}/FIELDS during one hour near first perihelion (2018 November 6, 0500-0600 UT) at radius 36 $R_\odot$, at the full sampling frequency of 299 Hz.  
Short-term fluctuations are mostly Alfv\'enic in the sense of conserving $|B|$.
Domains of nearly constant $|B|$ are often separated by minute-scale changes in $|B|$ and sharp, major jumps in the components of {\bf B}.  
We argue that these separate domains of magnetic pressure-balanced Alfv\'enic fluctuations could correspond to flocculation mixing layers.
\label{fig:Bflucts}}
\end{figure}

It is also useful to take a closer look at the {\it PSP} data near perihelion to motivate the more detailed analysis that follows. 
Figure \ref{fig:Bflucts} provides an example of magnetic field data for one hour near the first perihelion.
Here we see that the magnetic field components show large fluctuations. The radial component is predominantly negative, but shows sporadically large clusters of fluctuations \citep{ChhiberEA2020ApJS}. Meanwhile, the magnetic field 
displays a structure consisting of regions of relatively constant magnetic field strength separated by sharp changes. At 
\add{high} temporal cadence we can 
\add{identify} a few regions in which the main (radial) magnetic field changes polarity for brief times; 
these are the switchbacks. 

\begin{figure}
    \centering
    \includegraphics[width=.75\linewidth]{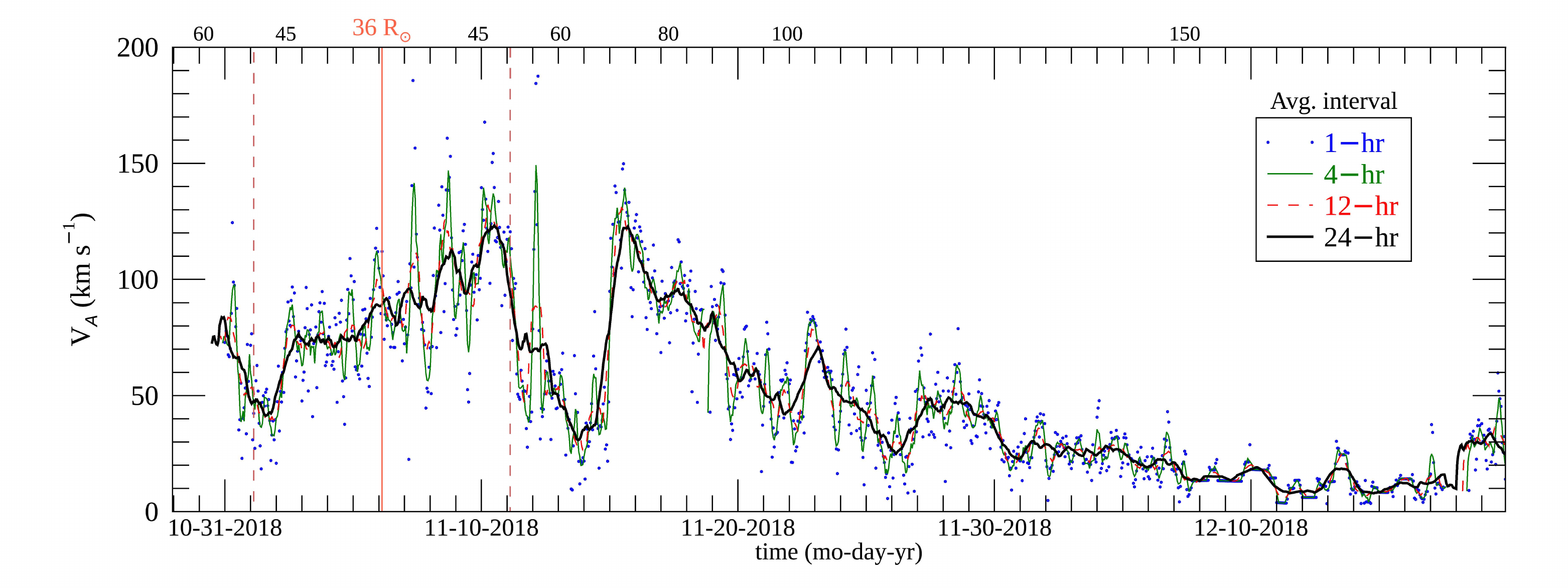}
    \includegraphics[width=.75\linewidth]{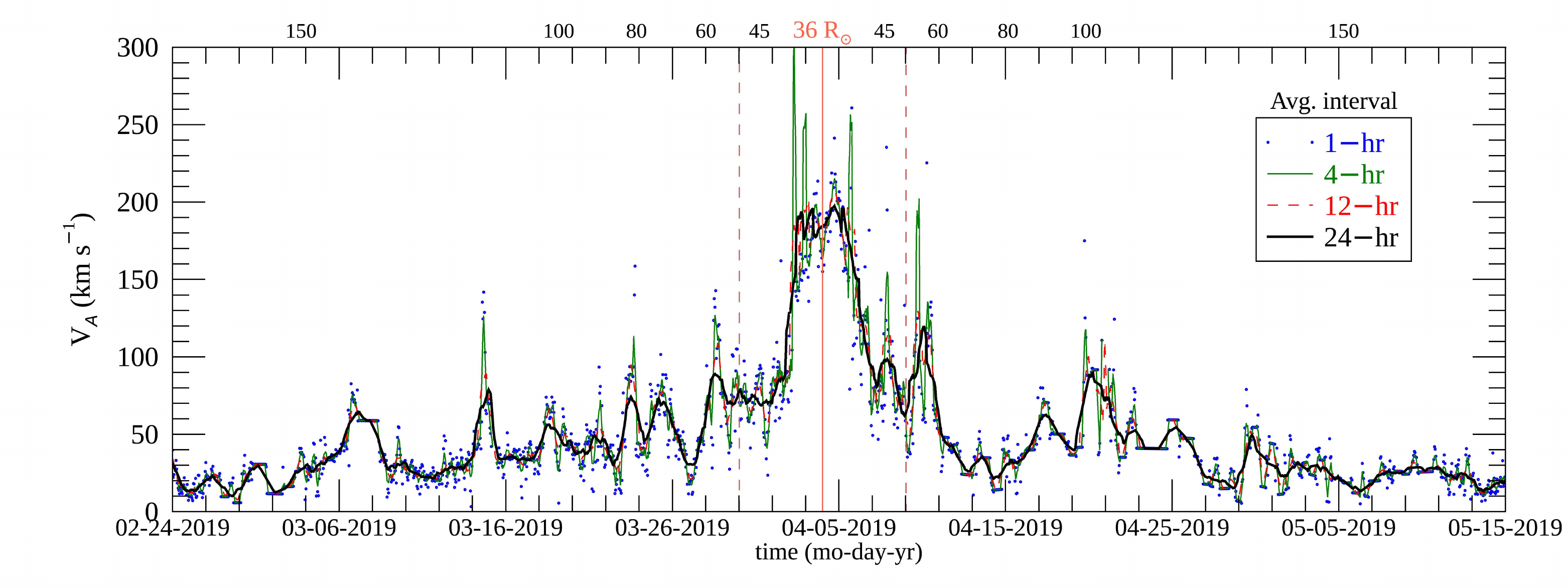}
    \caption{Alfv\'en speed vs.\ time computed at four levels of coarse graining during parts of the (top) first  and
     (bottom) second {\it PSP} orbit. The data are plotted at 1-hr cadence in each case, while averaging is performed over a moving window of specified duration. The red vertical lines mark the respective perihelia, and the dashed vertical lines demarcate a period of 10 days centered on the perihelia. Selected heliocentric distances of {\it PSP} are marked above the upper horizontal axes.}
    \label{fig:Va-E12}
\end{figure}

\begin{figure}
    \centering
         \includegraphics[width=.75\linewidth]{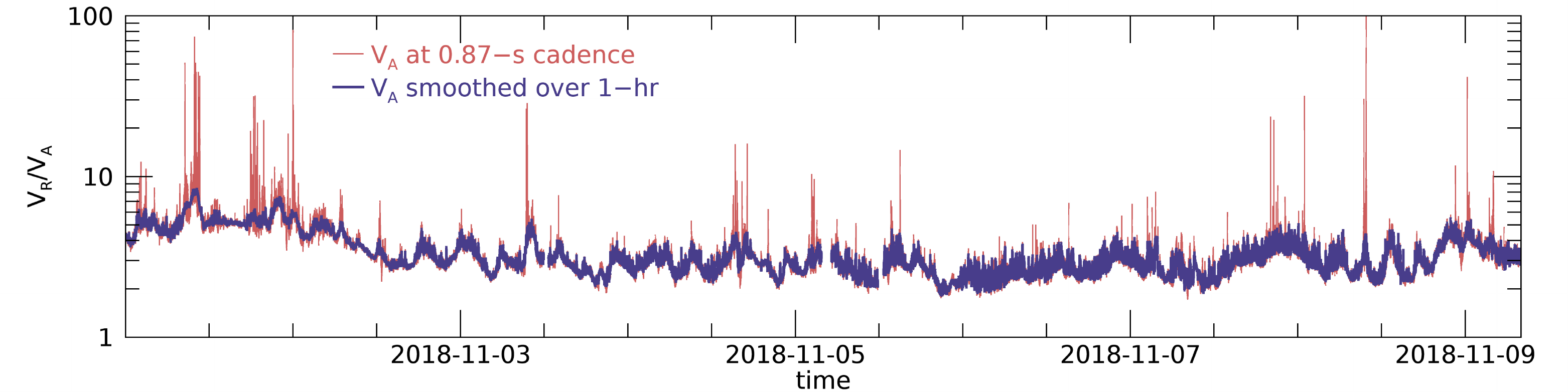}
        \includegraphics[width=.75\linewidth]{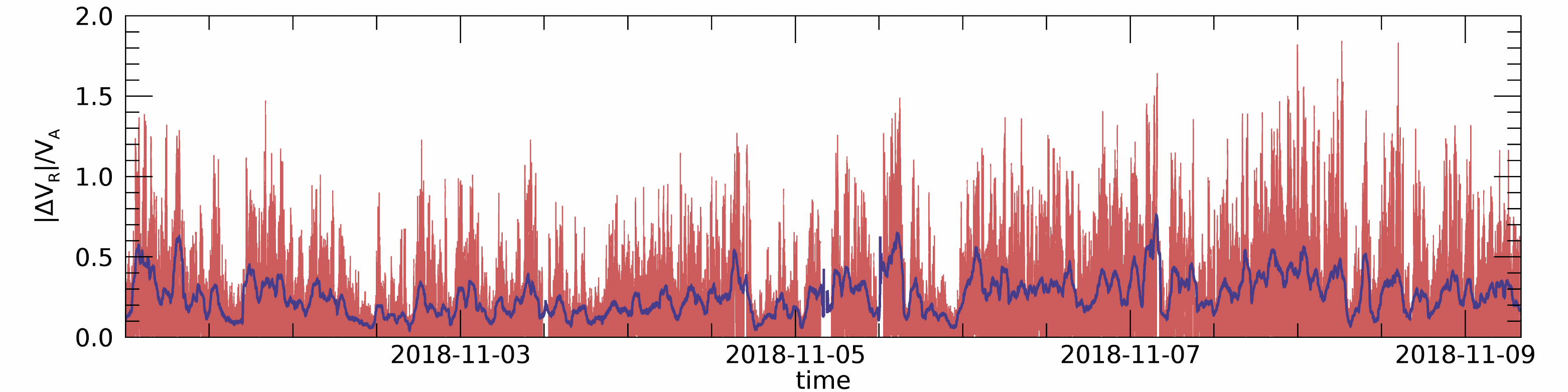}
     \caption{(Top) Radial velocity of solar wind protons in Alfv\'en speed units, near the first {\it PSP} perihelion. Radial speed is sampled at a cadence of 1 NYs $\approx 0.87$ s, and normalized either by the 1 NYs Alfv\'en speed or the 1-hour running average of the Alfv\'en speed,
     as indicated. (Bottom) Red curve shows the absolute value of increments $\Delta V_R = V_R(t+\tau) - V_R(t)$ 
     of the radial velocity $V_R$ computed from 0.87 s data and for a time lag of $\tau=10$ min, approximately the correlation time in this part of the {\it PSP} orbit \citep[see][]{ParasharEA20ApJS}. Also shown is the 1-hour rms of $\Delta V_R$ (blue curve). The increments are normalized by the 1-hour moving average of \(V_A\). It is apparent
     that there are many intermittently distributed $V_R$-increments 
     that exceed the local (smoothed) Alfv\'en speed.
     }
    \label{fig:VaVr-E1}
\end{figure}

\subsection{Radial Velocity Shear, Alfv\'en Speed and Conditions Near  
Shear Instability}

A key element of the shear-driving hypothesis is the conversion of initially more ordered but inhomogeneous
flows into more randomized flows,
a conversion that must occur in the presence 
of ordered magnetic fields. 
This is a 
scenario that can explain the transition from striation to flocculation
in the {\it STEREO} images
\citep{DeForestEA16}.  
As discussed in 
\S \ref{hypothesis}, 
a uniform magnetic
field will resist Kelvin-Helmholtz-like rollups when the velocity differential is less than the local 
Alfv\'en speed.
Therefore we need
to examine quantitative features of the 
plasma flows in 
comparison with the local Alfv\'en speed.
To establish the 
context, we begin by computing the
Alfv\'en speed from the {\it PSP} data, which we show for the first two encounters (E1 and E2) 
in Figure \ref{fig:Va-E12}.
Four levels of averaging are shown, over 1, 4, 12, and 24 hours.
The salient feature is that there is considerable variation of Alfv\'en speed in both encounters, 
with values around 80 to 110 km s$^{-1}$ near first perihelion, with 
higher values close to 200 km s$^{-1}$ near second perihelion, and with values as low as $\sim10$ km s$^{-1}$ at 
greater distance from the Sun.

As a next step we examine the radial velocity, normalized to the Alfv\'en speed, for E1. 
This is motivated by Chandrasekhar's 
condition for suppression, $V_A > \Delta V$, where $\Delta V$ is an appropriate measure of the velocity contrast across a shear layer. 
\add{(In the current complex environment, we view that a physically relevant value of the Alfv\'en speed $V_A$ would be 
an appropriate regional average.)} 
However we emphasize that 
here we are not looking for conditions for subsequent linear instability, 
since the additional signatures we 
examine would indicate a {\it past} instability closer to the Sun rather than an imminent instability. 
Nevertheless, we do not rule out that subsequent instability may take place, particularly because, as mentioned above, polarity reversals are observed further from the Sun 
and there is evidence that their 
frequency may actually increase 
with increasing radial distance \citep{MacneilEA20,OwensEA20}. 
The presence of velocity changes over relatively short distances that exceed the local Alfv\'en speed is an indication that the criterion for Kelvin-Helmholtz rollup is likely to be reached in this region (or it may even be in progress as we observe it)---something that cannot be ascertained with single-point measurements.

In Figure \ref{fig:VaVr-E1} (top)
we show the radial plasma velocity of solar wind protons.
Two resolutions are shown, one in which $V_A$ is computed at 0.87-s resolution,
and the other using one-hour smoothed (running averaged) $V_A$. 
There are numerous variations of 
$V_R$ that are larger than one or a few Alfv\'en speeds, but we must ask at what scales these occur.
To that end, we compute the increments of the observed radial component of plasma velocity, normalized in an analogous way to the 1-hour moving average of $V_A$. 
The increment is defined as 
$\Delta V_R = V_R(t+\tau) - V_R(t)$ 
with time lag $\tau$.
The value 
$\tau=10$ minutes is selected,
corresponding to the typical 
measured correlation time in the first encounter 
\citep{ParasharEA20ApJS}, which is 
expected to be a typical large scale magnetic flux tube size.
Therefore, the measured increments are estimates of velocity contrasts 
$\Delta U$ between adjacent flux tubes as suggested in Figure 
\ref{fig:cartoon}. 
We see that
$\Delta V_R$  frequently exceeds
$V_A$. 
This is a way to assess the likelihood
of nearby nonlinear K-H activity.
We conclude that the case for the development of a mixing layer is reasonably well supported.

\begin{figure}
    \centering
    \includegraphics[width=.75\linewidth]{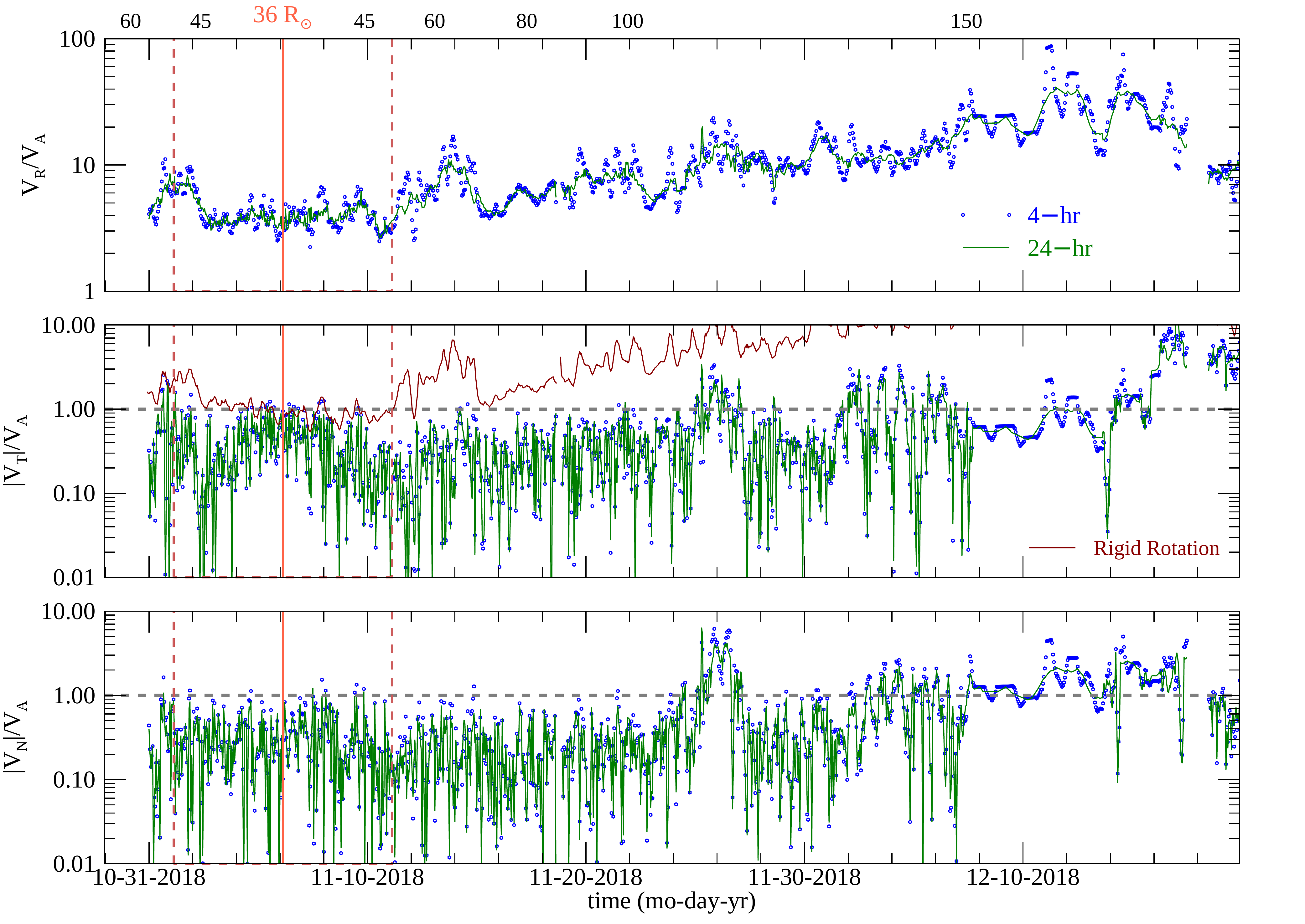}
        \caption{Proton velocity components in (coarse-grained) Alfv\'en speed units during the first {\it PSP} orbit.
        Two levels of coarse-grained Alfv\'en speed are used (4- and 24-hour moving averages), while the proton velocities are plotted at 1-hour cadence. The brown curve in the middle panel shows the speed of rigid rotation in units of the 4-hour moving average of \(V_A\). This would be the tangential speed of the plasma if it were corotating with the Sun, with angular speed corresponding to the sidereal rotation period of 24.47 days. Selected heliocentric distances of {\it PSP} are marked above the upper horizontal axis.}
    \label{fig:V_Va-E1}
\end{figure}

\begin{figure}
    \centering
    \includegraphics[width=.75\columnwidth]{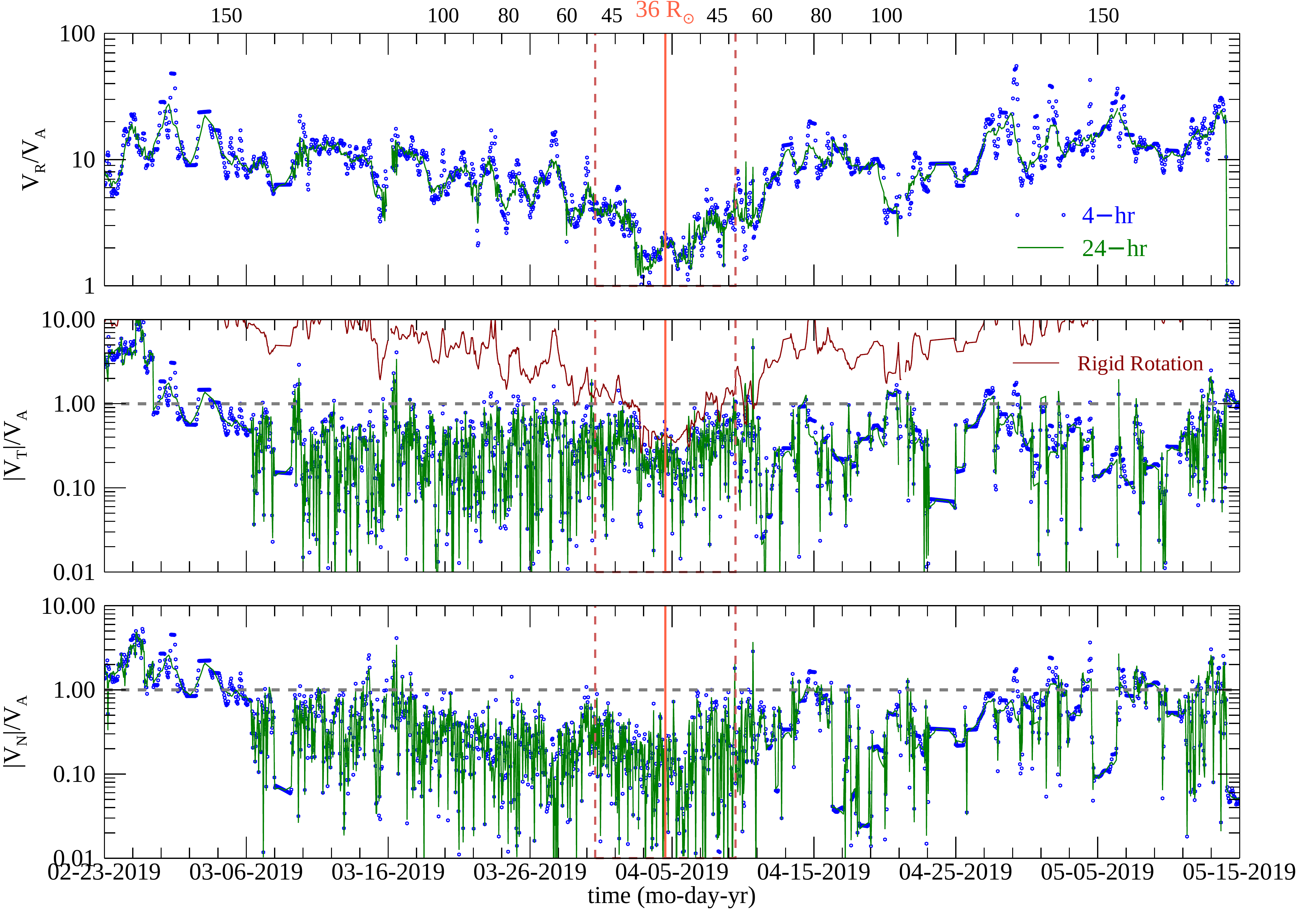}
        \caption{Proton velocity components in (coarse-grained) Alfv\'en speed units during the second {\it PSP} orbit. 
        See caption of Figure \ref{fig:V_Va-E1} for more details.}
    \label{fig:V_Va-E2}
\end{figure}

\begin{figure}
\includegraphics[width=.3\columnwidth]{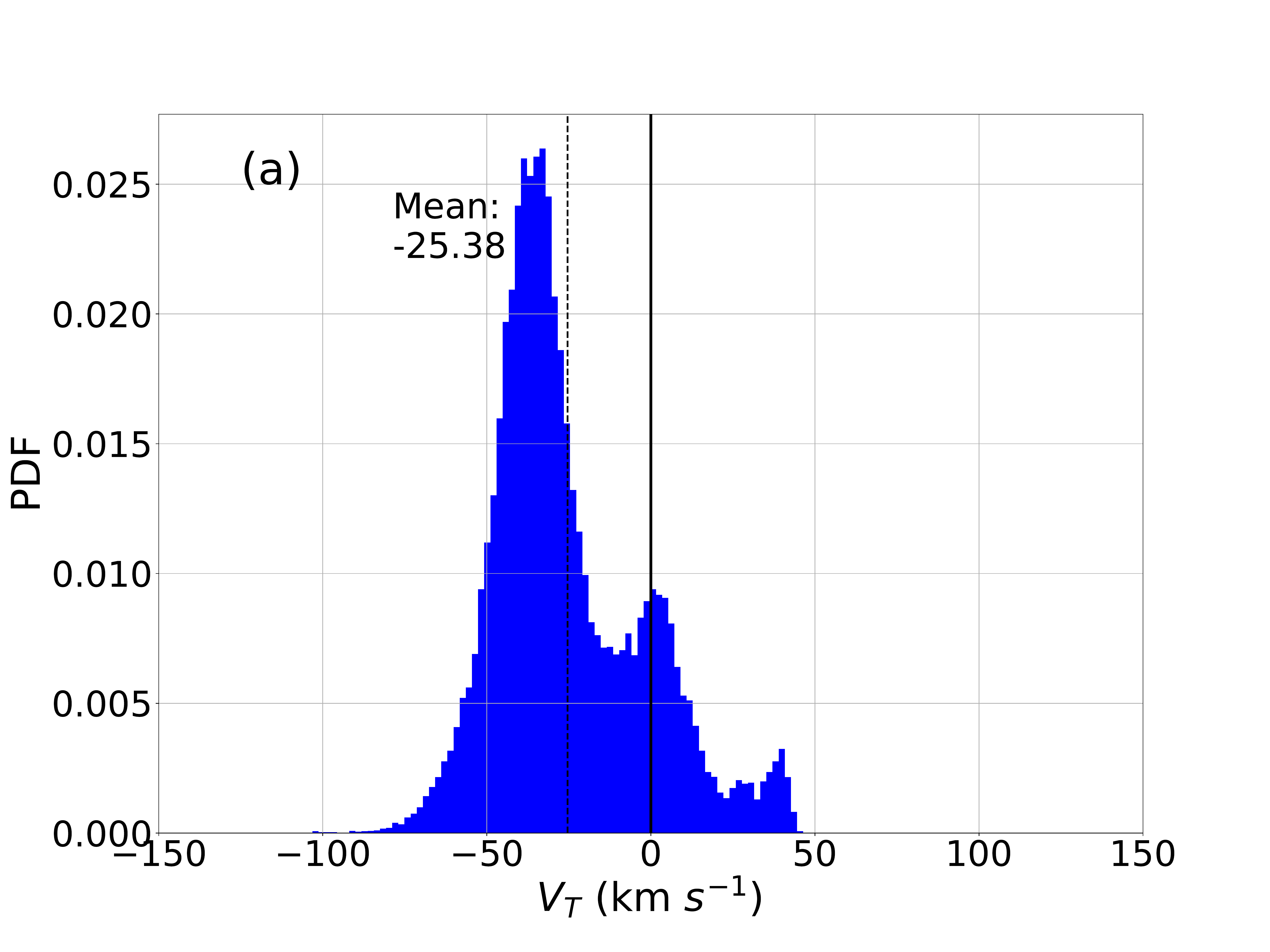} \includegraphics[width=0.3\columnwidth]{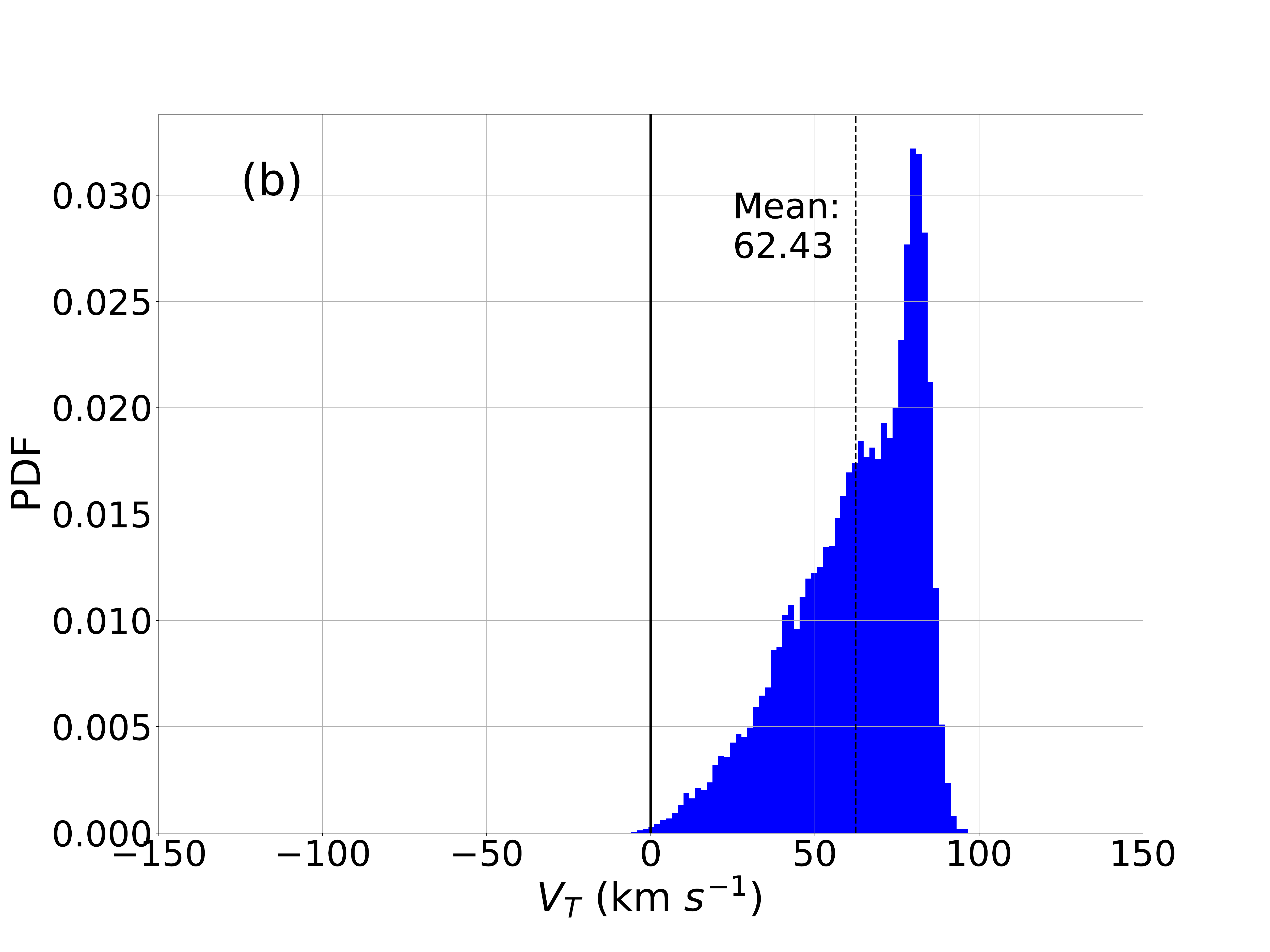} \includegraphics[width=0.3\columnwidth]{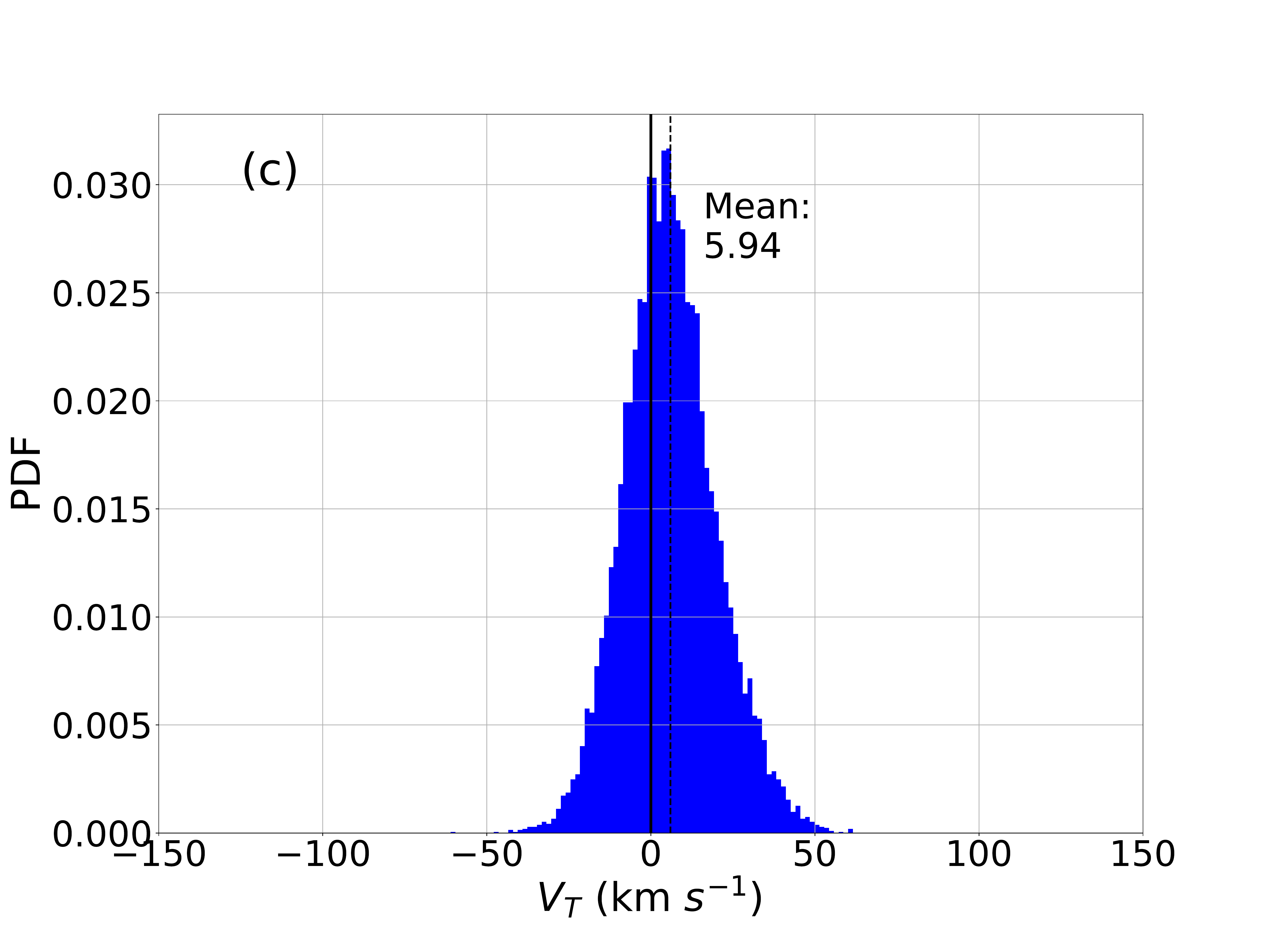}
\caption{(a) 
Probability distribution function (PDF) of longitudinal solar wind velocity $V_T$ (in km s$^{-1}$) 
as measured by {\it PSP} with a cadence of 1 NYs ($\approx0.87$ s) during 2018 Oct 31, 0800-1600 UT, about 6 days before first perihelion.  Vertical dashed line indicates mean value.  During this and other time periods far from perihelion, $V_T$ is randomly distributed around zero, with sub-distributions around positive or negative values for up to a few hours (as seen here mostly at negative values).
(b) Similar PDF for
2018 Nov 6, 0000-0800 UT, including the time of first perihelion.  
Near perihelion, $V_T$ is clearly biased toward positive values, indicating partial corotation with the Sun.  
(c) Similar PDF for 2018 Nov 11, 0000-0800 UT, or 5 days after first perihelion.  
Far from perihelion, $V_T$ is again randomly distributed around zero.}
\label{fig:VT}
\end{figure}

\subsection{Transverse Velocity and Fluctuation Components}

The behavior of the transverse velocity components is also significant and may exhibit signatures of plasma rollup, a process that also involves convection of the magnetic field.
In the idealized case, the magnetic field resists the rollup,  and 
is amplified as it is distorted by the velocity shear
\citep{Miura82,GoldsteinSSPP89,RobertsJGR92,MalagoliEA96}. If there were little or no transverse velocity initially, one would expect that the maximum excursion of the transverse velocity would be, roughly speaking, bounded by the local (amplified) Alfv\'en speed, according to the typical condition 
of equipartition of 
energy
between magnetic and flow energies in the solar wind frame.

To examine the excursion of the transverse velocities, 
Figures \ref{fig:V_Va-E1} and \ref{fig:V_Va-E2}
\add{show} the three Cartesian components of velocity normalized to the locally-averaged Alfv\'en speed during the first and second orbits, respectively. 
It is apparent that the two transverse components $V_T$ and $V_N$ are almost always nicely bounded 
by the local Alfv\'en speed. 
To be specific, among 1-s values of $|V_T|/V_A$ and $|V_N|/V_A$ from the first encounter, about 7\% exceeded 1 and none exceeded 4. Note that a 2-hour moving average of \(V_A\) was used to obtain these percentages.

Figures \ref{fig:V_Va-E1}(b) and \ref{fig:V_Va-E2}(b) also indicate the speed of rigid rotation with the Sun at the sidereal rotation period of 24.47 d \citep{Pecseli20}, which is plotted in units of the 4-h moving average of $V_A$ (brown curves).  
It can be seen that at times near both the first and second perihelia, $V_T$ was comparable to the speed of corotation with the Sun.

The probability distributions of the longitudinal velocity component $V_T$ for 
E1
are shown at three positions along the orbit 
in Figure \ref{fig:VT}.
We note that 
the distribution for
the inbound orbit, 6 days prior to perihelion, shows a multi-component distribution with several distinct peaks (Figure \ref{fig:VT}(a)). 
Each peak covers a spread in $V_T$ that
resembles a separate 
sub-distribution. 
From the time series (not shown) these are seen to result from time periods in which $V_T$ fluctuates about positive or negative values for durations from a fraction of an hour to a few hours; longer durations are more common at greater distance from the Sun. 
The 8-hour time period shown in Figure \ref{fig:VT}(a) happens to have more negative values.
Near perihelion, the distribution shows
a strong bias towards positive $V_T$ (Figure \ref{fig:VT}(b)) at speeds comparable to the corotation velocity of around 70 km s$^{-1}$. 
There is again a broad distribution, in this case skewed towards smaller values. 
Five days after perihelion, for the example period shown in Figure \ref{fig:VT}(c), the distribution is centered roughly about $V_T=0$ with a single strong maximum. 
The distributions of the latitudinal component $V_N$ for E1 (not shown) are qualitatively similar except they do not exhibit a bias toward positive values near perihelion.

In the corona, the nascent solar wind is expected to be channeled along magnetic flux tubes that corotate with the Sun in the longitudinal direction.  
Thus the {\it PSP} observations near first perihelion are consistent with partial corotation in that the longitudinal solar wind velocity $V_T$ fluctuates around the corotation speed while the latitudinal component $V_N$ fluctuates around zero.  
These observed patterns are also evident in Figure \ref{fig:E1B}, where the global 
simulation variables $(u_\theta, u_\varphi)$ 
correspond to {\it PSP} velocity components 
$(-V_N,V_T)$.
We refer to partial corotation because, according to Figure \ref{fig:VT}(b), most of the solar wind has $V_T$ below the corotational value of $\approx 70$ km s$^{-1}$.  
Indeed, such ``slippage'' of solar wind elements from corotation is expected to occur beyond the Alfv\'en critical zone 
where 
the magnetic field no longer controls the solar wind flow.
Therefore, we interpret the observation of partial corotation near the first perihelion as evidence that {\it PSP} was already close to the Alfv\'en critical zone.  
Farther from the Sun, there is no apparent corotation of the solar wind (see Figures \ref{fig:E1B} and \ref{fig:VT}) and such slippage becomes complete.  
However, near first perihelion the partial corotation indicates partial slippage, and suggests that neighboring magnetic flux tubes could have substantially different $V_T$ values.  
In other words, in addition to the radial velocity shear suggested in Figure \ref{fig:cartoon} and Section 4.2, there could also be longitudinal velocity shear associated with partial corotation.

Intriguingly, in Figure \ref{fig:VT}(b) part of the $V_T$ distribution is actually faster than the corotational speed, which
could be attributed to Kelvin-Helmholtz rollups in the mixing layer outside the Alfv\'en critical zone.
These interpretations of large reported $V_T$ need to be viewed as tentative, given that modeling has so far not been able to reproduce $V_T$ values as large as those discussed here.

\subsection{Domains and Anisotropy of Alfv\'enic Fluctuations}

Ever since the seminal work of \citet{BelcherDavis71}, it has been recognized that magnetic and velocity fields in the solar wind tend to fluctuate together, which has been attributed to an Alfv\'en mode  
\begin{equation}
    {\bf v}=\pm \frac{\bf b}{\sqrt{\mu_0\rho}},
    \label{eq:alf}
\end{equation}
where ${\bf v}\equiv{\bf V}-{\bf V}_0$ and ${\bf b}\equiv{\bf B}-{\bf B}_0$, 
subtracting any   large-scale (mean) fields ${\bf V}_0$ and ${\bf B}_0$, $\rho$ is the mass density, and the right hand side is the magnetic fluctuation expressed in terms of the Alfv\'en speed. 
Even at large amplitudes such fluctuations 
are solutions of the incompressible 
MHD equations \citep{Moffatt}.
If there is a mean magnetic field ${\bf B}_0$, 
then the 
the ``+'' sign indicates propagation along $-{\bf B}_0$ and the ``$-$'' sign indicates propagation along ${\bf B}_0$.


\begin{figure}
\includegraphics[width=.3\columnwidth]{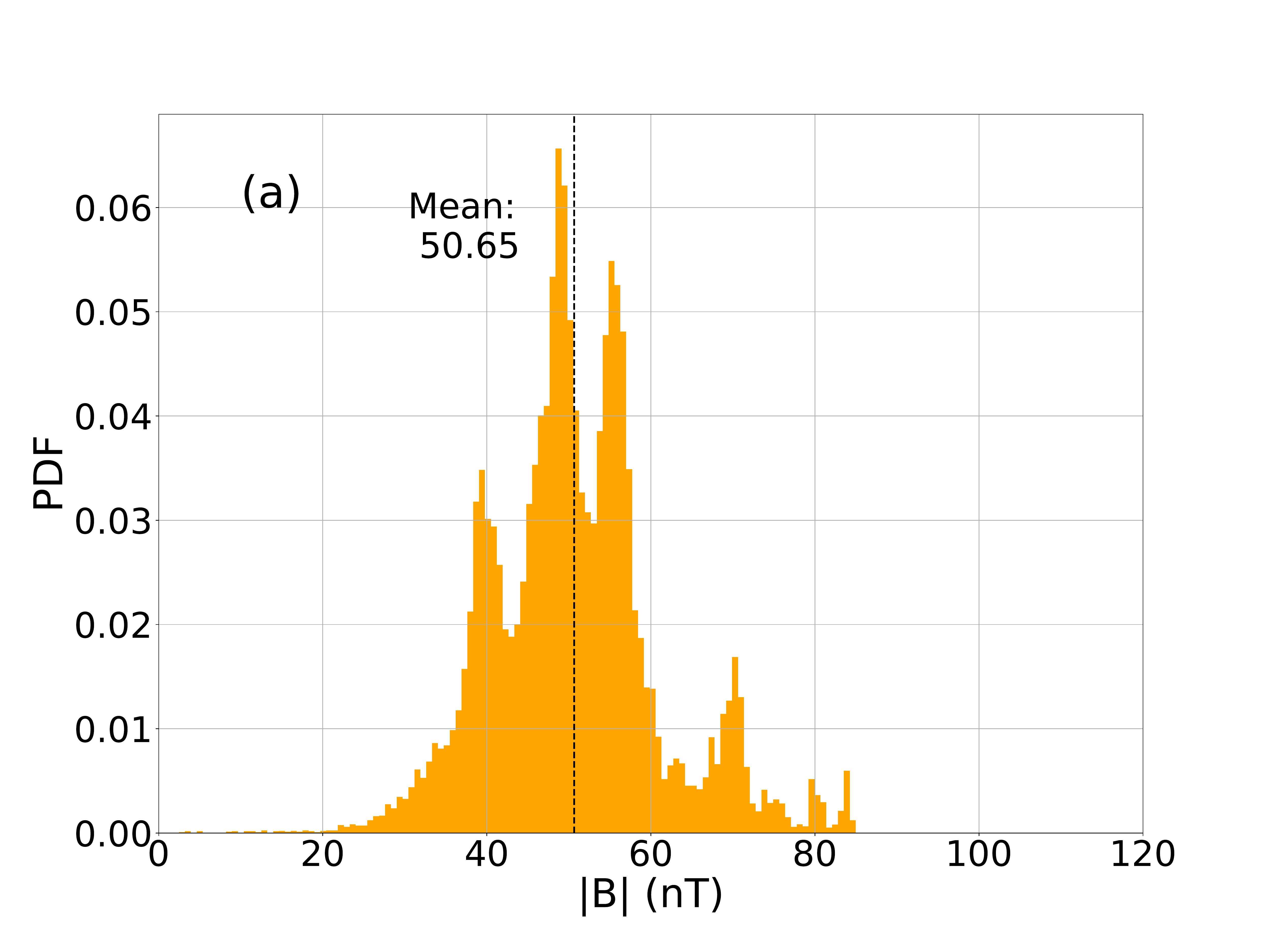} \includegraphics[width=0.3\columnwidth]{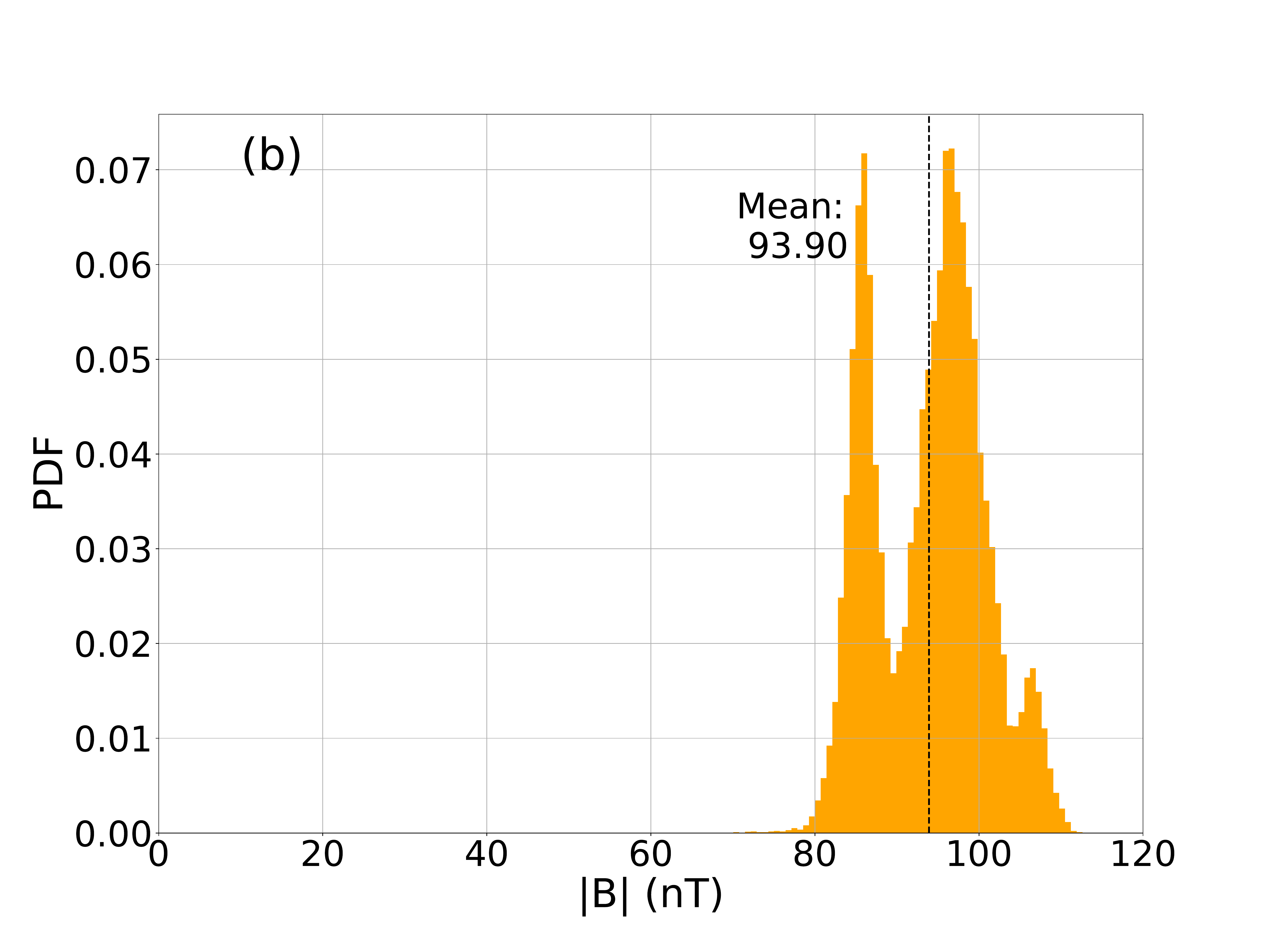} \includegraphics[width=0.3\columnwidth]{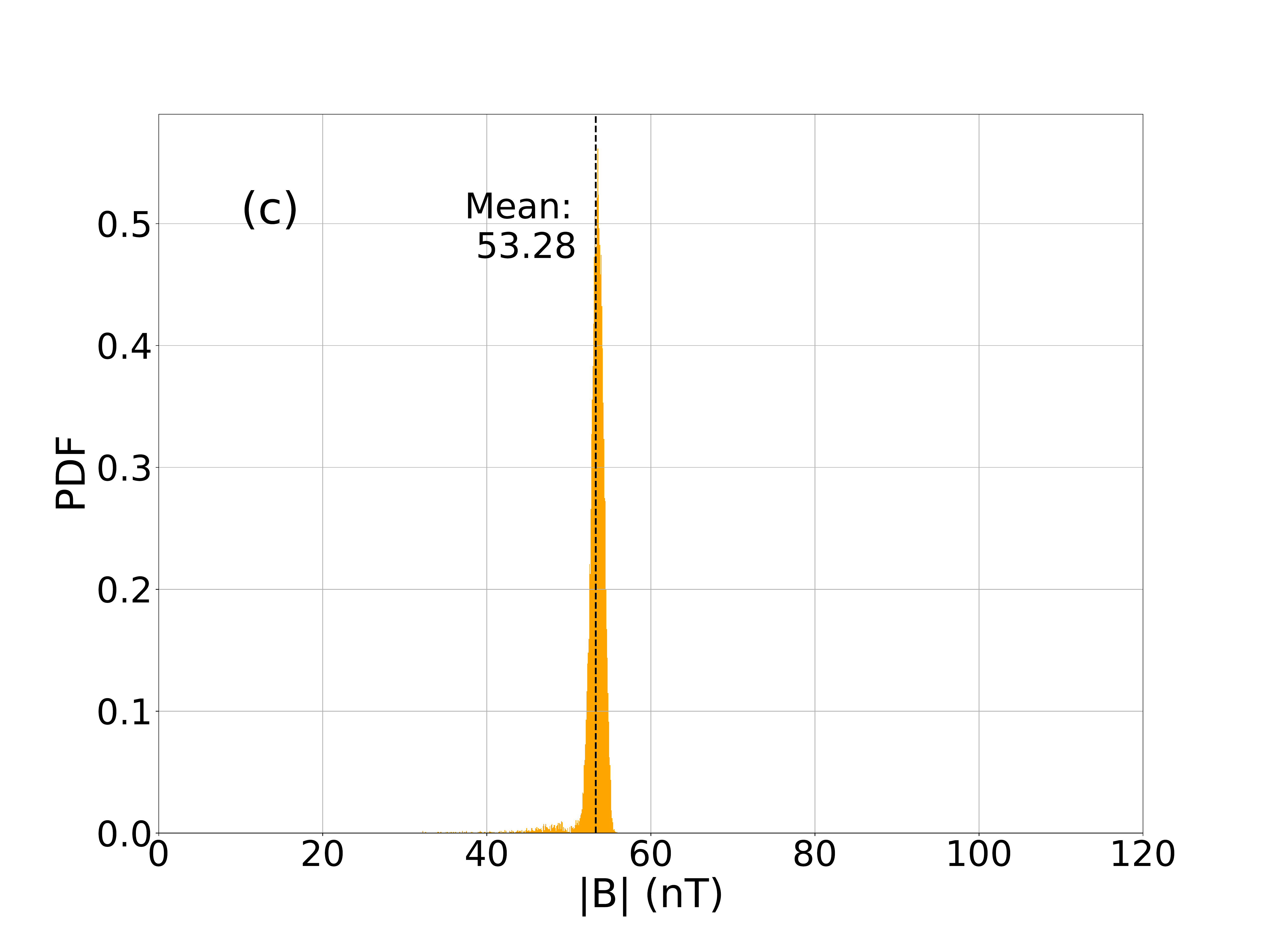}
\caption{(a) PDF of magnetic field magnitude $|B|$ (in nT) as measured by {\it PSP} sampled at a cadence of 1 NYs ($\approx0.87$ s) during 2018 Oct 31, 0800-1600 UT, about 6 days before first perihelion.  
Vertical dashed line indicates mean value.  
Clumps in the distribution of $|B|$ correspond to local flux tubes. 
(b) Similar PDF for
2018 Nov 6, 0000-0800 UT, including the time of first perihelion.  The mean of $|B|$ was generally larger when {\it PSP} was closer to the Sun.
(c) Similar PDF for 2018 Nov 11, 0000-0800 UT, or 5 days after first perihelion.  At this location, the entire 8-h period effectively comprises a single domain. 
\label{fig:B}}
\end{figure}

For compressible MHD, large-amplitude propagating 
solutions exist for the Alfv\'en mode so long as the total field magnitude
$|B| = |({\bf B}_0 + {\bf b})|$ is uniform
\citep{GoldsteinEA74,BarnesHollweg74,Barnes79a}. However, note that the divergence requirement on the magnetic field limits the spatial region over which this ``incompressible'' mode can exist \citep{Barnes79a}.
From the work of \citet{BelcherDavis71} and many others, {\it in situ} measurements of solar wind fluctuations throughout the heliosphere have indicated that such Alfv\'enic fluctuations are predominantly outward.
In {\it PSP} data, such fluctuations are common and frequently of large amplitude, with 
$|b|\sim|B|$ \citep[see, e.g., Figure 2 of][]{KasperEA19Nature}.
\citet{ParasharEA20ApJS} and \citet{HorburyEA20}
recently described several measures of Alfv\'enicity as applied to the {\it PSP} E1 data.
Other aspects of Alfv\'enic fluctuations
have also gained recent attention \citep{MatteiniEA18,MatteiniEA19,DAmicisEA20}.

The nearly constant magnetic field magnitude $|B|$ (a distinctive property of Alfv\'enic fluctuations) 
is often evident in {\it PSP} 
data as illustrated in Figure \ref{fig:Bflucts}.
For such cases, 
in terms of its components, the vector ${\bf B}$ is randomly walking on a sphere of nearly constant $|B|$, as described by \citet{Barnes81}.
In turbulence, 
constant magnetic pressure 
may be associated 
with 
rapid, local
relaxation processes that also 
favor patches of 
flow-field alignment, as
in Eq.\ (\ref{eq:alf}) \citep{MatthaeusEA08,OsmanEA11-align}. 
\citet{MatteiniEA15}
offer an alternative 
view of 
constancy of $|B|$, namely, that it is associated with 
conservation of ion kinetic energy
in the reference frame of observed alpha particle motion.

As can be seen from Figure \ref{fig:Bflucts},
{\it PSP} data from the first encounter reveal 
Alfv\'enic domains with nearly constant $|B|$ that are often separated 
by sharp, major jumps in the components of ${\bf B}$, as necessary to 
preserve the divergence condition $\nabla\cdot{\bf B}=0$ when $\nabla\cdot{\bf V}\ne0$. 
Another indication of the domain structure comes from probability distribution 
functions (PDFs) of $|B|$ as
shown in Figure \ref{fig:B}. 
It is clear that eight hour samples 
near perihelion are likely to contain 
one to a few regions of nearly constant magnetic 
field. 
This is consistent with Alfv\'enic turbulence, 
but in addition, it is consistent with mixing layer dynamics, as we shall see below. 

At the interface between two plasma flows with relative shear, once the condition $\Delta V>V_A$ is met, Kelvin-Helmholtz dynamics are possible.  
This develops into a mixing layer that eventually includes rollups and magnetic polarity reversals (switchbacks) \citep{MalagoliEA96}, and the mixing layer is expected to grow with distance along the flow, or in this case with distance from the Sun.
Now, in our hypothesis (see Figure \ref{fig:cartoon}), there are numerous magnetic flux tubes in the nascent solar wind. 
Some of the interfaces between these should develop the Kelvin-Helmholtz instability and mixing layers.
These mixing layers should grow until they come into contact.
When they do, it is possible that they merge in the sense that the shear-driven dynamics (i.e., flocculation) homogenizes the magnetic pressure within the merged region.
We interpret the domains of Alfv\'enic turbulence with nearly constant $|B|$ as such (possibly merged) mixing layers, and they exhibit sharp boundaries as topological defects across which the dynamics have not yet balanced the magnetic pressure.
{\it PSP} data also provide some evidence that these domains become larger with increasing heliocentric distance $r$, as expected for mixing layers that grow and merge.
In Figure \ref{fig:B}(c) we see a case 5 days after first perihelion when a very narrow distribution of nearly constant $|B|$ was observed for an entire 8-h period, in contrast with the 8-h period near first perihelion in which multiple distributions were observed (Figure \ref{fig:B}(b)).
As described earlier for $V_T$, sub-distributions in the time series for $|B|$ that last several hours are more common at increased $r$ several days away from perihelion, while the variation in {\it PSP} travel speed was relatively minor.

In Figure \ref{fig:Bflucts} we see that at some times within a domain of nearly constant $|B|$,
in association with particularly strong magnetic fluctuations, $|B|$ temporarily decreases for up to a few minutes before returning back to the same nearly constant level.
At such times the magnetic pressure balance is temporarily disrupted within the domain.
Sudden drops in $|B|$ have previously been reported during switchbacks, i.e., reversals in $B_R$ \citep{BaleEA19Nature,KasperEA19Nature}.
Here we note that a temporary decrease in $|B|$ can occur together with strong fluctuations in {\it any} of the magnetic field components, e.g., in Figure \ref{fig:Bflucts} at hour 5.37 or 5.63, and are not particular to switchbacks.  
This is consistent with the view that many switchbacks belong to a continuum of fluctuations that can occur in all field components as part of {\it in situ} dynamics in the solar wind.



To examine the Alfv\'enic magnetic fluctuations in more detail, we calculated statistics of magnetic increments ${\bf \Delta B} = {\bf B}(t+\tau)-{\bf B}(t)$ for time lags $\tau$ of 1 s, 10 s, 1 min, 10 min, 1 h, and 6 h.  
(Note that the FIELDS instrument typically samples the magnetic field 
at 299 Hz so even the 1-s lag is much longer than the instrumental resolution.)
In order to study the fluctuation anisotropy, we decomposed the magnetic increments into a parallel component $\Delta B_\parallel$ along the magnetic field ${\bf B}(t)$ and two components along basis vectors perpendicular to ${\bf B}(t)$, in the R-T plane ($\Delta B_1$, a roughly longitudinal increment) and perpendicular to the R-T plane ($\Delta B_2$, a roughly latitudinal increment).
We calculated the variances (mean squares) of these quantities as a measure of the scale-dependent fluctuation energy in these components.
To measure the conservation of $|B|$, we also calculated the variance of increments in $|B|$.

\begin{figure}
\centering
\includegraphics[width=.45\columnwidth]{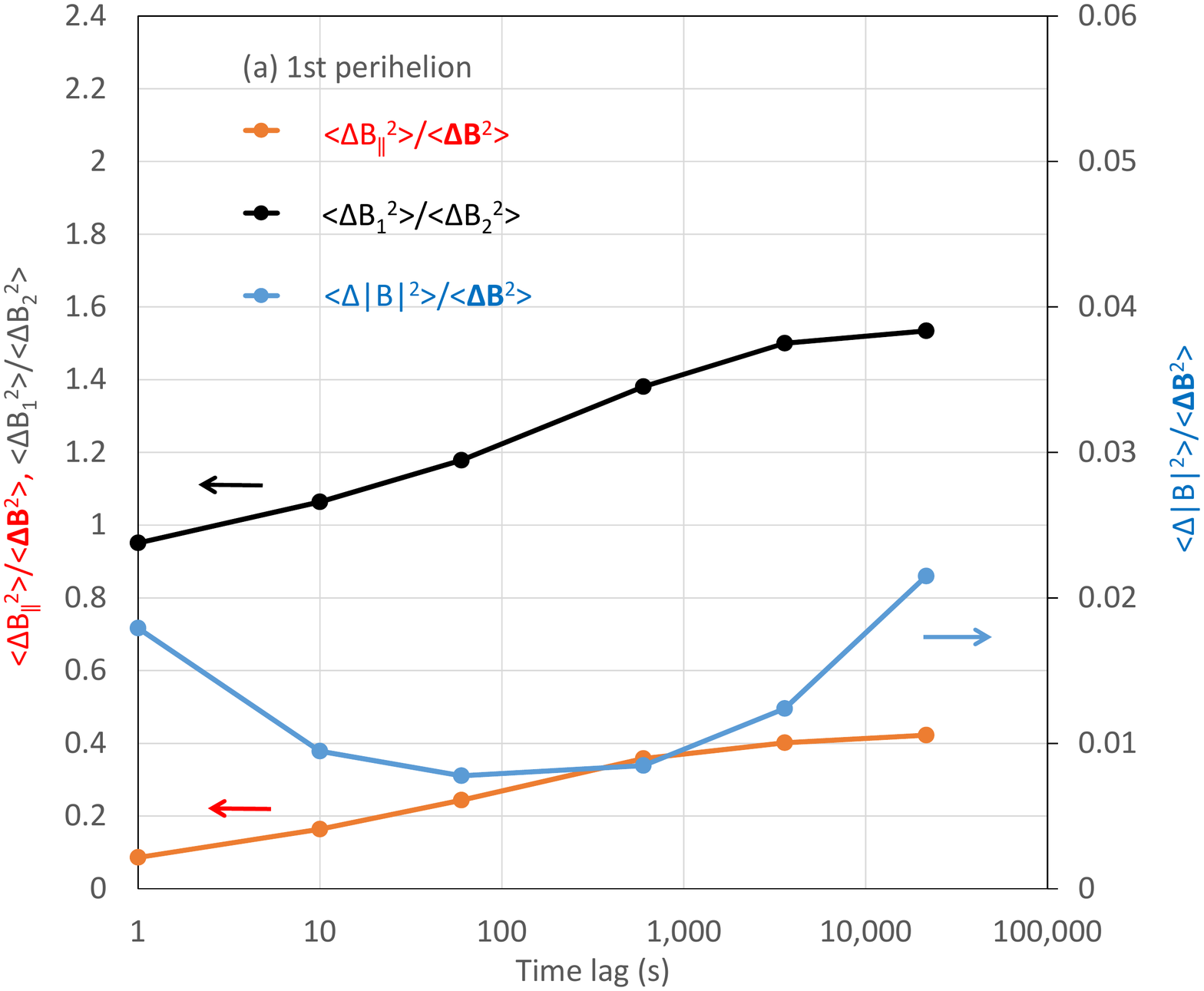} 
\includegraphics[width=.45\columnwidth]{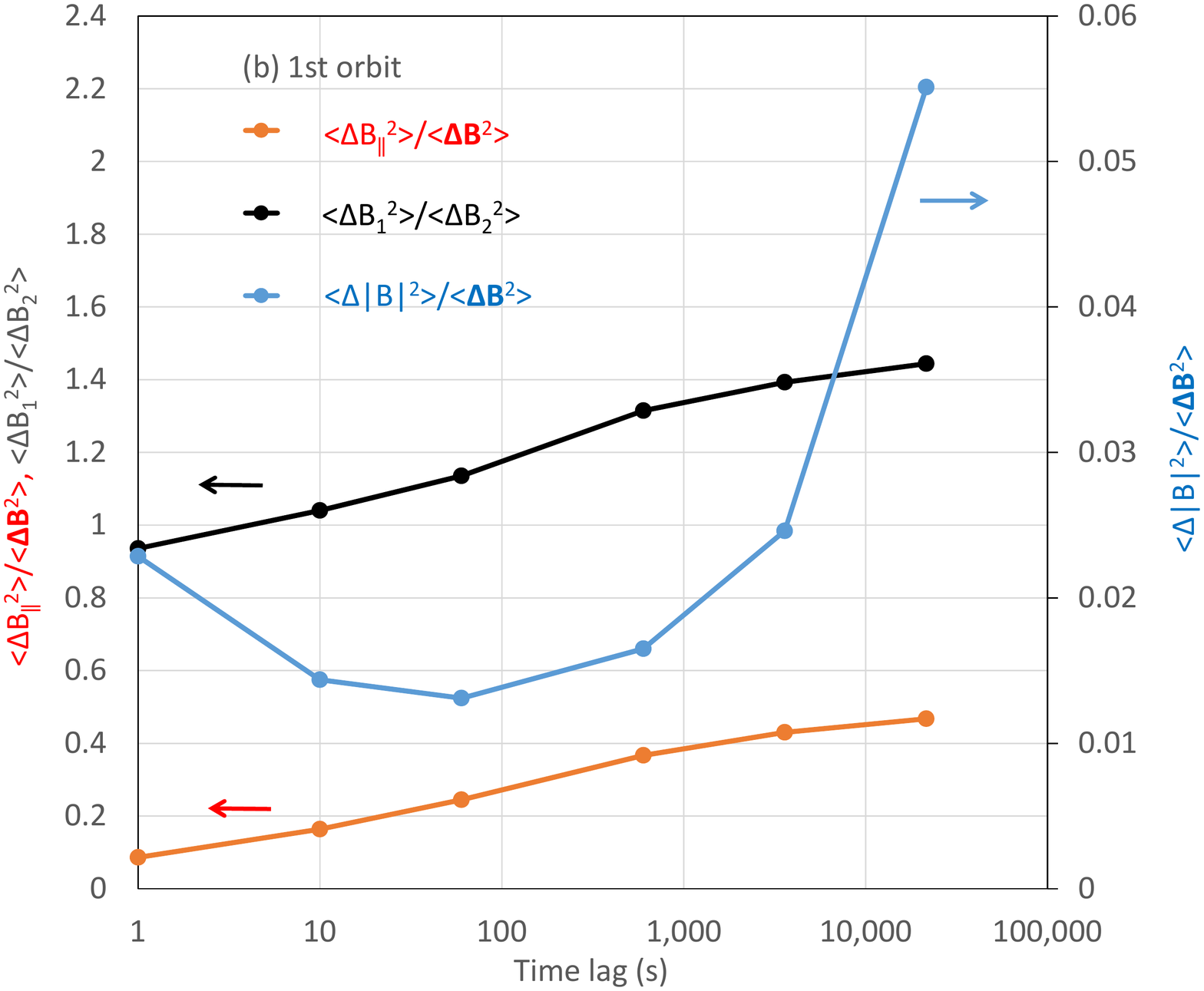} 
\includegraphics[width=.45\columnwidth]{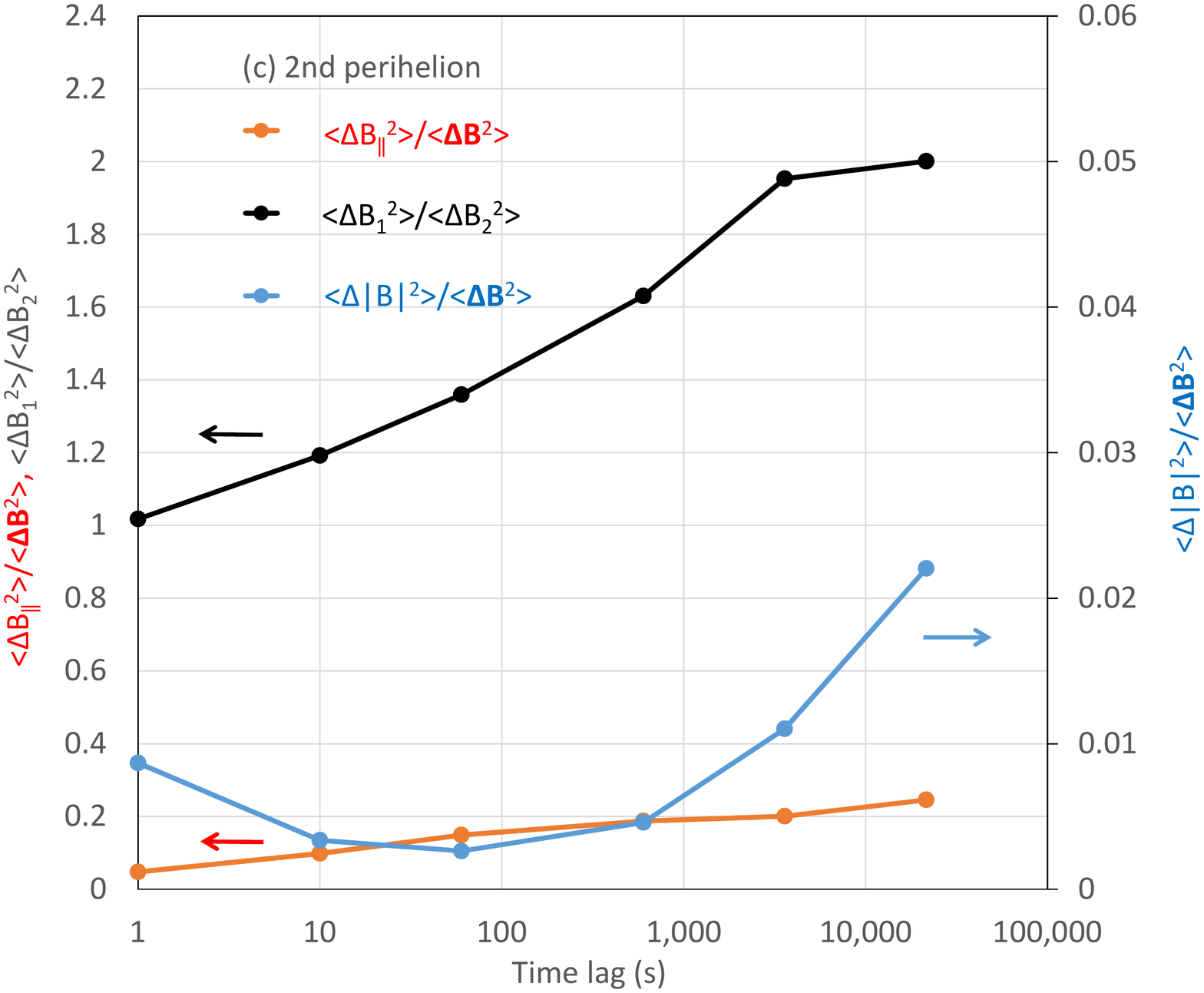} 
\includegraphics[width=.45\columnwidth]{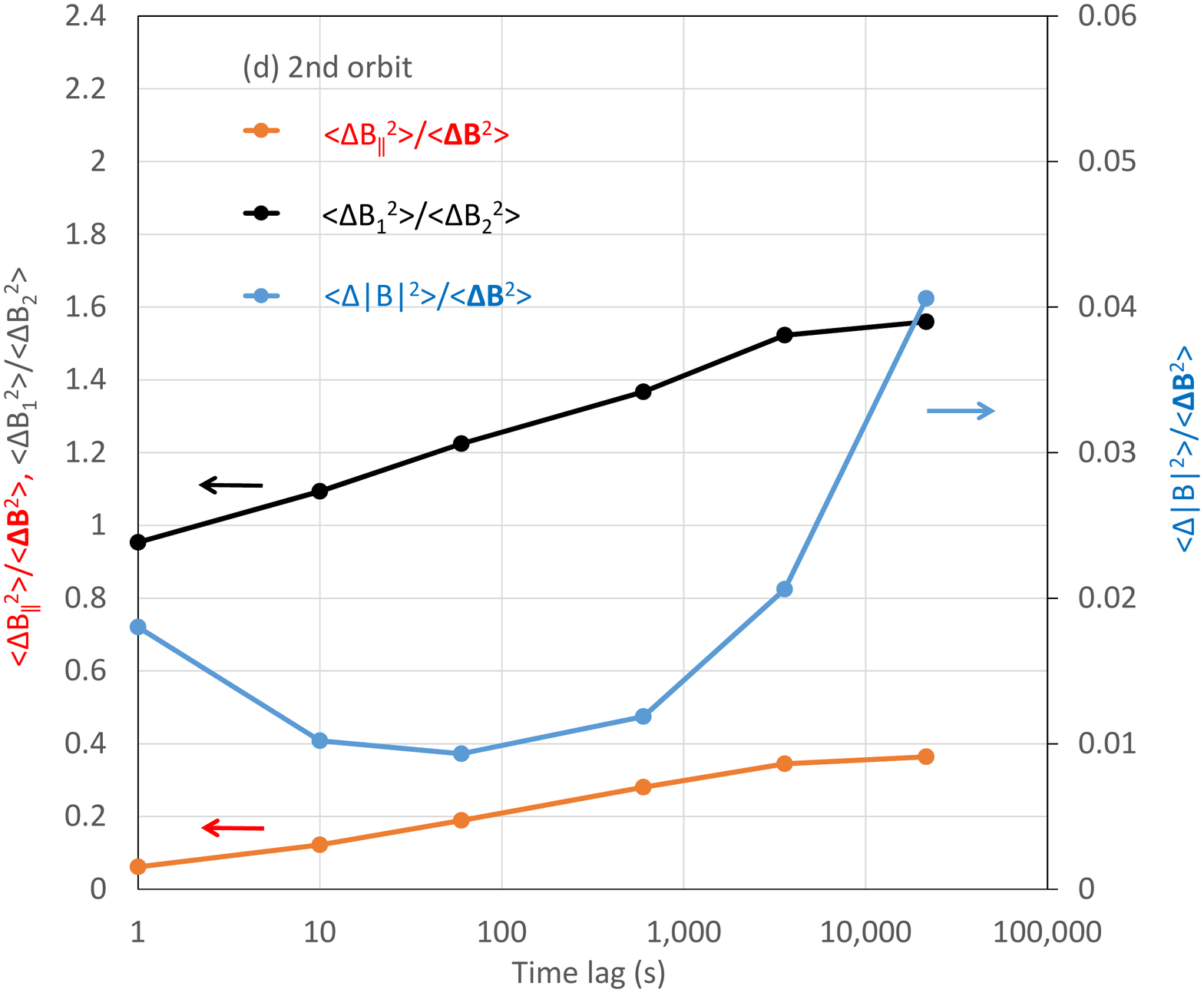}
\caption{
Ratios of variances of magnetic field increments measured by {\it PSP} (a) near 1st perihelion, 2018 Nov 3-10, (b) during 1st orbit, from 2018 Oct 6 to 2018 Dec 19, (c) near 2nd perihelion, 2019 Apr 3-6, and (d) during 2nd orbit, from 2019 Feb 20 to 2019 May 15.  
For varying time lags $\tau$, vector magnetic increments ${\bf \Delta B} = {\bf B}(t+\tau)-{\bf B}(t)$ are calculated and decomposed into a component $\Delta B_\parallel$ parallel to ${\bf B}(t)$  (mostly radial) and two perpendicular components, $\Delta B_1$ in the R-T plane (mostly longitudinal) and $\Delta B_2$ out of the R-T plane (mostly latitudinal).  
The magnitude increment ($\Delta|B|$) is also calculated.  
The low variance ratios of the magnitude increment and parallel increment to the total increment (blue and red curves, respectively)
indicate the near constancy of $|B|$ and magnetic pressure, a characteristic of Alfv\'enic fluctuations.  
The variance ratio of longitudinal to latitudinal increments (black curves) is between 1.4 and 2 for the longest (6-h) increments but decreases to about 1 for the shortest (1-s) increments. 
The anisotropy of longer-time increments is unlikely to originate in or below the inner corona, and can be attributed to longitudinal velocity shear near the Alfv\'en critical zone due to partial corotation, leading to perpendicular field increments that are predominantly longitudinal over large scales and isotropize after a turbulent cascade to smaller scales.
\label{fig:incr}}
\end{figure}

The variance ratio of the magnitude increment to the total increment, $\langle(\Delta|B|)^2\rangle/\langle({\bf \Delta B})^2\rangle$, usually remained below 0.05 throughout both the first and second {\it PSP} orbits, only rarely exceeding 0.2, for all values of $\tau$.
This confirms the basically Alfv\'enic nature of the fluctuations.
Throughout both orbits, there were some special time periods with an unusually low ratio, i.e., especially good magnetic pressure balance.
It turns out that such special time periods occurred near both the first and second perihelia, i.e., during 2018 Nov 3-10 and during 2019 Apr 3-6.  
We display results for this ratio and ratios between variances of increment components, as a function of time lag $\tau$, for these special times near the first and second perihelia and also for the entire data sets of the first orbit (2018 Oct 6 to 2018 Dec 19) and second orbit (2019 Feb 20 to 2019 May 15) in Figure \ref{fig:incr}.

This figure shows that the variance ratio of the magnitude increment to the total increment
(blue curves) is indeed lower for the special periods near first perihelion (a) and second perihelion (c) compared with the full orbits (b and d).  
Yet even for data from the full orbits, the ratio for lags up to 1 h remains below 0.025, confirming the near constancy of $|B|$ and magnetic pressure over such time scales. 
At $\tau=6$ h, the ratio increases, indicating that this time scale is frequently greater than the domain duration; even so the ratio remains below 0.06.

The ratio of the parallel increment variance to total increment variance (red curves) is also quite low ($<0.1$) for the 1-s time lags, and it grows larger for longer time lags according to the near constancy of $|B|$ and the increase in the increment amplitude $|\Delta B|$ for increasing lag $\tau$.
For a small amplitude ($|\Delta B|\ll|B|$), we would expect $\Delta B_\parallel\approx\Delta|B|$.  
However, for large-amplitude Alfv\'en mode fluctuations that maintain constant $|B|$, i.e., for spherical polarization in which ${\bf B}$ remains on a sphere in its component space,  $\Delta B_\parallel$ is directly related to the fluctuation amplitude.  
Here this geometric effect dominates over the actual magnitude fluctuations, with $\langle(\Delta B_\parallel)^2\rangle\gg\langle(\Delta|B|)^2\rangle$ even for our smallest (1-s) lags.

A surprising result from this analysis is an anisotropy between the two perpendicular components of the magnetic field increment.
The variance ratio of roughly longitudinal to latitudinal perpendicular increments, $\langle(\Delta B_1)^2\rangle/\langle(\Delta B_2)^2\rangle$, ranges from 1.4 to 2 for the longest (6-h) lags while decreasing to about 1 for the shortest (1-s) lags.  
The transition seems to relate to the correlation time of several minutes.
This anisotropy for long lags persists throughout both orbits, though on average it is particularly strong during time periods with better magnetic pressure balance such as the times close to perihelia.
As such it does not appear to be related to the direction of the {\it PSP} orbital motion, which varies strongly and systematically throughout the orbit.

This anisotropy of magnetic increments for long $\tau$ is unlikely to originate in or below the inner corona, in which the latitudinal and longitudinal directions are not strongly distinguished.
It can be understood in terms of velocity shear above the Alfv\'en critical zone between flux tubes with varying degrees of longitudinal corotation, leading to perpendicular field increments that are predominantly longitudinal over large scales and then isotropize after a turbulent cascade to smaller scales.

\subsection{Cross Helicity and Signatures of Velocity Shear}

In terms of a volume average, here designated as
$\langle \cdots \rangle$, the cross helicity may be defined as
\begin{equation}
    H_c \equiv \langle {\bf v} \cdot {\bf b} \rangle
    = \frac14 \left( Z^2_+ - Z^2_-\right ) 
    \label {eq:Hc}
\end{equation}
where the fluctuations in magnetic field
are computed in Alfv\'en units as 
${\bf b} = ({\bf B} - {\bf B}_0)/\sqrt{\mu_0\rho}$,
and the Els\"asser energies are 
$Z^2_+ = \langle |{\bf v} + {\bf b}|^2 \rangle$
and $Z^2_- = \langle |{\bf v} - {\bf b}|^2 \rangle$.
The traditional view is that the $\pm$ Els\"asser
fluctuations ${\bf z}^\pm = {\bf v} \pm {\bf b}$
comprise wave packets that propagate either 
along the ${\bf B}_0$ direction (${\bf z}^-$),
or opposed to it (${\bf z}^+$).
This definition corresponds to and generalizes 
the large amplitude eigenmodes described in the previous 
section (cf. Eq. \ref{eq:alf}). $H_c$ is an ideal invariant of the 
incompressible MHD system and has significance 
whether or not a mean magnetic field ${\bf B}_0$ is present.

\add{
It is well known that in the inner heliosphere, solar wind fluctuations have a strong cross-helicity in the sense that propagation is dominantly outward \citep{BelcherDavis71}.
In our simple cartoon (Figure \ref{fig:cartoon}), if all the 
magnetic
field is outward and the fast outward streams are located in the 
flux tubes with 
weaker magnetic field, 
then the 
cross-helicity of long-wavelength fluctuations
will be negative and they will travel outward except in 
switchback 
regions.
If the prevailing magnetic polarity is inward (as it is during E1 and E2), 
in the context of our cartoon the faster streams should still be 
in the weaker flux tubes, 
to give a positive cross helicity and the observed outward propagation. In a more complete description of the solar wind there is also likely to be a broadband spectrum of more standard outward propagating Alfv\'enic fluctuations.
}

In {\it PSP} observations close to the Sun, the 
cross helicity measured by the ratio
$\sigma_c = (Z_+^2 - Z_-^2)/( Z_+^2 + Z_-^2)$
is generally quite large, suggestive of 
a preponderance of outward traveling Alfv\'en waves. 
There are departures from Alfv\'enicity for increments at small lags, including reduced cross-helicity, as reported by \citet{ParasharEA20ApJS}.
At the same time, our present analysis shows that the magnetic magnitude increment is quite small for small time lags (Figure \ref{fig:incr}), which is consistent with mostly Alfv\'en-mode fluctuations. 
This could be because above the conventional Alfv\'en point velocity shears can begin supplying turbulence energy \citep{ZankEA96,BreechEA08} that remains
nearly incompressible but not entirely outward-directed.


\section{Results: 3D Compressible MHD Simulation of Mixing Layer Dynamics}

Having examined several 
plasma and magnetic field diagnostics in the {\it PSP} data,
we now turn to the results of 
more local compressible MHD simulations,
emphasizing points of comparison with 
the observations.
We seek to examine further possible 
points of consistency 
with plasma dynamics driven by 
nonlinear mixing layer dynamics as 
envisioned in \S \ref{hypothesis}
for the transition between striation and flocculation outside
the Alfv\'en critical zone.

\begin{figure}
    \centering
    \includegraphics[width=.45\columnwidth]{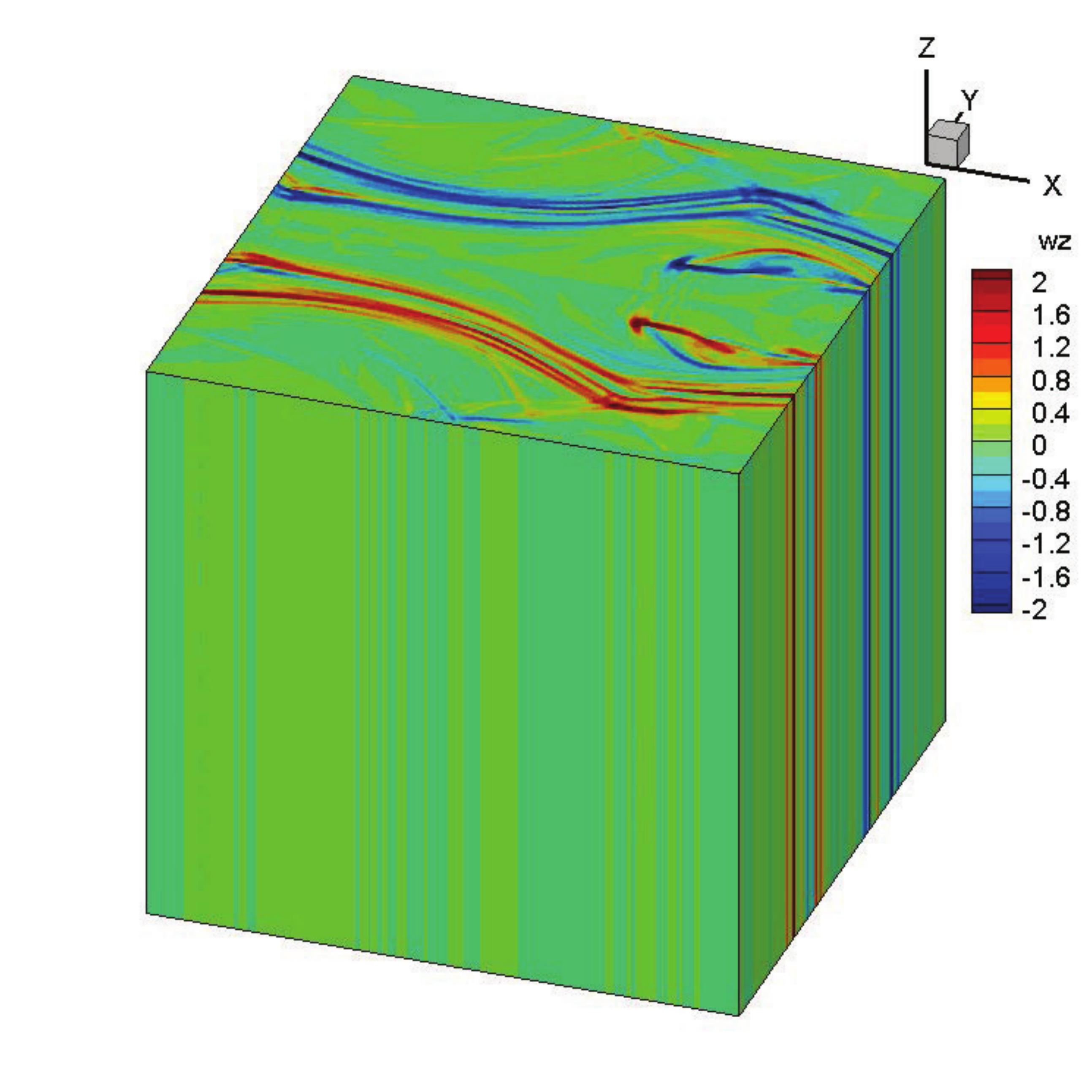} \includegraphics[width=.45\columnwidth]{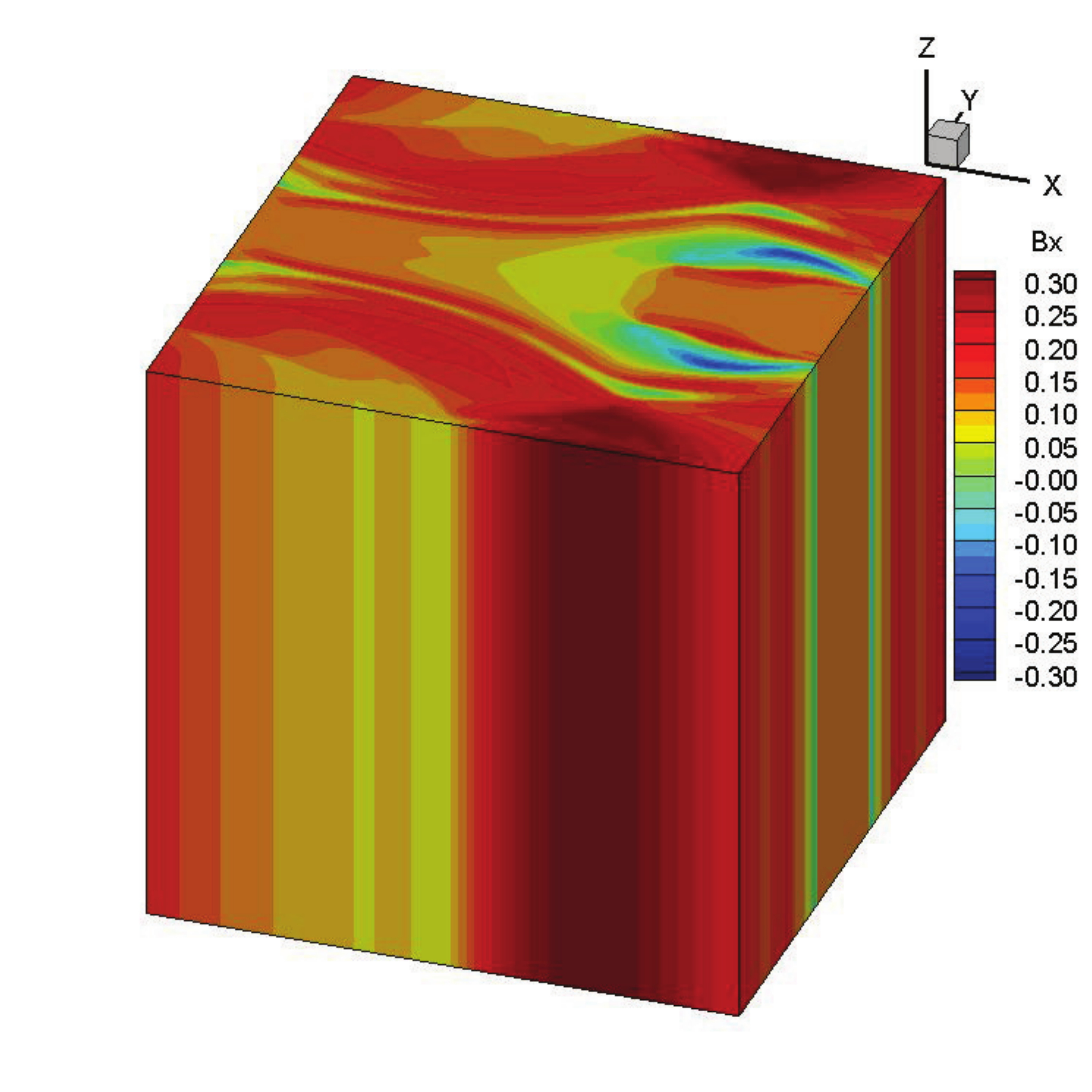}
            \caption{(Left) Vorticity in the $z$-direction and (right) magnetic field in the $x$-direction (shown by color scales) from 3D compressible MHD simulation at $t=120$.
            Vortex rollup, inhibited by the magnetic field, is just beginning  to take effect. 
            Initially flow reversal occurred across two thin planes at $y=L_y/4$ and $3L_y/4$ and the magnetic field was initially entirely in the $+x$-direction, stronger outside those two planes and weaker between them.
            }
    \label{fig:simt120}
\end{figure}

\begin{figure}
    \centering
    \includegraphics[width=.45\columnwidth]{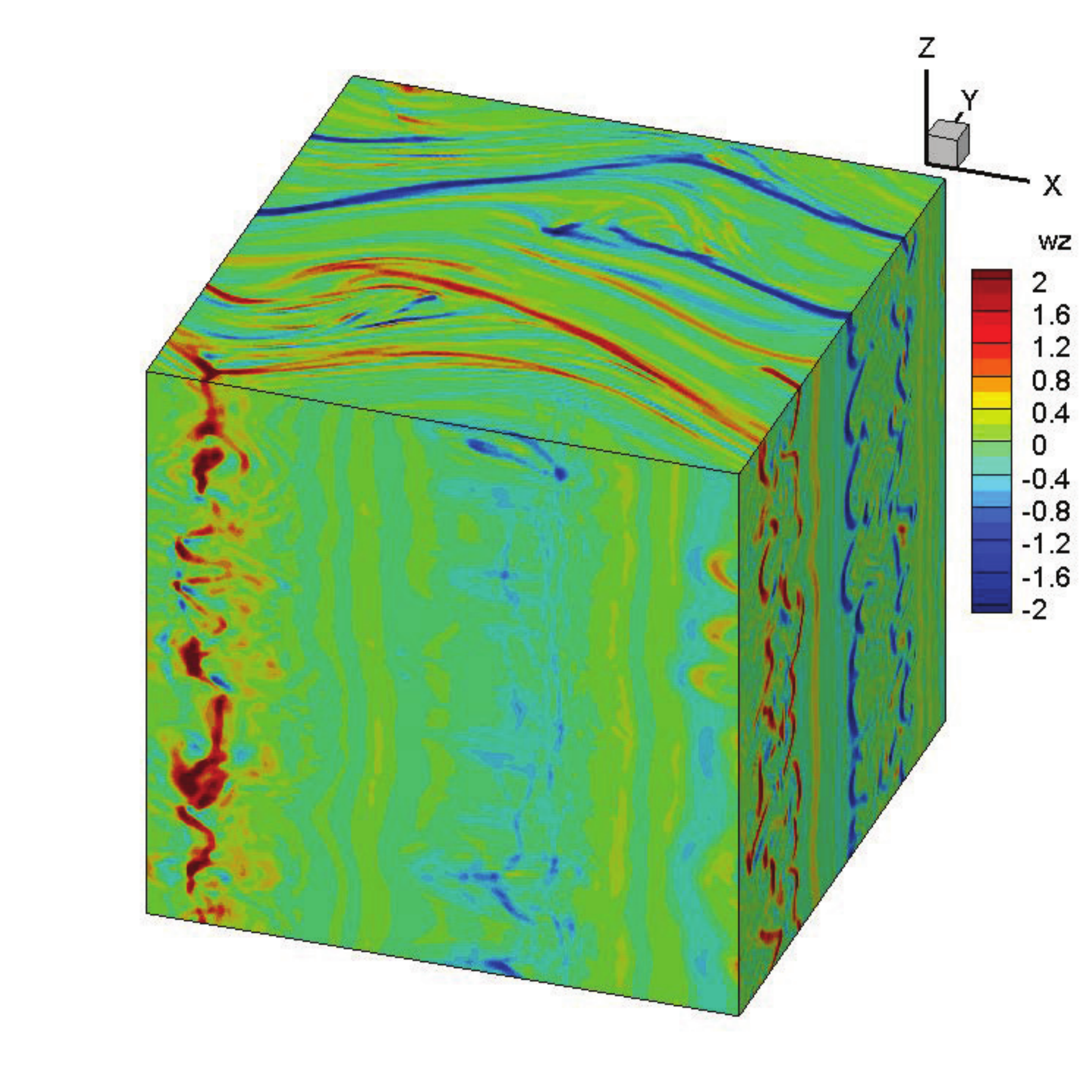} \includegraphics[width=.45\columnwidth]{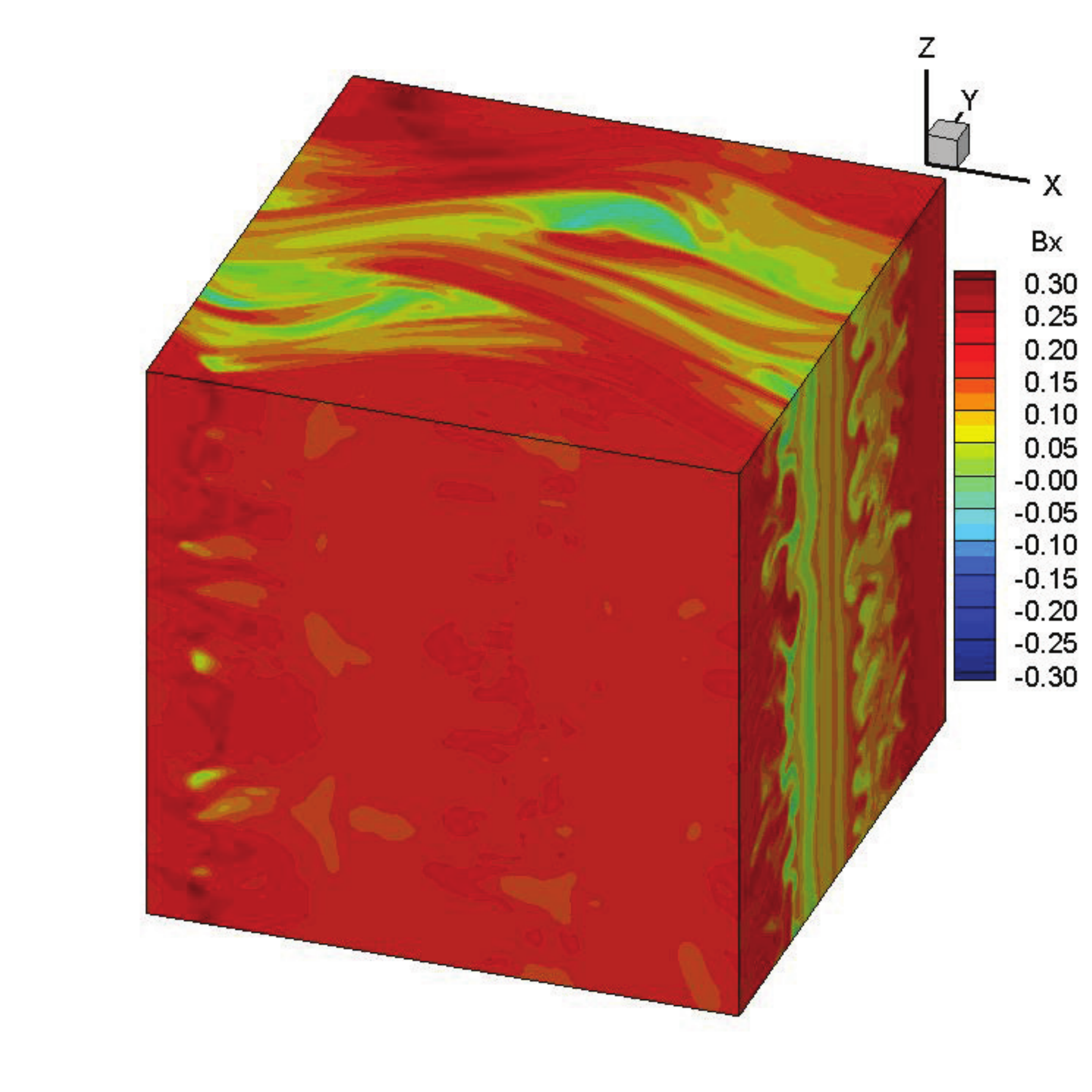}
            \caption{(Left) Vorticity in the $z$-direction and (right) magnetic field in the $x$-direction (shown by color scales) from 3D compressible MHD simulation at a later time, $t=290$.  Vortex rollup is now well developed, producing two switchback regions, which recur intermittently throughout the simulation run. 
            }
    \label{fig:simt290}
\end{figure}

\subsection{Compressible 
MHD simulation results}

We carried out a number of compressible MHD runs using the approach outlined in 
\S \ref{mhdsims}.
The unperturbed initial state in all cases is based
on two periodic planar shear layers between co-linear magnetic field regions that change strength 
across the same regions as the 
shear layers. 
Therefore the sheetlike regions of 
electric current density approximately coincide with 
layers of vorticity. This is to emulate in a simple way the juxtapositioning of parallel 
flux tubes with varying axial velocity fields, as suggested in Figure \ref{fig:cartoon}.
We choose this simple shear configuration to test our hypothesis in a simple form: Can organized flows and magnetic fields that are initially oriented in one direction give rise to Kelvin-Helmholtz dynamics and plasma signatures such as those observed by {\it PSP}?
In the actual solar wind, as we have pointed out there is at least partial corotation in the longitudinal direction (see Section 4.3) and, we believe, also some slippage of individual flux tubes and longitudinal velocity shear as well (see Section 4.4).

The baseline parameters corresponding to the results shown here
were given in Table \ref{tab:parameters}.
Different runs (not shown)
were done in two dimensions and in three dimensions, with 
varying velocity contrast 
$\Delta U$ 
across the shear layers, 
and several values of  
uniform $B_x$ in the strong and weak magnetic field regions. 
Within the range of 
parameters that were varied, the results were all similar; therefore 
we show just one case in the diagnostics here, as 
described in Table \ref{tab:parameters}.

Beginning from 
the initial state described above, the dynamics proceed 
along the lines of a hydrodynamic mixing layer.
The initially planar vorticity layers distort 
due to the early stages of vortex rollup.
The magnetic field is too weak to 
prevent the distortion from reaching macroscopic dimensions. A snapshot of this state is shown in the two panels of Figure 
\ref{fig:simt120},
where the breakup 
of the vortex layers begins along with large magnetic field  directional deflections
and small regions 
of weak field in which 
the polarity reverses. 
Figure \ref{fig:simt290} describes the state of the system later, at simulation time $t=290$, when the shear-driven 
dynamics are more fully developed and clearly in a nonlinear stage.   In particular, the phenomenon of rollups has noticeably emerged, where the vortex layers have folded.  
Shocklets have formed at locations at which the flow direction change is relatively abrupt. Perhaps most importantly, there are now large transverse velocities in both the positive and negative $\hat y$ (vertical)
directions. The large deflected motions have carried along the magnetic field (right panel), which also exhibits large transverse deflections. 
Examination of the sign and magnitude of the streamwise $\hat x$-direction magnetic field  
(indicated by the color legend)
indicates the presence of regions
of polarity reversals, or ``switchbacks''. Here these are entirely caused by nonlinear instability 
driven by the initial shear layers. 

\begin{figure}
\begin{center}
\includegraphics[width=0.45\columnwidth]{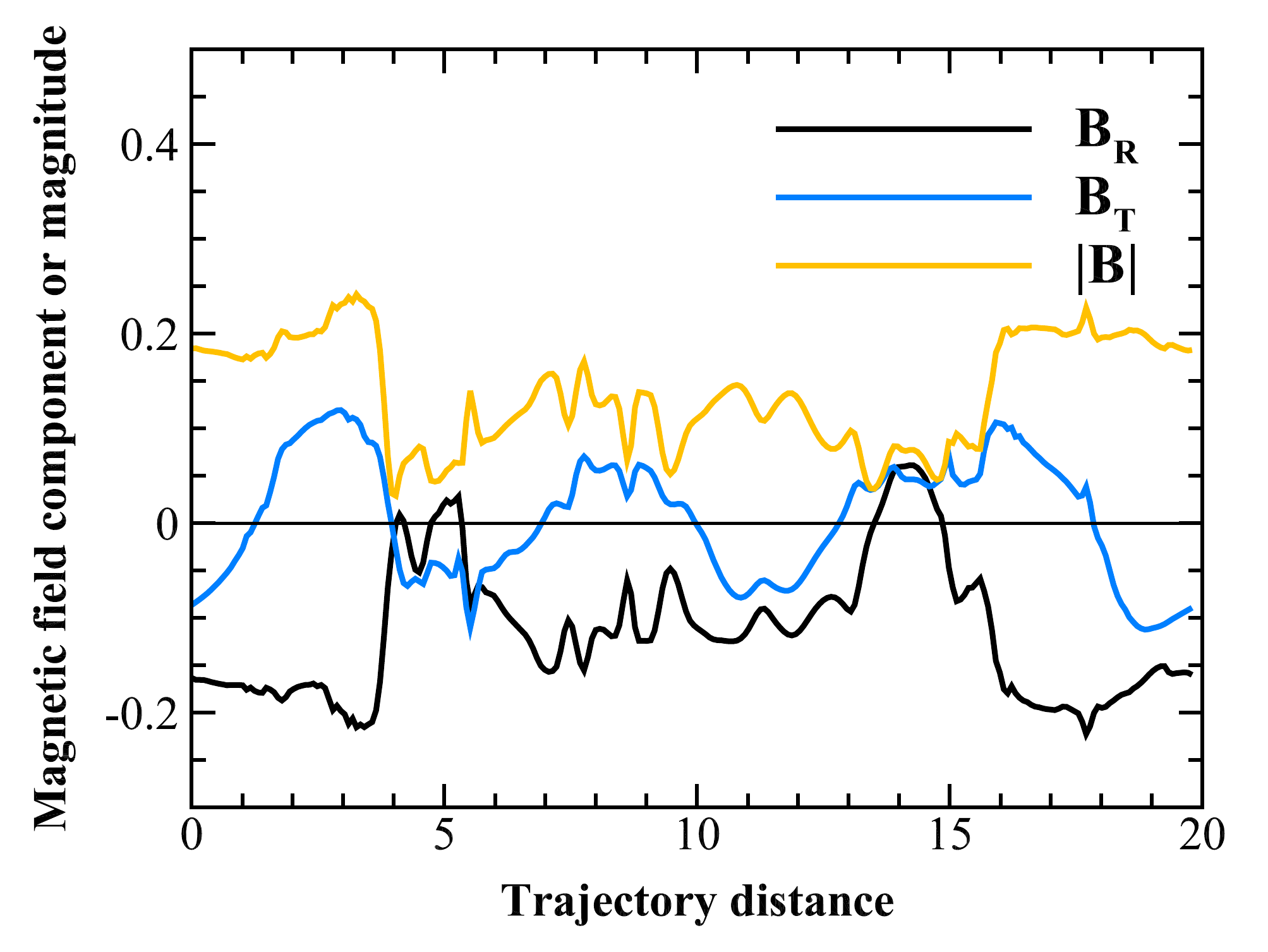}
\includegraphics[width=0.38\columnwidth]{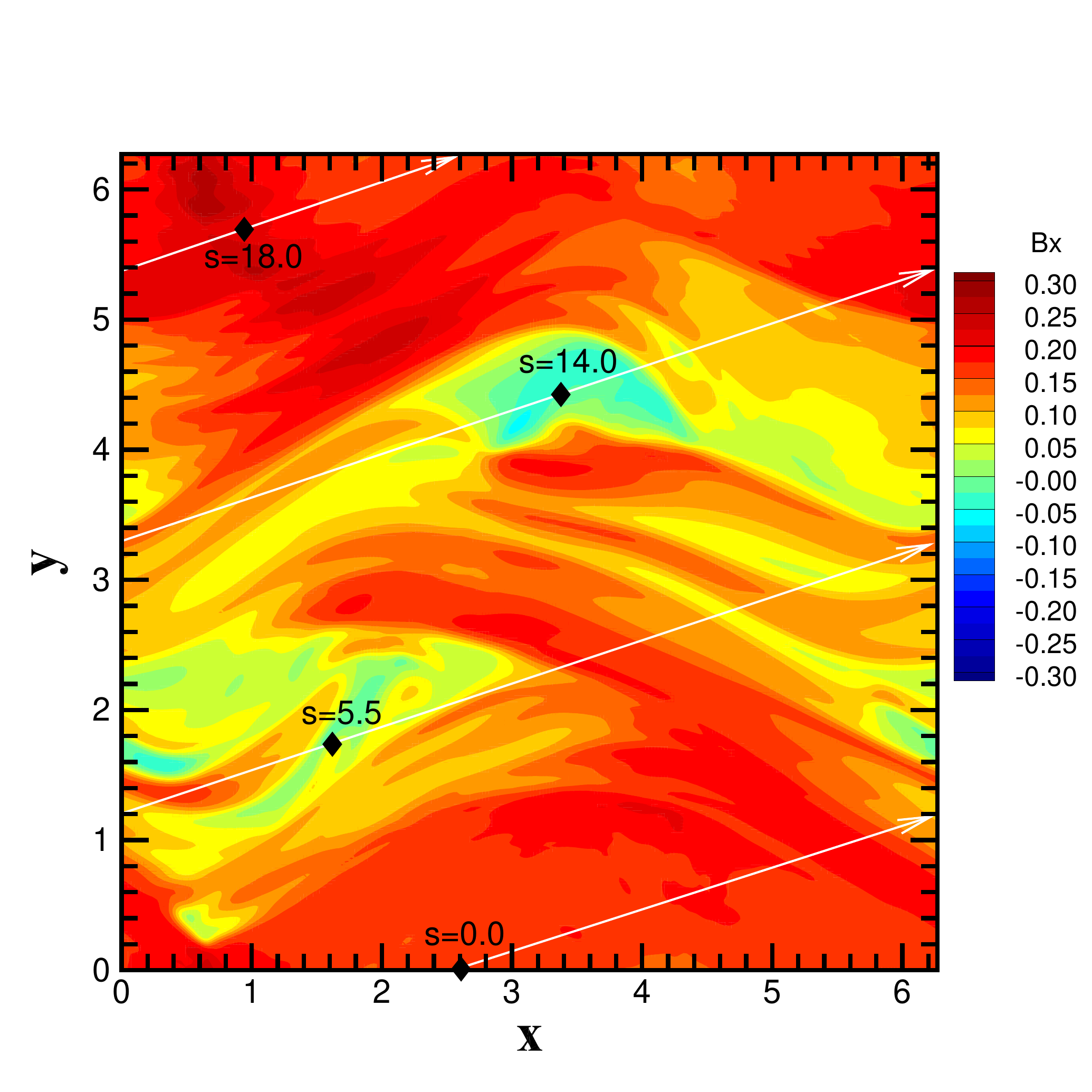}
\caption{\add{(left)} Magnetic field components and magnitude from the 3D compressible MHD simulation at time $t=290$. 
Note the presence of ``switchbacks'', i.e., reversals of $B_R$, as well as large symmetric fluctuations of transverse
$B_T$. The magnetic field
magnitude $|{\bf B}|$
is relatively constant within regions delineated by the proximity of switchbacks.
This figure can be qualitatively compared with Figure \ref{fig:Bflucts}.
\add{(right) The simulation plane and trajectory employed to obtain the data in the left panel.
The trajectory is drawn, annotated with reference distances $s$.} 
\label{fig:Bsimcut}}
\end{center}
\end{figure}

A complementary graph of the magnetic field components is shown in 
Figure \ref{fig:Bsimcut}
for time $t=290$ of the MHD simulation.
The spatial structure is sampled as a function of distance 
\add{$s$} along a trajectory at an 
\add{$18^\circ$} 
angle relative to the axes of the box that threads through the (periodic) box several times, to produce a one-dimensional series that spans about ten correlation lengths, similar to the {\it PSP} data sample shown in 
Figure \ref{fig:Bflucts}.
\add{For reasons to be discussed shortly, we associate the $x$-direction along the mean field in the simulation with the $-$R direction for {\it PSP} measurements, and the $y$-direction with the T direction, so the left panel uses $B_R=-B_x$ and $B_T=B_y$.}
Note the region of approximately constant $|B|$ and the presence of switchbacks in the simulation plot, in qualitative accord with the {\it PSP} data.

We will now show several 
diagnostics that permit 
a quantitative comparison 
of several features of the simulation and the observations by 
{\it PSP}. 
We do not expect precise correspondence because the simulation setup is a vast oversimplification of the interplanetary physical system. But if our conjectures are correct concerning the basic physics that drives the evolution in the solar wind, then we might find some consistency in the comparisons.


\begin{figure}
    \centering
    \includegraphics[width=.5\columnwidth]{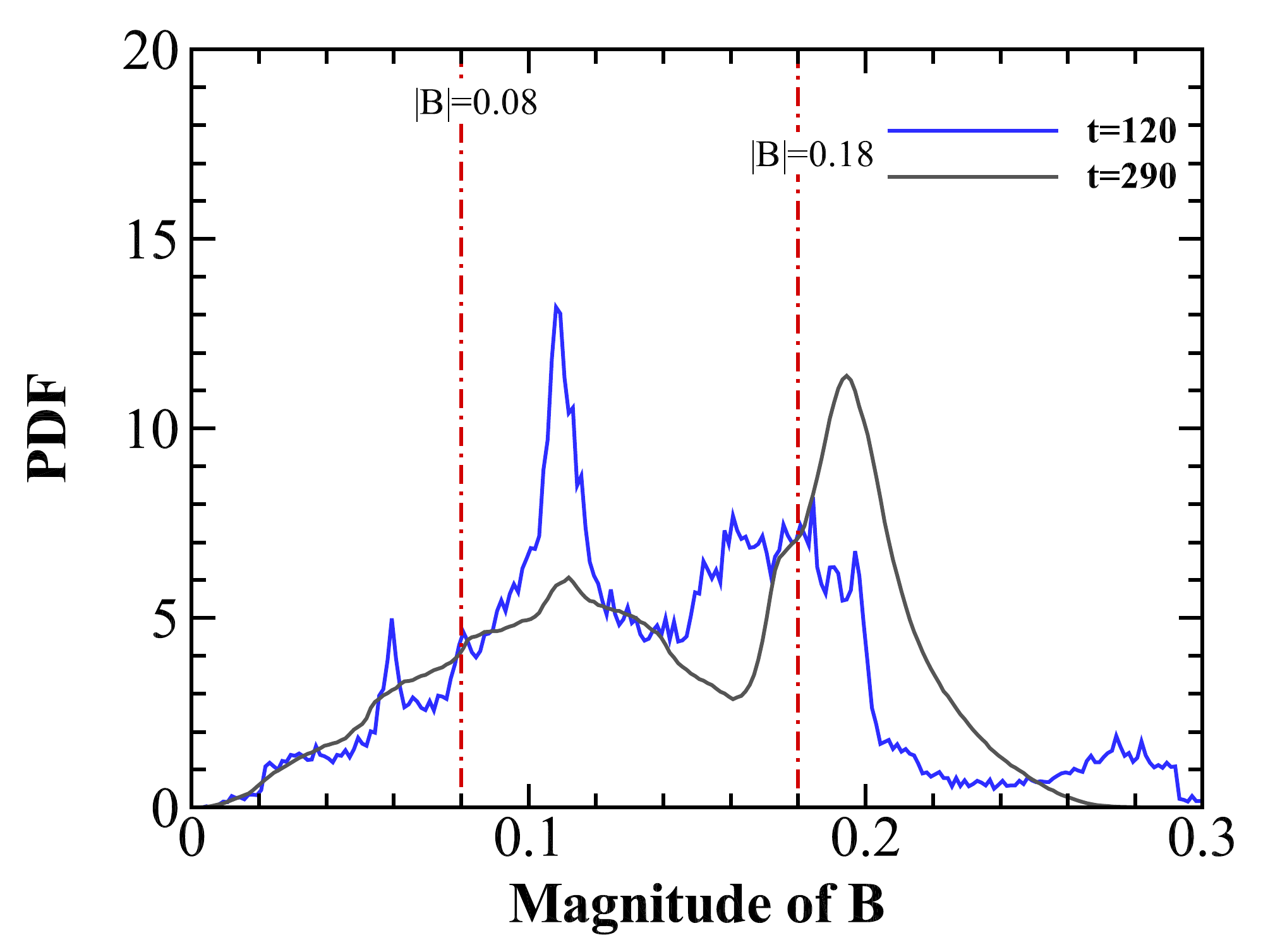}
        \caption{
         PDF of magnetic field magnitude from the 3D compressible MHD simulation at simulation times $t=120$ and $t=290$. Initial magnetic field values are annotated.  This figure can be qualitatively compared with Figure \ref{fig:B}(b). 
        }
    \label{fig:Bmagsim}
\end{figure}

\subsection{Comparison with {\it PSP} data: 
Magnetic Field Magnitude}

A special property of 
large amplitude Alfv\'en 
waves, a constant magnitude $|B|$, 
is apparently a familiar property in MHD 
turbulence. Constant magnitude patches or regions
have also been observed in {\it PSP} data, 
as shown for the first {\it PSP} encounter in Figures \ref{fig:Bflucts} and \ref{fig:B}.
An analysis of 
the shear-driven MHD simulation results
also shows a similar distribution of 
magnetic field magnitude, as illustrated 
in Figure \ref{fig:Bmagsim}.
Note that at the earlier time, $t=120$, the distribution has peaks associated with the initial conditions, in which $|B|$ was concentrated at two initial values, shown by vertical dot-dashed lines at 0.08 and 0.18.
These peaks have smoothed somewhat by the later time $t=290$, representing a more developed dynamical state that we consider comparable to the solar wind at initial {\it PSP} perihelia, somewhat downstream of the Alfv\'en critical zone.  
Even at the later simulation time one observes in the 
PDF of $|B|$ the presence of preferred values of 
$|B|$ or sub-distributions as seen in the {\it PSP} data.

\begin{figure}
    \centering
    \includegraphics[width=.46\columnwidth]{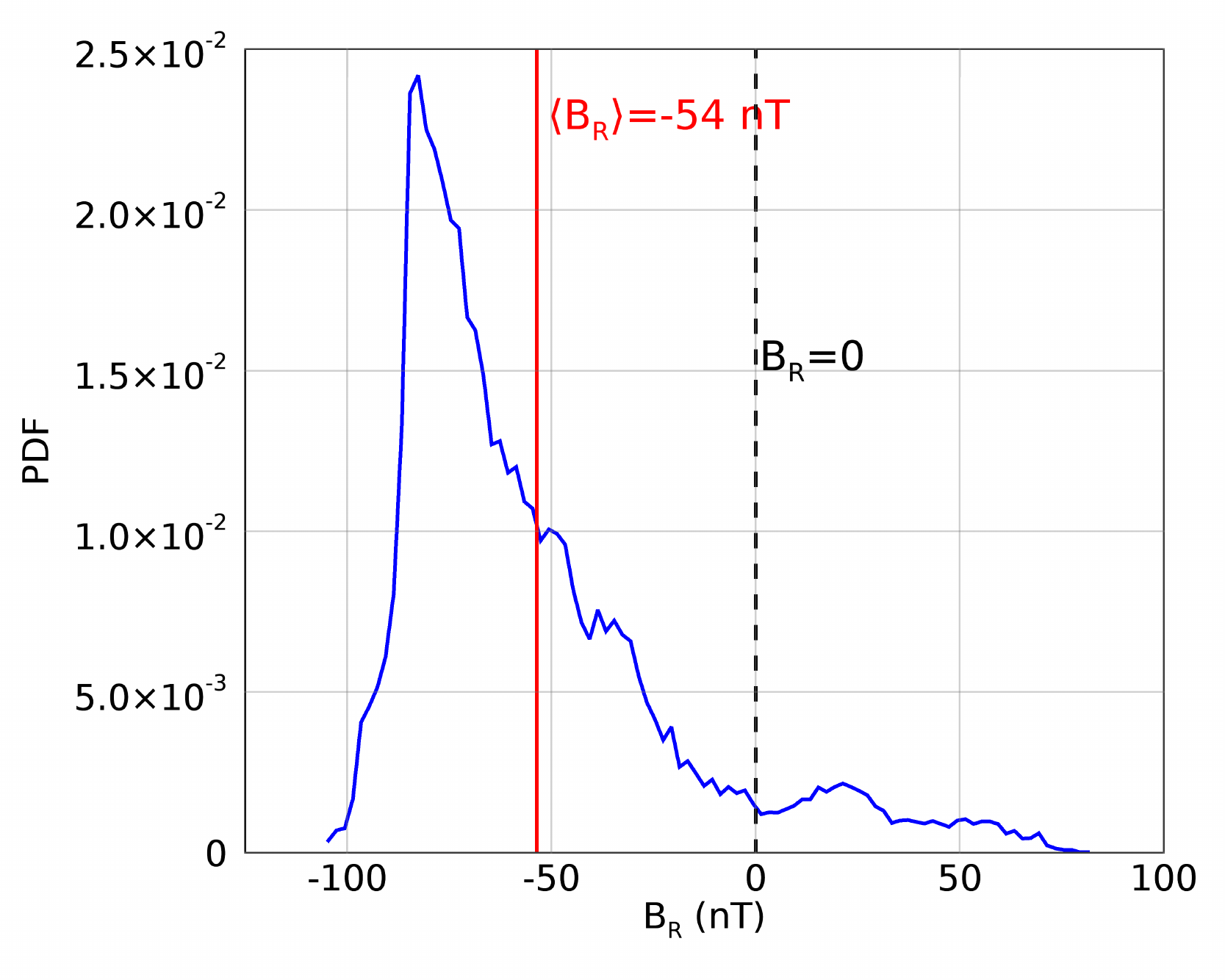}
    \includegraphics[width=0.46\columnwidth]{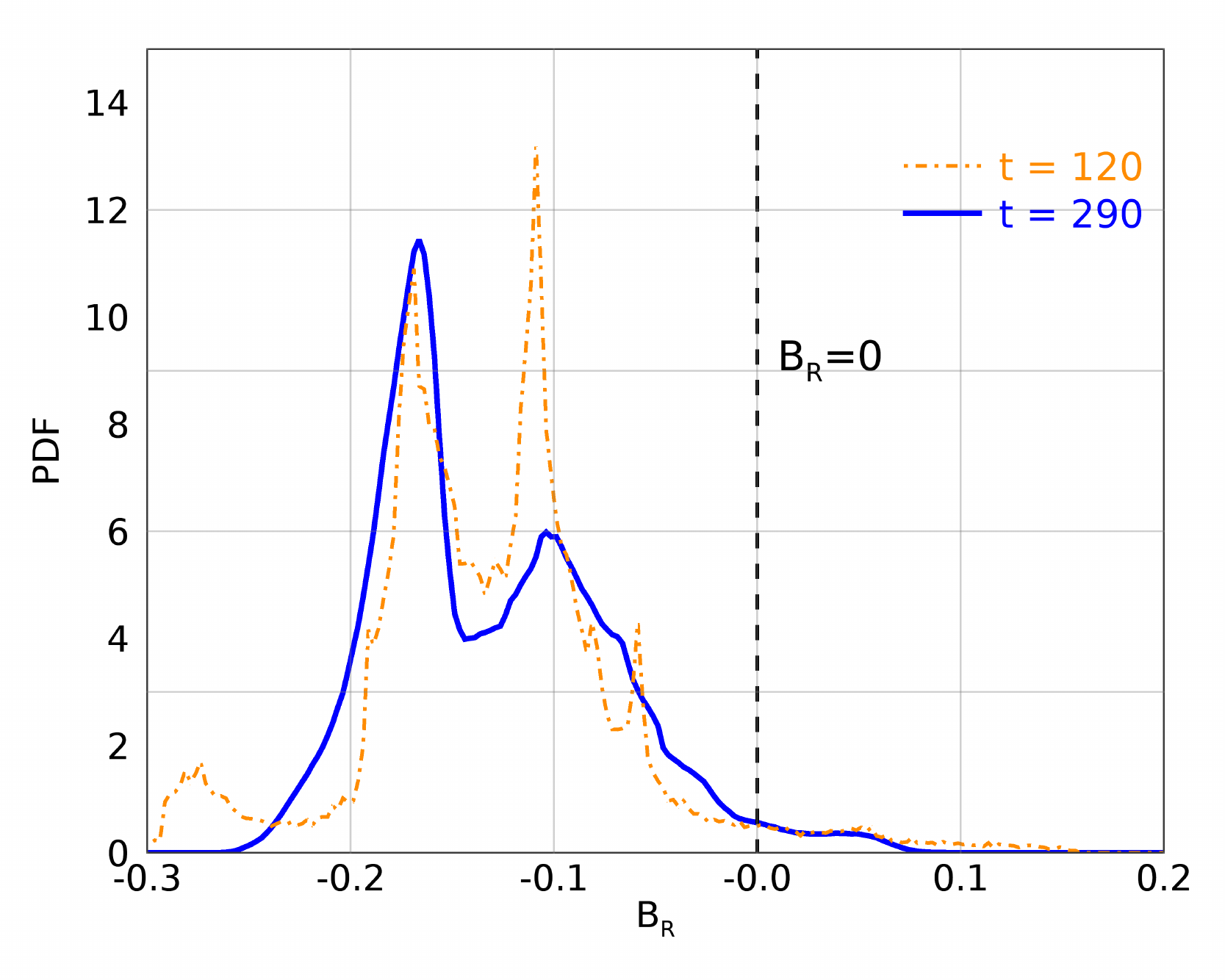}
            \caption{(Left) PDF of radial magnetic field from 8 hours of data that include {\it PSP}'s 1st perihelion; (right) PDF of ``radial'' $x$-component of magnetic field at two times from the shear-driven 3D compressible MHD simulation.
            }
    \label{fig:pdfswitchbacks}
\end{figure}

\subsection{Comparison with {\it PSP} data: 
Magnetic Switchbacks}

Considerable attention has been given to the appearance  
of switchbacks in the {\it PSP} data \citep{BaleEA19Nature, DudokDeWitEA20}
where, as mentioned earlier, they are seen more dramatically 
near perihelion than in the more distant solar wind.
While many switchbacks 
are seen in the first orbit, as in Figure \ref{fig:stacked},
a close up look at one hour of data, as in Figure \ref{fig:Bflucts},
shows that most of the solar wind is filled with unipolar negative radial field. 

For the present purposes it is of interest to compare the 
frequency of occurrence of reversed polarity magnetic fields in the presence of a dominant magnetic 
polarity and strong shear. 
In this way one can compare    
switchbacks from {\it PSP} with 
magnetic polarity reversals in our standard MHD simulation, 
noting that for most of E1 and E2 the large-scale magnetic field at {\it PSP} was nearly in the $-$R direction, which we associate with the $x$-direction along the mean field in the simulation, whereas a {\it PSP} measurement of a transverse component such as T corresponds to the component along $y$, the direction that cuts across shear layers in the simulation.
To this end we compute 
the distribution of the radial magnetic field
component $B_R$ from 8 hours of {\it PSP} data near first perihelion (2018 Nov 6, 0000-0800 UT) and compare this to 
the distribution of $B_R\equiv - B_x$ from the simulation at two times. 
This comparison is presented in the two panels of Figure \ref{fig:pdfswitchbacks}.

The qualitative features of these distributions are quite similar: 
The {\it PSP} data show one strong peak at a dominant negative polarity, with a shelf-like distribution that extends to positive polarity 
values, indicating switchbacks. 
In the simulation there are two preferred values of dominant polarity at the earlier time shown, and a single strong peak at the later time that we believe better represents the more developed dynamics downstream of the Alfv\'en critical zone. 
At both times the simulation shows a relatively flat, low-level, shelf-like distribution of reversed polarity, much like the observed distribution in the left panel. 

The simulation data in Figure \ref{fig:pdfswitchbacks}, as well as analysis of several other  simulations we have carried out,
demonstrate that switchbacks occur with similar frequencies in 
MHD mixing layer dynamics and in the {\it PSP} data near first 
perihelion.

\subsection{Comparison with {\it PSP} data:  Transverse Velocities}

\begin{figure}
    \centering
    \includegraphics[width=.45\columnwidth]{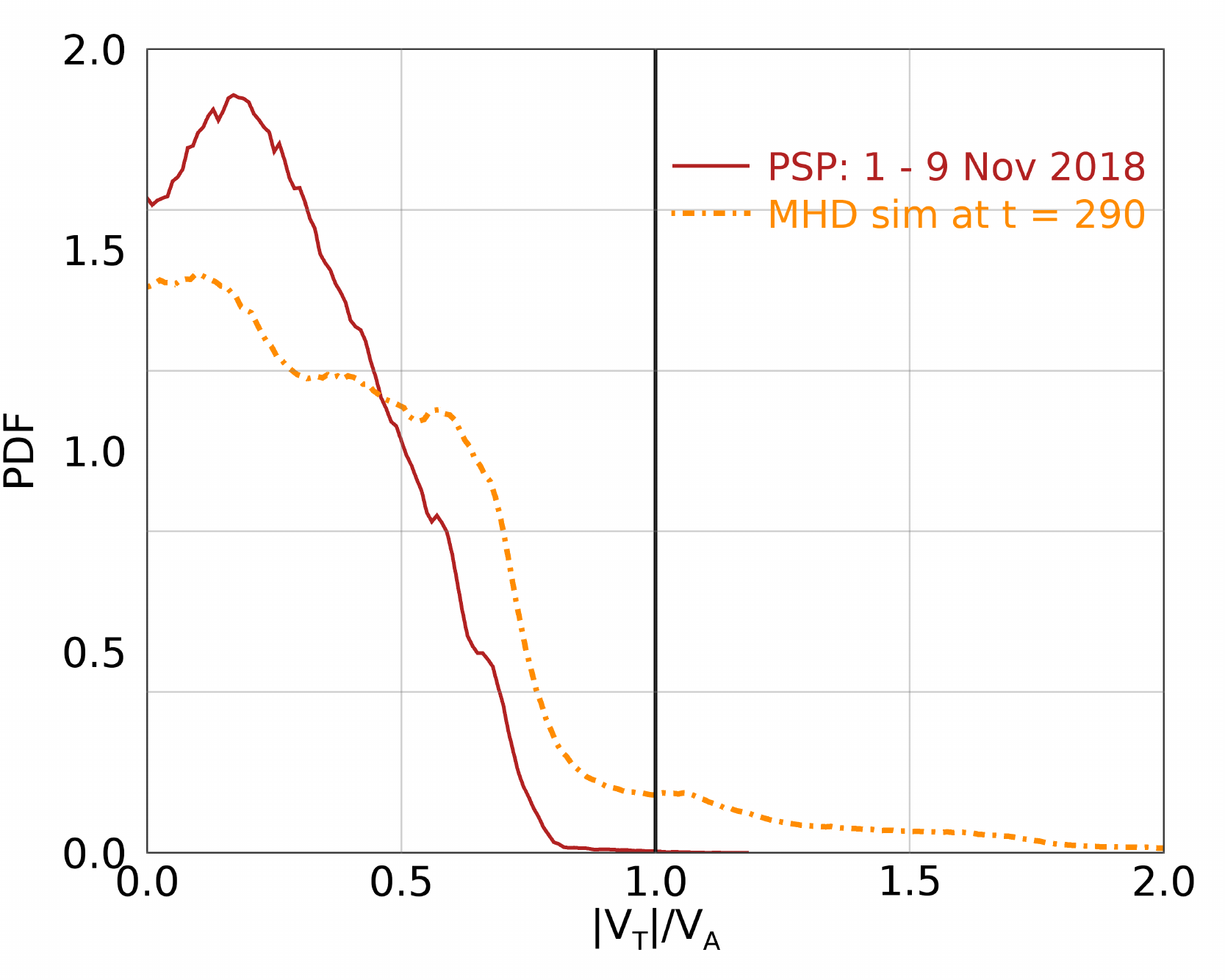}
            \caption{
            \add{Solid red curve shows the PDF} of the transverse velocity normalized to the
            \add{1-hour running average of the}
            Alfv\'en speed
            during the first {\it PSP} encounter. 
            \add{Dash-dotted orange curve shows the PDF of the transverse velocity} from the shear-driven 3D compressible MHD 
            simulation, \add{normalized to the local Alfv\'en speed}.
            Compare with Figures \ref{fig:V_Va-E1} and \ref{fig:V_Va-E2}.
              }
    \label{fig:pdfVT}
\end{figure}

The {\it PSP} results shown above demonstrate that the transverse 
velocity components near first {\it PSP} perihelion
are essentially bounded by the local or 
neighborhood value of 
the Alfv\'en speed. 
It is also interesting,
using the same normalization, 
to compare the distribution of 
the $V_T$ component of velocity 
in the {\it PSP} data with 
the corresponding ``non-radial'' (transverse) component 
$V_y$ of the plasma velocity in the MHD simulation
data.
 One can see in Figure \ref{fig:pdfVT}
that 
the local Alfv\'en speed  represents
an approximate limit that 
constrains the 
dynamics in both the simulation and {\it PSP} observations, which is consistent with the reasoning above (see Section 4.3). 


\section{Discussion and Conclusions}

Motivated by {\it STEREO} observations \citep{DeForestEA16}, 
we have examined
the possibility that shear driven dynamics 
drive  
the transition from elongated, or striated, structures in the 
lower corona to more isotropic, or flocculated, structures above the 
the Alfv\'en critical zone.  The associated release of 
energy in the sheared flows represents 
an additional source of energy over and above the preexisting 
Alfv\'enic turbulence originating at lower coronal altitudes.
If the above hypothesis is correct, this transition
signals an enhancement in outer scale 
turbulence energy that persists to much larger heliocentric distances.
This hypothesis has been examined here beginning with 
clues from imaging, and further
motivated by the 
large turbulence amplitudes
seen in global MHD
simulations. 
{\it Parker Solar Probe} provides an opportunity to begin detailed analysis 
of the consequences of the shear driving hypothesis. 
The first results of this analysis have been 
presented in some detail here.

\begin{figure}
    \centering
    \includegraphics[width=.55\columnwidth]{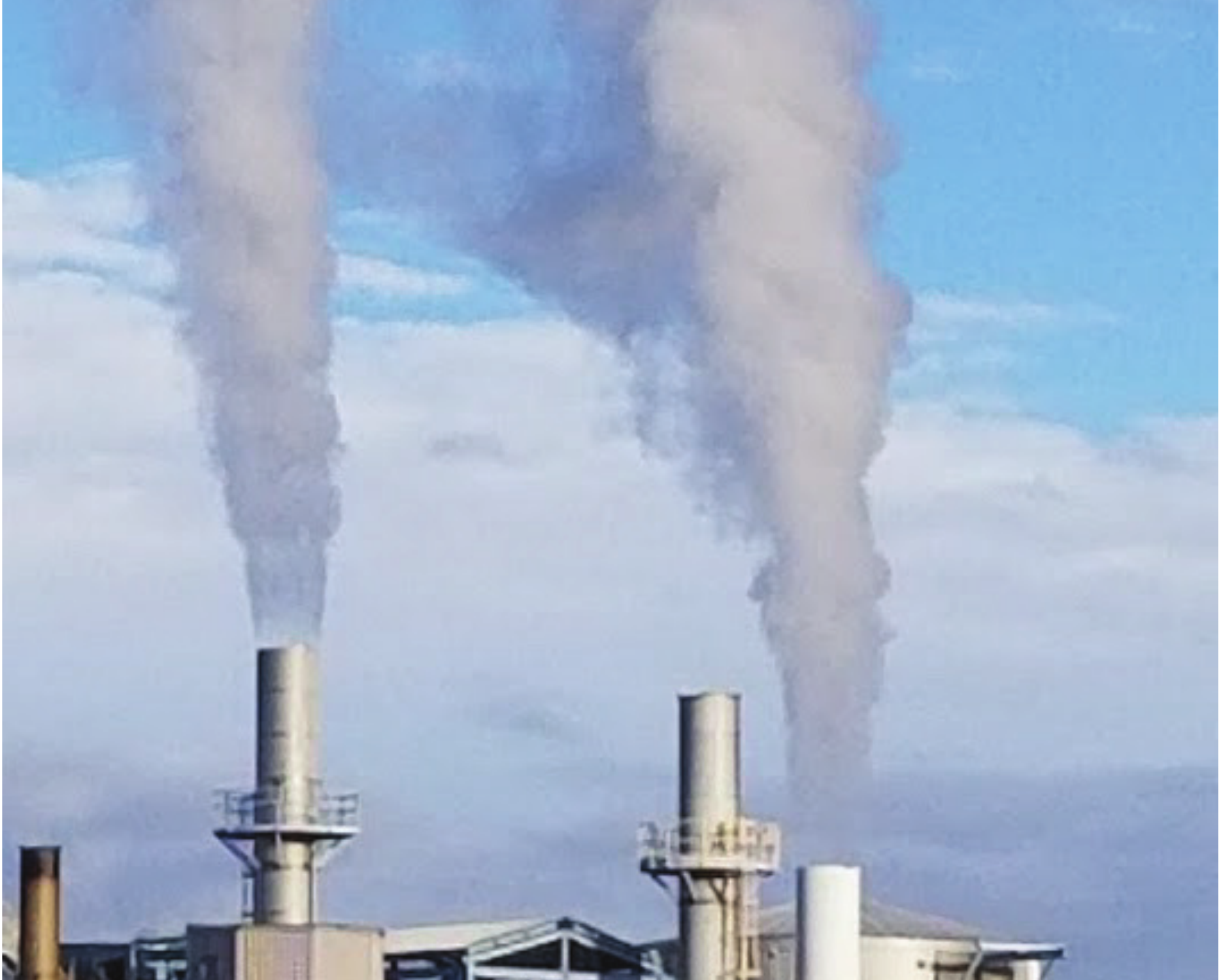}
              \caption{
            Images of plumes above smokestacks in Newark,
            Delaware, USA with escaping vapors exhibiting a sequence of changes
analogous to what is envisioned for the shear driven dynamics of the young solar wind.
Upon escape from the constraining smokestack, the plume is initially 
well-collimated. Roll-ups are initiated near the edges due to shear. 
At greater distances the plumes become wider and more isotropic.  
              }
    \label{fig:twostacks}
\end{figure}

The basic picture is that of a magnetized parallel mixing layer in which the velocity contrasts are large enough to cause significant deflections and even reversals of the magnetic field. 
The salient features of the 
mixing layer are well known in hydrodynamic, engineering 
and practical applications; a common example is shown 
in Figure \ref{fig:twostacks}.
The magnetic field parallel to the flow 
presents a complication in that it resists deflection.
However, as anticipated in theory \citep{Chandrasekhar,MiuraPritchett82}
and demonstrated in simulations \citep{GoldsteinSSPP89,RobertsJGR92,
MalagoliEA96,LandiEA06},
a sufficiently strong shear in comparison with the ambient Alfv\'en 
speed will produce the typical Kelvin-Helmholtz-type rollups.
We show that, in principle,
episodic switchbacks can be accounted for by {\it in situ} solar wind dynamics, not requiring nonradial inputs from the Sun or its inner corona.

The 
heliospheric simulations that confirm the likelihood 
of such conditions in the solar wind are 
consistent with a number of 
previous observations \citep{Borovsky16,UsmanovEA18,HorburyEA18,LockwoodEA19}.
Indeed the expected amplitude of turbulent fluctuations
inferred from turbulence modeling suggests that 
large amplitude departures from the 
laminar state, including switchbacks, 
may be anticipated as {\it PSP} perihelia migrate 
inward towards the Alfv\'en critical zone.
It is also important to recognize that 
the fluctuations that propagate and convect outward from the lower corona and into 
the critical zone from below may not 
be uniform or homogeneous. 
In fact, it is well established in 
observations that the vertical velocity of 
fluctuations may vary considerably, with typical variations due to 
type II spicules \citep{DePontieuEA09} in the range of 50-100 km s$^{-1}$. 
Similar contrasts in radial velocity are seen throughout the inner corona in 
analysis of deep exposure {\it STEREO-A}/COR2 coronagraph images \citep{DeForestEA18}. These types of fluctuations may also cause magnetic reversals \citep{SamantaEA19}
that might propagate upwards and possibly survive to the Alfv\'en critical
zone. If they do survive to the Alfv\'en critical zone, 
these fluctuations would be expected, under the right detailed 
conditions, to contribute to
driving of turbulence through nonlinear instability.  

{\it PSP} now has made several passes through the outer 
parts of this region, and provides substantial data relevant to the present suggestions.  
Here we have examined the time series, and the distributions
of magnetic field magnitude, radial magnetic field and transverse 
velocities. All of these appear to be consistent with expectations and simulation results for
a shear-driven dynamics scenario.  
Furthermore, the domains of Alfv\'enic fluctuations are consistent with mixing layers that grow and/or merge with distance from the Alfv\'en critical zone, and we find anisotropy among the perpendicular magnetic increments with stronger longitudinal increments at large scales, which seems unlikely to arise from deep in the solar corona but could be explained in terms of longitudinal velocity shear associated with partial corotation. 

As an additional step to test this hypothesis,
we carried 
out 3D
high Mach number compressible MHD simulations
as driven by an initial  planar velocity
shear layer with a parallel sheared 
magnetic field and a velocity contrast 
of three times the Alfv\'en speed.
Vortex rollup and 
nonlinear Kelvin-Helmholtz activity is anticipated and 
observed. 
As expected, the magnetic field is deflected, 
sometimes through large angles. Reversals of the field direction, 
corresponding to the phenomenon of
``switchbacks'' in the {\it PSP} observations,
are seen with similar frequency in the simulations and in the 
observations. 
The distributions of transverse velocity,
radial magnetic field, and  magnetic field magnitude all show similarities between 
the simulation results and {\it PSP} observations from the first encounter. 

Several features of the {\it PSP} observations
are related to the familiar appearance of 
high cross-helicity states (Alfv\'enicity) in the inner heliosphere
\citep{BelcherDavis71,BrunoCarbone13}, as well as their 
familiar sense of polarization associated with 
a predominantly outward propagating character. 
Large-amplitude Alfv\'en waves of pure polarization are 
required to be ``spherically polarized'' with the magnetic 
amplitude wandering on a constant magnitude surface \citep{BarnesHollweg74}.
Ordinarily one would associate such states 
with incompressibility. Indeed, even though the 
magnetosonic Mach numbers exceed unity here, the density fluctuations 
are observed to be small \citep{KruparEA20},
presumably 
because the Alfv\'encity is large enough to prohibit 
proliferation of compressive modes. 
Furthermore we observed that the magnitude 
of the magnetic field is relatively constant in patches. 
This may be an additional indicator of 
large scale flux tube structures originating in the 
lower corona and subsequent mixing layers, where transverse pressure balance is approximately 
realized due to quasi-two-dimensional turbulent 
relaxation \citep{ServidioEA08-depress}.
 
Another interesting feature of 
the observations relates to the outward-propagating 
polarization of the  cross-helicity,
viewed in the context of the 
idealized 
configuration illustrated in Figure \ref{fig:cartoon}.
Below the Alfv\'en critical zone,
substantial cross-helicity can be present in the vertical (axial) 
magnetic field of the flux tubes and the excess radial (axial) 
velocity found in some magnetic flux tubes. Both of these
features are inferred from coronagraph observations \citep{DeForestEA18}.
It is interesting that to maintain a sense of outward propagation
at the scale of the (model) flux tubes in a unipolar region, 
the faster flowing flux tubes must 
coexist with weaker axial magnetic fields, while 
flux tubes with slower radial speeds
would have stronger radial magnetic fields. 
This sense of correlation remains the same whether the unipolar region
has positive or negative radial magnetic field.
This sense may dominate for other reasons; for example, strong closed fields may inhibit acceleration of the nascent solar wind.
Apparently the regions with opposite sense of polarization, i.e., 
those that correspond to 
inward propagation,
are not present at significant levels 
during the {\it PSP} first perihelion, which lies 
outside the Alfv\'en critical zone. 
This situation may 
change when {\it PSP} passes through or below the critical zone. 
For example, the inward-type modes may build up in the critical zone
due to sharp Alfv\'en speed gradients 
or due to stagnation. 

It appears that velocity differences between flux tubes 
may be available, under the right conditions, 
to drive large amplitude fluctuations that are responsible for the
transition between striation in the sub-Alfv\'enic inner corona and flocculation
in the super-Alfv\'enic outer corona. This 
phenomenon may be characteristic of what 
we may reasonably call the 
``young solar wind''.  
The region in which this appears to occur is 
outside the Alfv\'en critical zone and near the first $\beta=1$ zone. 
This is where pressure fluctuations become large enough 
to overcome the rigidity of the magnetic field. Vortex rollups and 
large deflections or switchbacks of the magnetic field become possible.
From the first detailed examination of the relevant 
evidence presented here, 
it appears that this hypothesis is reasonable, or at least 
not ruled out. 

The injection of additional turbulence energy due to shear-induced rollup
may set the scale of the energy-containing eddies 
in the region of injection, thus determining
the turbulence 
correlation length observed from about 40 $R_\odot$ outward to 
1 au and beyond. 
In this regard the potential for a significant additional 
injection of energy outside the Alfv\'en critical zone
may act as an ``afterburner'' 
that further boosts heating and acceleration. 

%
While the evidence summarized above appears to 
support the mixing layer hypothesis, it does not  
diminish the potential importance of 
large amplitude fluctuations that originate in the 
lower corona, propagate outward, 
and survive 
into the region where mixing layer dynamics occurs. 
Such fluctuations 
could be generated by field line stirring and reconnection in the
photospheric ``furnace''
that produces braiding of field lines, nanoflares, and a turbulent cascade that 
is probably responsible for heating the corona and accelerating
the wind to supersonic and super-Alfv\'enic speeds
\citep{McKenzieEA95,AxfordEA99,MattEA99-ch,CranmerEA07,VerdiniEA10-accn}. 
There could also be
ejecta from large scale interchange reconnection \citep{FiskKasper20} or 
wavelike fluctuations launched from spicules 
\citep{SamantaEA19}. 
Some fluctuations of these 
types originating from lower altitudes may also 
produce local large-angle magnetic deflections. 
We suspect that it will be difficult to rule out contributions
to large angular deflections due to several 
potential sources. 
In any case, such fluctuations, upon arrival in the Alfv\'en critical zone, 
would contribute to the perturbations that unleash 
the nonlinear magnetized mixing layer phenomena that we describe here. 


{\it Predictions for future perihelia.}
As perihelia move closer and then enter the Alfv\'en critical zone, we expect to observe a further increase in both the mean magnetic field and the amplitude of broadband turbulence.  
As we move closer to the region of ``striations'', the more random fluctuations seen due to rollups should
give way to
more organized patterns of near-radially aligned flux tubes.
These striations contain the velocity shears and magnetic shears that provide the energy for the rollups further along.  
Approaching these more organized magnetic structures, 
we expect the frequency of switchbacks to decrease, and the sharpness of 
velocity jumps to increase as the sub-Alfv\'enic region is approached, assuring greater confinement, and the suppression of Kelvin-Helmholtz-like mixing layer dynamics. 
The amplitude and frequency of occurrence of large fluctuating tangential velocity 
should decrease, while periods of corotation should become more frequent as the plasma 
at lower altitudes
comes under increasing control of the more rigid lower coronal magnetic field. 
In this same region {\it PSP} may begin to see other signatures, for example, component reconnection between adjacent striated flux tubes, or indications of helical field lines within them. It is possible that if {\it PSP} perihelia
lie deep enough in the corona, the striated flux tubes may  display properties such as Beltrami, Alfv\'enic, and force-free signatures that are indicative
of approach to a generalized relaxed state of turbulence
\citep{ServidioEA08-prld}. 
This is directly analogous to our compressible MHD simulation results at very early times 
and the behavior of familiar smoke plumes near the confining 
smoke pipes as seen in Figure \ref{fig:twostacks}.

In this paper we have presented 
a hypothesis regarding the 
role of shear driven dynamics in the region currently explored by {\it PSP} as well as supporting evidence.
As a logical consequence, we develop a set of expectations for what {\it PSP} will observe as its perihelia explore deeper in the Alfv\'en critical zone and below. 
These predictions 
will soon be tested. 
In any case, 
approaching these regions
for first-ever 
{\it in situ} observation, {\it PSP} is 
expected to reveal 
important features of the plasma, electromagnetic, and energetic particle environment 
in the solar corona 
that shape the entire heliosphere.

\acknowledgments 
The authors are grateful to Wiwithawin Charoenngam for plotting assistance.
This work utilizes data produced collaboratively between AFRL/ADAPT and NSO/NISP.
This research has been supported 
 in part by grant RTA6280002 from Thailand Science Research and Innovation and the Parker Solar Probe mission under the 
 ISOIS project 
 (contract NNN06AA01C) and a subcontract 
 to University of Delaware from
 Princeton University (SUB0000165).
 M.L.G. acknowledges support from the Parker Solar Probe FIELDS MAG team.
 \add{Y.Y. and M.W. acknowledge support from NSFC (Grants 11672123,11902138, and 91752201).}
 Additional support is acknowledged from the  NASA LWS program  (NNX17AB79G) and the HSR program (80NSSC18K1210 \& 80NSSC18K1648).



\begin{thebibliography}{}
\expandafter\ifx\csname natexlab\endcsname\relax\def\natexlab#1{#1}\fi
\providecommand{\url}[1]{\href{#1}{#1}}
\providecommand{\dodoi}[1]{doi:~\href{http://doi.org/#1}{\nolinkurl{#1}}}
\providecommand{\doeprint}[1]{\href{http://ascl.net/#1}{\nolinkurl{http://ascl.net/#1}}}
\providecommand{\doarXiv}[1]{\href{https://arxiv.org/abs/#1}{\nolinkurl{https://arxiv.org/abs/#1}}}

\bibitem[{{Arge} {et~al.}(2010){Arge}, {Henney}, {Koller}, {Compeau}, {Young},
  {MacKenzie}, {Fay}, \& {Harvey}}]{Arge_et_al_2010}
{Arge}, C.~N., {Henney}, C.~J., {Koller}, J., {et~al.} 2010, in American
  Institute of Physics Conference Series, Vol. 1216, Twelfth International
  Solar Wind Conference, ed. M.~{Maksimovic}, K.~{Issautier},
  N.~{Meyer-Vernet}, M.~{Moncuquet}, \& F.~{Pantellini}, 343--346

\bibitem[{{Axford} \& {McKenzie}(1992)}]{AxfordMcKenzie92}
{Axford}, W.~I., \& {McKenzie}, J.~F. 1992, in Solar Wind Seven Colloquium, ed.
  E.~{Marsch} \& R.~{Schwenn}, 1--5

\bibitem[{{Axford} {et~al.}(1999){Axford}, {McKenzie}, {Sukhorukova},
  {Banaszkiewicz}, {Czechowski}, \& {Ratkiewicz}}]{AxfordEA99}
{Axford}, W.~I., {McKenzie}, J.~F., {Sukhorukova}, G.~V., {et~al.} 1999, \ssr,
  87, 25, \dodoi{10.1023/A:1005197529250}

\bibitem[{{Bale} {et~al.}(2016){Bale}, {Goetz}, {Harvey}, {Turin}, {Bonnell},
  {Dudok de Wit}, {Ergun}, {MacDowall}, {Pulupa}, {Andre}, {Bolton},
  {Bougeret}, {Bowen}, {Burgess}, {Cattell}, {Chandran}, {Chaston}, {Chen},
  {Choi}, {Connerney}, {Cranmer}, {Diaz-Aguado}, {Donakowski}, {Drake},
  {Farrell}, {Fergeau}, {Fermin}, {Fischer}, {Fox}, {Glaser}, {Goldstein},
  {Gordon}, {Hanson}, {Harris}, {Hayes}, {Hinze}, {Hollweg}, {Horbury},
  {Howard}, {Hoxie}, {Jannet}, {Karlsson}, {Kasper}, {Kellogg}, {Kien},
  {Klimchuk}, {Krasnoselskikh}, {Krucker}, {Lynch}, {Maksimovic}, {Malaspina},
  {Marker}, {Martin}, {Martinez-Oliveros}, {McCauley}, {McComas}, {McDonald},
  {Meyer-Vernet}, {Moncuquet}, {Monson}, {Mozer}, {Murphy}, {Odom},
  {Oliverson}, {Olson}, {Parker}, {Pankow}, {Phan}, {Quataert}, {Quinn},
  {Ruplin}, {Salem}, {Seitz}, {Sheppard}, {Siy}, {Stevens}, {Summers}, {Szabo},
  {Timofeeva}, {Vaivads}, {Velli}, {Yehle}, {Werthimer}, \&
  {Wygant}}]{BaleEA16}
{Bale}, S.~D., {Goetz}, K., {Harvey}, P.~R., {et~al.} 2016, \ssr, 204, 49,
  \dodoi{10.1007/s11214-016-0244-5}

\bibitem[{{Bale} {et~al.}(2019){Bale}, {Badman}, {Bonnell}, {Bowen}, {Burgess},
  {Case}, {Cattell}, {Chandran}, {Chaston}, {Chen}, {Drake}, {de Wit},
  {Eastwood}, {Ergun}, {Farrell}, {Fong}, {Goetz}, {Goldstein}, {Goodrich},
  {Harvey}, {Horbury}, {Howes}, {Kasper}, {Kellogg}, {Klimchuk}, {Korreck},
  {Krasnoselskikh}, {Krucker}, {Laker}, {Larson}, {MacDowall}, {Maksimovic},
  {Malaspina}, {Martinez-Oliveros}, {McComas}, {Meyer-Vernet}, {Moncuquet},
  {Mozer}, {Phan}, {Pulupa}, {Raouafi}, {Salem}, {Stansby}, {Stevens}, {Szabo},
  {Velli}, {Woolley}, \& {Wygant}}]{BaleEA19Nature}
{Bale}, S.~D., {Badman}, S.~T., {Bonnell}, J.~W., {et~al.} 2019, \nat, 576,
  237, \dodoi{10.1038/s41586-019-1818-7}

\bibitem[{Balogh {et~al.}(1999)Balogh, Forsyth, Lucek, Horbury, \&
  Smith}]{BaloghEA99}
Balogh, A., Forsyth, R.~J., Lucek, E.~A., Horbury, T.~S., \& Smith, E.~J. 1999,
  Geophys.\ Res.\ Lett., 26, 631

\bibitem[{Barnes(1979)}]{Barnes79a}
Barnes, A. 1979, in Solar System Plasma Physics, vol. {I}, ed. E.~N. Parker,
  C.~F. Kennel, \& L.~J. Lanzerotti (Amsterdam: North-Holland), 251

\bibitem[{Barnes(1981)}]{Barnes81}
Barnes, A. 1981, J.\ Geophys.\ Res., 86, 7498

\bibitem[{Barnes \& Hollweg(1974)}]{BarnesHollweg74}
Barnes, A., \& Hollweg, J.~V. 1974, J.\ Geophys.\ Res., 79, 2302

\bibitem[{Bavassano {et~al.}(1982)Bavassano, Dobrowolny, Mariani, \&
  Ness}]{BavassanoEA82a}
Bavassano, B., Dobrowolny, M., Mariani, F., \& Ness, N.~F. 1982, J.\ Geophys.\
  Res., 87, 3617, \dodoi{10.1029/JA087iA05p03617}

\bibitem[{Belcher \& Davis~Jr.(1971)}]{BelcherDavis71}
Belcher, J.~W., \& Davis~Jr., L. 1971, J.\ Geophys.\ Res., 76, 3534,
  \dodoi{10.1029/JA076i016p03534}

\bibitem[{{Borovsky}(2016)}]{Borovsky16}
{Borovsky}, J.~E. 2016, Journal of Geophysical Research (Space Physics), 121,
  5055, \dodoi{10.1002/2016JA022686}

\bibitem[{Breech {et~al.}(2008)Breech, Matthaeus, Minnie, Bieber, Oughton,
  Smith, \& Isenberg}]{BreechEA08}
Breech, B., Matthaeus, W.~H., Minnie, J., {et~al.} 2008, J.\ Geophys.\ Res.,
  113, \dodoi{10.1029/2007JA012711}

\bibitem[{{Bruno} \& {Carbone}(2013)}]{BrunoCarbone13}
{Bruno}, R., \& {Carbone}, V. 2013, Living Reviews in Solar Physics, 10, 2,
  \dodoi{10.12942/lrsp-2013-2}

\bibitem[{Chandrasekhar(1981)}]{Chandrasekhar}
Chandrasekhar, S. 1981, Hydrodynamic and Hydromagnetic Stability (New York:
  Dover)

\bibitem[{{Chhiber} {et~al.}(2018){Chhiber}, {Usmanov}, {DeForest},
  {Matthaeus}, {Parashar}, \& {Goldstein}}]{ChhiberEA18-global-floc}
{Chhiber}, R., {Usmanov}, A.~V., {DeForest}, C.~E., {et~al.} 2018, \apjl, 856,
  L39, \dodoi{10.3847/2041-8213/aab843}

\bibitem[{{Chhiber} {et~al.}(2019{\natexlab{a}}){Chhiber}, {Usmanov},
  {Matthaeus}, \& {Goldstein}}]{ChhiberEA19-1}
{Chhiber}, R., {Usmanov}, A.~V., {Matthaeus}, W.~H., \& {Goldstein}, M.~L.
  2019{\natexlab{a}}, Astrophys. J. Suppl., 241, 11,
  \dodoi{10.3847/1538-4365/ab0652}

\bibitem[{{Chhiber} {et~al.}(2019{\natexlab{b}}){Chhiber}, {Usmanov},
  {Matthaeus}, {Parashar}, \& {Goldstein}}]{ChhiberEA19-2}
{Chhiber}, R., {Usmanov}, A.~V., {Matthaeus}, W.~H., {Parashar}, T.~N., \&
  {Goldstein}, M.~L. 2019{\natexlab{b}}, \apjs, 242, 12,
  \dodoi{10.3847/1538-4365/ab16d7}

\bibitem[{{Chhiber} {et~al.}(2020){Chhiber}, {Goldstein}, {Maruca}, {Chasapis},
  {Matthaeus}, {Ruffolo}, {Band yopadhyay}, {Parashar}, {Qudsi}, {de Wit},
  {Bale}, {Bonnell}, {Goetz}, {Harvey}, {MacDowall}, {Malaspina}, {Pulupa},
  {Kasper}, {Korreck}, {Case}, {Stevens}, {Whittlesey}, {Larson}, {Livi},
  {Velli}, \& {Raouafi}}]{ChhiberEA2020ApJS}
{Chhiber}, R., {Goldstein}, M.~L., {Maruca}, B.~A., {et~al.} 2020, \apjs, 246,
  31, \dodoi{10.3847/1538-4365/ab53d2}

\bibitem[{Cranmer {et~al.}(2007)Cranmer, {van}~{Ballegooijen}, \&
  Edgar}]{CranmerEA07}
Cranmer, S.~R., {van}~{Ballegooijen}, A.~A., \& Edgar, R.~J. 2007, Astrophys.\
  J.\ Suppl. Ser, 171, 520, \dodoi{10.1086/518001}

\bibitem[{Crooker {et~al.}(2004)Crooker, Kahler, Larson, \& Lin}]{Crooker_2004}
Crooker, N.~U., Kahler, S.~W., Larson, D.~E., \& Lin, R.~P. 2004, Journal of
  Geophysical Research: Space Physics, 109, \dodoi{10.1029/2003JA010278}

\bibitem[{Dahlburg {et~al.}(1998)Dahlburg, Boncinelli, \&
  Einaudi}]{DahlburgEA98}
Dahlburg, R.~B., Boncinelli, P., \& Einaudi, G. 1998, Physics of Plasmas, 5,
  79, \dodoi{10.1063/1.872677}

\bibitem[{{D'Amicis} {et~al.}(2020){D'Amicis}, {Matteini}, {Bruno}, \&
  {Velli}}]{DAmicisEA20}
{D'Amicis}, R., {Matteini}, L., {Bruno}, R., \& {Velli}, M. 2020, \solphys,
  295, 46, \dodoi{10.1007/s11207-020-01606-2}

\bibitem[{Davies \& Gather(1993)}]{Davies1993Hampel}
Davies, L., \& Gather, U. 1993, Journal of the American Statistical
  Association, 88, 782, \dodoi{10.1080/01621459.1993.10476339}

\bibitem[{{De Pontieu} {et~al.}(2009){De Pontieu}, {McIntosh}, {Hansteen}, \&
  {Schrijver}}]{DePontieuEA09}
{De Pontieu}, B., {McIntosh}, S.~W., {Hansteen}, V.~H., \& {Schrijver}, C.~J.
  2009, \apjl, 701, L1, \dodoi{10.1088/0004-637X/701/1/L1}

\bibitem[{{DeForest} {et~al.}(2018){DeForest}, {Howard}, {Velli}, {Viall}, \&
  {Vourlidas}}]{DeForestEA18}
{DeForest}, C.~E., {Howard}, R.~A., {Velli}, M., {Viall}, N., \& {Vourlidas},
  A. 2018, Astrophys. J., 862, 18, \dodoi{10.3847/1538-4357/aac8e3}

\bibitem[{{DeForest} {et~al.}(2016){DeForest}, {Matthaeus}, {Viall}, \&
  {Cranmer}}]{DeForestEA16}
{DeForest}, C.~E., {Matthaeus}, W.~H., {Viall}, N.~M., \& {Cranmer}, S.~R.
  2016, Astrophys. J., 828, 66, \dodoi{10.3847/0004-637X/828/2/66}

\bibitem[{{Dudok de Wit} {et~al.}(2020){Dudok de Wit}, {Krasnoselskikh},
  {Bale}, {Bonnell}, {Bowen}, {Chen}, {Froment}, {Goetz}, {Harvey},
  {Jagarlamudi}, {Larosa}, {MacDowall}, {Malaspina}, {Matthaeus}, {Pulupa},
  {Velli}, \& {Whittlesey}}]{DudokDeWitEA20}
{Dudok de Wit}, T., {Krasnoselskikh}, V.~V., {Bale}, S.~D., {et~al.} 2020,
  \apjs, 246, 39, \dodoi{10.3847/1538-4365/ab5853}

\bibitem[{{Einaudi} {et~al.}(1999){Einaudi}, {Boncinelli}, {Dahlburg}, \&
  {Karpen}}]{EinaudiEA99}
{Einaudi}, G., {Boncinelli}, P., {Dahlburg}, R.~B., \& {Karpen}, J.~T. 1999,
  \jgr, 104, 521, \dodoi{10.1029/98JA02394}

\bibitem[{{Fisk} \& {Kasper}(2020)}]{FiskKasper20}
{Fisk}, L.~A., \& {Kasper}, J.~C. 2020, \apjl, 894, L4,
  \dodoi{10.3847/2041-8213/ab8acd}

\bibitem[{Fox {et~al.}(2016)Fox, Velli, Bale, Decker, Driesman, Howard, Kasper,
  Kinnison, Kusterer, Lario, {et~al.}}]{FoxSSR16}
Fox, N., Velli, M., Bale, S., {et~al.} 2016, Space Science Reviews, 204, 7

\bibitem[{{Goldstein} {et~al.}(1974){Goldstein}, {Klimas}, \&
  {Barish}}]{GoldsteinEA74}
{Goldstein}, M.~L., {Klimas}, A.~J., \& {Barish}, F.~D. 1974, in Solar Wind
  Three, ed. C.~T. {Russell}, 385--387

\bibitem[{Goldstein {et~al.}(1987)Goldstein, Roberts, Ghosh, \&
  Matthaeus}]{GoldsteinEA87-eslab}
Goldstein, M.~L., Roberts, D.~A., Ghosh, S., \& Matthaeus, W.~H. 1987, in
  Proceedings of the 21st ESLAB Symposium, Bolkesi{\o}, Norway, ed. B.~Battrick
  \& E.~J. Rolfe (Eur.\ Space Agency Spec.\ Publ.\ ESA SP-275), 115

\bibitem[{Goldstein {et~al.}(1989{\natexlab{a}})Goldstein, Roberts, \&
  Matthaeus}]{GoldsteinEA89}
Goldstein, M.~L., Roberts, D.~A., \& Matthaeus, W.~H. 1989{\natexlab{a}}, in
  Solar {S}ytem {P}lasma {P}hysics, Geophys. Monogr. Ser. Bolkesi{\o}, Norway,
  ed. J.~H. Waite, J.~L. Burch, \& R.~L. Moore, Vol.~54 (Washington, D.C.:
  AGU), 113

\bibitem[{Goldstein {et~al.}(1989{\natexlab{b}})Goldstein, Roberts, \&
  Matthaeus}]{GoldsteinSSPP89}
Goldstein, M.~L., Roberts, D.~A., \& Matthaeus, W.~H. 1989{\natexlab{b}}, Solar
  System Plasma Physics, 54, 113

\bibitem[{Hickmann {et~al.}(2015)Hickmann, Godinez, Henney, \&
  Arge}]{Hickmann_et_al_2015}
Hickmann, K.~S., Godinez, H.~C., Henney, C.~J., \& Arge, C.~N. 2015, Solar
  Phys., 290, 1105, \dodoi{10.1007/s11207-015-0666-3}

\bibitem[{Hollweg(1974)}]{Hollweg74}
Hollweg, J.~V. 1974, J.\ Geophys.\ Res., 79, 1539

\bibitem[{{Horbury} {et~al.}(2018){Horbury}, {Matteini}, \&
  {Stansby}}]{HorburyEA18}
{Horbury}, T.~S., {Matteini}, L., \& {Stansby}, D. 2018, \mnras, 478, 1980,
  \dodoi{10.1093/mnras/sty953}

\bibitem[{{Horbury} {et~al.}(2020){Horbury}, {Woolley}, {Laker}, {Matteini},
  {Eastwood}, {Bale}, {Velli}, {Chandran}, {Phan}, {Raouafi}, {Goetz},
  {Harvey}, {Pulupa}, {Klein}, {Dudok de Wit}, {Kasper}, {Korreck}, {Case},
  {Stevens}, {Whittlesey}, {Larson}, {MacDowall}, {Malaspina}, \&
  {Livi}}]{HorburyEA20}
{Horbury}, T.~S., {Woolley}, T., {Laker}, R., {et~al.} 2020, \apjs, 246, 45,
  \dodoi{10.3847/1538-4365/ab5b15}

\bibitem[{{Kahler} {et~al.}(1996){Kahler}, {Crooker}, \&
  {Gosling}}]{KahlerEA96}
{Kahler}, S.~W., {Crooker}, N.~U., \& {Gosling}, J.~T. 1996, \jgr, 101, 24373,
  \dodoi{10.1029/96JA02232}

\bibitem[{{Kasper} {et~al.}(2016){Kasper}, {Abiad}, {Austin}, {Balat-Pichelin},
  {Bale}, {Belcher}, {Berg}, {Bergner}, {Berthomier}, {Bookbinder}, {Brodu},
  {Caldwell}, {Case}, {Chand ran}, {Cheimets}, {Cirtain}, {Cranmer}, {Curtis},
  {Daigneau}, {Dalton}, {Dasgupta}, {DeTomaso}, {Diaz-Aguado}, {Djordjevic},
  {Donaskowski}, {Effinger}, {Florinski}, {Fox}, {Freeman}, {Gallagher},
  {Gary}, {Gauron}, {Gates}, {Goldstein}, {Golub}, {Gordon}, {Gurnee}, {Guth},
  {Halekas}, {Hatch}, {Heerikuisen}, {Ho}, {Hu}, {Johnson}, {Jordan},
  {Korreck}, {Larson}, {Lazarus}, {Li}, {Livi}, {Ludlam}, {Maksimovic},
  {McFadden}, {Marchant}, {Maruca}, {McComas}, {Messina}, {Mercer}, {Park},
  {Peddie}, {Pogorelov}, {Reinhart}, {Richardson}, {Robinson}, {Rosen},
  {Skoug}, {Slagle}, {Steinberg}, {Stevens}, {Szabo}, {Taylor}, {Tiu}, {Turin},
  {Velli}, {Webb}, {Whittlesey}, {Wright}, {Wu}, \& {Zank}}]{KasperEA16}
{Kasper}, J.~C., {Abiad}, R., {Austin}, G., {et~al.} 2016, \ssr, 204, 131,
  \dodoi{10.1007/s11214-015-0206-3}

\bibitem[{{Kasper} {et~al.}(2019){Kasper}, {Bale}, {Belcher}, {Berthomier},
  {Case}, {Chandran}, {Curtis}, {Gallagher}, {Gary}, {Golub}, {Halekas}, {Ho},
  {Horbury}, {Hu}, {Huang}, {Klein}, {Korreck}, {Larson}, {Livi}, {Maruca},
  {Lavraud}, {Louarn}, {Maksimovic}, {Martinovic}, {McGinnis}, {Pogorelov},
  {Richardson}, {Skoug}, {Steinberg}, {Stevens}, {Szabo}, {Velli},
  {Whittlesey}, {Wright}, {Zank}, {MacDowall}, {McComas}, {McNutt}, {Pulupa},
  {Raouafi}, \& {Schwadron}}]{KasperEA19Nature}
{Kasper}, J.~C., {Bale}, S.~D., {Belcher}, J.~W., {et~al.} 2019, \nat, 576,
  228, \dodoi{10.1038/s41586-019-1813-z}

\bibitem[{Krupar {et~al.}(2020)Krupar, Szabo, Maksimovic, Kruparova, Kontar,
  Balmaceda, Bonnin, Bale, Pulupa, Malaspina, {et~al.}}]{KruparEA20}
Krupar, V., Szabo, A., Maksimovic, M., {et~al.} 2020, The Astrophysical Journal
  Supplement Series, 246, 57

\bibitem[{{Landi} {et~al.}(2006){Landi}, {Hellinger}, \& {Velli}}]{LandiEA06}
{Landi}, S., {Hellinger}, P., \& {Velli}, M. 2006, \grl, 33, L14101,
  \dodoi{10.1029/2006GL026308}

\bibitem[{Landi {et~al.}(2005)Landi, Velli, \& Einaudi}]{LandiEA05}
Landi, S., Velli, M., \& Einaudi, G. 2005, Astrophys.\ J., 624, 392

\bibitem[{{Lau} \& {Liu}(1980)}]{LauLiu80}
{Lau}, Y.~Y., \& {Liu}, C.~S. 1980, Physics of Fluids, 23, 939,
  \dodoi{10.1063/1.863100}

\bibitem[{{Lockwood} {et~al.}(2019){Lockwood}, {Owens}, \&
  {Macneil}}]{LockwoodEA19}
{Lockwood}, M., {Owens}, M.~J., \& {Macneil}, A. 2019, \solphys, 294, 85,
  \dodoi{10.1007/s11207-019-1478-7}

\bibitem[{Lotova {et~al.}(1985)Lotova, Blums, \& Vladimirskii}]{LotovaEA85}
Lotova, N.~A., Blums, D.~F., \& Vladimirskii, K.~V. 1985, Astron.\ Astrophys.,
  150, 266

\bibitem[{{Lotova} {et~al.}(2011){Lotova}, {Vladimirskii}, \&
  {Obridko}}]{LotovaEA11}
{Lotova}, N.~A., {Vladimirskii}, K.~V., \& {Obridko}, V.~N. 2011, \solphys,
  269, 129, \dodoi{10.1007/s11207-010-9686-1}

\bibitem[{{Macneil} {et~al.}(2020){Macneil}, {Owens}, {Wicks}, {Lockwood},
  {Bentley}, \& {Lang}}]{MacneilEA20}
{Macneil}, A.~R., {Owens}, M.~J., {Wicks}, R.~T., {et~al.} 2020, \mnras, 494,
  3642, \dodoi{10.1093/mnras/staa951}

\bibitem[{Malagoli {et~al.}(1996)Malagoli, Bodo, \& Rosner}]{MalagoliEA96}
Malagoli, A., Bodo, G., \& Rosner, R. 1996, Astrophys. J., {\bf 456}, 708

\bibitem[{Matteini {et~al.}(2015)Matteini, Horbury, Pantellini, Velli, \&
  Schwartz}]{MatteiniEA15}
Matteini, L., Horbury, T.~S., Pantellini, F., Velli, M., \& Schwartz, S.~J.
  2015, ApJ, 802, 11, \dodoi{10.1088/0004-637x/802/1/11}

\bibitem[{{Matteini} {et~al.}(2018){Matteini}, {Stansby}, {Horbury}, \&
  {Chen}}]{MatteiniEA18}
{Matteini}, L., {Stansby}, D., {Horbury}, T.~S., \& {Chen}, C.~H.~K. 2018,
  \apjl, 869, L32, \dodoi{10.3847/2041-8213/aaf573}

\bibitem[{{Matteini} {et~al.}(2019){Matteini}, {Stansby}, {Horbury}, \&
  {Chen}}]{MatteiniEA19}
---. 2019, Nuovo Cimento C Geophysics Space Physics C, 42, 16,
  \dodoi{10.1393/ncc/i2019-19016-y}

\bibitem[{{Matthaeus} {et~al.}(2008){Matthaeus}, {Pouquet}, {Mininni},
  {Dmitruk}, \& {Breech}}]{MatthaeusEA08}
{Matthaeus}, W.~H., {Pouquet}, A., {Mininni}, P.~D., {Dmitruk}, P., \&
  {Breech}, B. 2008, Physical Review Letters, 100, 085003,
  \dodoi{10.1103/PhysRevLett.100.085003}

\bibitem[{Matthaeus {et~al.}(1999)Matthaeus, Zank, Oughton, Mullan, \&
  Dmitruk}]{MattEA99-ch}
Matthaeus, W.~H., Zank, G.~P., Oughton, S., Mullan, D.~J., \& Dmitruk, P. 1999,
  Astrophys.\ J., 523, L93

\bibitem[{Mc{Comb}(1990)}]{McComb}
Mc{Comb}, W.~D. 1990, The Physics of Fluid Turbulence (New York: Oxford
  Univeristy Press)

\bibitem[{Mc{K}enzie {et~al.}(1995)Mc{K}enzie, Banaszkiewicz, \&
  Axford}]{McKenzieEA95}
Mc{K}enzie, J.~F., Banaszkiewicz, M., \& Axford, W.~I. 1995, Astron.\
  Astrophys., 303, L45

\bibitem[{Michel(1967)}]{Michel_1967}
Michel, F.~C. 1967, J. Geophysical Res., 72, 1

\bibitem[{{Miura}(1982)}]{Miura82}
{Miura}, A. 1982, \prl, 49, 779, \dodoi{10.1103/PhysRevLett.49.779}

\bibitem[{{Miura} \& {Pritchett}(1982)}]{MiuraPritchett82}
{Miura}, A., \& {Pritchett}, P.~L. 1982, \jgr, 87, 7431,
  \dodoi{10.1029/JA087iA09p07431}

\bibitem[{Moffatt(1978)}]{Moffatt}
Moffatt, H.~K. 1978, Magnetic Field Generation in Electrically Conducting
  Fluids (New York: Cambridge University Press)

\bibitem[{{Mozer} {et~al.}(2020){Mozer}, {Agapitov}, {Bale}, {Bonnell}, {Case},
  {Chaston}, {Curtis}, {Wit}, {Goetz}, {Goodrich}, {Harvey}, {Kasper},
  {Korreck}, {Krasnoselskikh}, {Larson}, {Livi}, {MacDowall}, {Malaspina},
  {Pulupa}, {Stevens}, {Whittlesey}, \& {Wygant}}]{MozerEA20}
{Mozer}, F.~S., {Agapitov}, O.~V., {Bale}, S.~D., {et~al.} 2020, \apjs, 246,
  68, \dodoi{10.3847/1538-4365/ab7196}

\bibitem[{M{\"u}ller {et~al.}(2013)M{\"u}ller, Marsden, St.~Cyr, Gilbert, \&
  {The Solar Orbiter Team}}]{Muller2013SP}
M{\"u}ller, D., Marsden, R.~G., St.~Cyr, O.~C., Gilbert, H.~R., \& {The Solar
  Orbiter Team}. 2013, Solar Physics, 285, 25,
  \dodoi{10.1007/s11207-012-0085-7}

\bibitem[{Osman {et~al.}(2011)Osman, Wan, Matthaeus, Breech, \&
  Oughton}]{OsmanEA11-align}
Osman, K.~T., Wan, M., Matthaeus, W.~H., Breech, B., \& Oughton, S. 2011,
  Astrophys.\ J., 741, 75, \dodoi{10.1088/0004-637X/741/2/75}

\bibitem[{{Owens} {et~al.}(2020){Owens}, {Lockwood}, {Macneil}, \&
  {Stansby}}]{OwensEA20}
{Owens}, M., {Lockwood}, M., {Macneil}, A., \& {Stansby}, D. 2020, \solphys,
  295, 37, \dodoi{10.1007/s11207-020-01601-7}

\bibitem[{{Parashar} {et~al.}(2020){Parashar}, {Goldstein}, {Maruca},
  {Matthaeus}, {Ruffolo}, {Bandyopadhyay}, {Chhiber}, {Chasapis}, {Qudsi},
  {Vech}, {Roberts}, {Bale}, {Bonnell}, {de Wit}, {Goetz}, {Harvey},
  {MacDowall}, {Malaspina}, {Pulupa}, {Kasper}, {Korreck}, {Case}, {Stevens},
  {Whittlesey}, {Larson}, {Livi}, {Velli}, \& {Raouafi}}]{ParasharEA20ApJS}
{Parashar}, T.~N., {Goldstein}, M.~L., {Maruca}, B.~A., {et~al.} 2020, \apjs,
  246, 58, \dodoi{10.3847/1538-4365/ab64e6}

\bibitem[{P\'ecseli(2020)}]{Pecseli20}
P\'ecseli, H.~L. 2020, Waves and Oscillations in Plasmas, Series in Plasma
  Physics (CRC Press).
\newblock \url{https://books.google.co.th/books?id=HMjhDwAAQBAJ}

\bibitem[{{Rappazzo} {et~al.}(2005){Rappazzo}, {Velli}, {Einaudi}, \&
  {Dahlburg}}]{RappazzoEA05}
{Rappazzo}, A.~F., {Velli}, M., {Einaudi}, G., \& {Dahlburg}, R.~B. 2005, \apj,
  633, 474, \dodoi{10.1086/431916}

\bibitem[{{Roberts} {et~al.}(1992){Roberts}, {Goldstein}, {Matthaeus}, \&
  {Ghosh}}]{RobertsJGR92}
{Roberts}, D.~A., {Goldstein}, M.~L., {Matthaeus}, W.~H., \& {Ghosh}, S. 1992,
  \jgr, 97, 17115, \dodoi{10.1029/92JA01144}

\bibitem[{Rogers \& Moser(1992)}]{RogersMoser92}
Rogers, M.~M., \& Moser, R.~D. 1992, Journal of Fluid Mechanics, 243,
  183–226, \dodoi{10.1017/S0022112092002696}

\bibitem[{{Samanta} {et~al.}(2019){Samanta}, {Tian}, {Yurchyshyn}, {Peter},
  {Cao}, {Sterling}, {Erd{\'e}lyi}, {Ahn}, {Feng}, {Utz}, {Banerjee}, \&
  {Chen}}]{SamantaEA19}
{Samanta}, T., {Tian}, H., {Yurchyshyn}, V., {et~al.} 2019, Science, 366, 890,
  \dodoi{10.1126/science.aaw2796}

\bibitem[{Servidio {et~al.}(2008)Servidio, Matthaeus, \&
  Dmitruk}]{ServidioEA08-depress}
Servidio, S., Matthaeus, W.~H., \& Dmitruk, P. 2008, Phys.\ Rev.\ Lett., 100,
  \dodoi{10.1103/PhysRevLett.100.095005}

\bibitem[{{Servidio} {et~al.}(2008){Servidio}, {Matthaeus}, \&
  {Dmitruk}}]{ServidioEA08-prld}
{Servidio}, S., {Matthaeus}, W.~H., \& {Dmitruk}, P. 2008, Physical Review
  Letters, 100, 095005, \dodoi{10.1103/PhysRevLett.100.095005}

\bibitem[{{Shi} {et~al.}(2020){Shi}, {Velli}, {Tenerani}, {Rappazzo}, \&
  {R{\'e}ville}}]{ShiEA20}
{Shi}, C., {Velli}, M., {Tenerani}, A., {Rappazzo}, F., \& {R{\'e}ville}, V.
  2020, \apj, 888, 68, \dodoi{10.3847/1538-4357/ab5fce}

\bibitem[{{Squire} {et~al.}(2020){Squire}, {Chandran}, \&
  {Meyrand}}]{SquireEA20}
{Squire}, J., {Chandran}, B.~D.~G., \& {Meyrand}, R. 2020, \apjl, 891, L2,
  \dodoi{10.3847/2041-8213/ab74e1}

\bibitem[{{Usmanov} {et~al.}(2014){Usmanov}, {Goldstein}, \&
  {Matthaeus}}]{UsmanovEA14}
{Usmanov}, A.~V., {Goldstein}, M.~L., \& {Matthaeus}, W.~H. 2014, Astrophys.
  J., 788, 43, \dodoi{10.1088/0004-637X/788/1/43}

\bibitem[{{Usmanov} {et~al.}(2016){Usmanov}, {Goldstein}, \&
  {Matthaeus}}]{UsmanovEA16}
---. 2016, Astrophys. J., 820, 17, \dodoi{10.3847/0004-637X/820/1/17}

\bibitem[{{Usmanov} {et~al.}(2018){Usmanov}, {Matthaeus}, {Goldstein}, \&
  {Chhiber}}]{UsmanovEA18}
{Usmanov}, A.~V., {Matthaeus}, W.~H., {Goldstein}, M.~L., \& {Chhiber}, R.
  2018, Astrophys. J., 865, 25, \dodoi{10.3847/1538-4357/aad687}

\bibitem[{Verdini {et~al.}(2010)Verdini, Velli, Matthaeus, Oughton, \&
  Dmitruk}]{VerdiniEA10-accn}
Verdini, A., Velli, M., Matthaeus, W.~H., Oughton, S., \& Dmitruk, P. 2010,
  Astrophys.\ J., 708, L116, \dodoi{10.1088/2041-8205/708/2/L116}

\bibitem[{{Yang} {et~al.}(2016){Yang}, {Wan}, {Shi}, {Yang}, \&
  {Chen}}]{YangEAJCP16}
{Yang}, Y., {Wan}, M., {Shi}, Y., {Yang}, K., \& {Chen}, S. 2016, Journal of
  Computational Physics, 306, 73, \dodoi{10.1016/j.jcp.2015.11.025}

\bibitem[{Zank {et~al.}(1996)Zank, Matthaeus, \& Smith}]{ZankEA96}
Zank, G.~P., Matthaeus, W.~H., \& Smith, C.~W. 1996, J.\ Geophys.\ Res., 101,
  17\,093

\end{thebibliography}

 \newcommand{\BIBand} {and} 
  \newcommand{\boldVol}[1] {\textbf{#1}} 
  \providecommand{\SortNoop}[1]{} 
  \providecommand{\sortnoop}[1]{} 
  \newcommand{\stereo} {\emph{{S}{T}{E}{R}{E}{O}}} 
  \newcommand{\au} {{A}{U}\ } 
  \newcommand{\AU} {{A}{U}\ } 
  \newcommand{\MHD} {{M}{H}{D}\ } 
  \newcommand{\mhd} {{M}{H}{D}\ } 
  \newcommand{\RMHD} {{R}{M}{H}{D}\ } 
  \newcommand{\rmhd} {{R}{M}{H}{D}\ } 
  \newcommand{\wkb} {{W}{K}{B}\ } 
  \newcommand{\alfven} {{A}lfv{\'e}n\ } 
  \newcommand{\alfvenic} {{A}lfv{\'e}nic\ } 
  \newcommand{\Alfven} {{A}lfv{\'e}n\ } 
  \newcommand{\Alfvenic} {{A}lfv{\'e}nic\ }

\end{document}